\documentclass[11pt,reqno]{amsart}

\usepackage{etex}
\usepackage{amsmath}
\usepackage{amssymb}
\usepackage{amsfonts}
\usepackage{amsthm}
\usepackage{a4wide}
\usepackage{bm}
\usepackage{graphics}
\usepackage{color}
\usepackage[sans]{dsfont}
\usepackage[toc,page]{appendix}

\usepackage{hyperref}
\usepackage{protosem}
\usepackage{linearA}
\usepackage{pgfkeys}
\usepackage{longtable}
\usepackage{pdflscape}
\usepackage{float}
\usepackage{cancel}
\usepackage{dsfont}
\usepackage{setspace}
\usepackage{array}
\usepackage{hyperref}
\usepackage{appendix}
\usepackage{amssymb}
\usepackage{epstopdf}
 \usepackage{multirow}
\usepackage{epsfig}
\usepackage{lscape}

\usepackage{yhmath}
\usepackage[english]{babel}
\usepackage[latin1]{inputenc}
\usepackage{wasysym}
\usepackage{tikz,pgflibraryshapes}
\usepackage{pstricks,pst-plot,pstricks-add}
\usepackage{enumerate}
\usetikzlibrary{arrows}
\usetikzlibrary{snakes}
\usetikzlibrary {shapes}
\usepackage{textcomp}
\usepackage{mathrsfs}
\usepackage[toc,page]{appendix}
\usepackage{enumitem}

\newtheorem{theorem}{Theorem}[subsection]
\newtheorem{lemma}{Lemma}[subsection]
\newtheorem{proposition}{Proposition}[subsection]

\newtheorem{corollary}{Corollary}[subsection]
\newtheorem{definition}{Definition}[subsection]

\newtheorem{assumption}[theorem]{Assumption}

\newtheorem{remark}{Remark}[subsection]

\numberwithin{equation}{section}

\numberwithin{equation}{section}

\begin{document}
\title{The preparation of states in quantum mechanics}

\author[J. Fr\"ohlich]{J\"urg Fr\"ohlich}
\address[J. Fr{\"o}hlich]{Institut f{\"u}r Theoretische Physik, ETH H{\"o}nggerberg, CH-8093 Z{\"u}rich, Switzerland}
\email{juerg@phys.ethz.ch}
\author[B. Schubnel]{Baptiste Schubnel}
\address[B. Schubnel]{Departement Mathematik, ETH Z{\"u}rich, CH-8092 Z{\"u}rich, Switzerland}
\email{baptiste.schubnel@math.ethz.ch}

\date{\today} 
\maketitle
\normalsize
\begin{abstract}
The important problem of how to prepare a quantum mechanical system, $S$, in a specific initial state of interest - e.g., for the purposes of some experiment - is addressed. Three distinct methods of state preparation are described. One of these methods has the attractive feature that it enables one to prepare $S$ in a preassigned initial state with certainty; i.e., the probability of success in preparing $S$ in a given state is unity. This method relies on coupling $S$ to an open quantum-mechanical environment, $E$, in such a way that the dynamics of $S \vee E$ pulls the state of $S$ towards an ``attractor", which is the desired initial state of $S$. This method is analyzed in detail.
\end{abstract}

\section{Aim of the Paper, Models and Summary of Results}
The main problem addressed in this paper is how one may go about preparing a given, spatially localized quantum-mechanical system, $S$, in a specific initial state of interest in performing some observation or experiment on $S$. It is our impression that it is difficult to find serious discussion and analysis of this important foundational problem in the literature. In particular, it appears to be ignored in most text books on introductory quantum mechanics. The purpose of our paper is  to make a modest contribution towards elucidating some solutions of this problem. 

After sketching several alternative techniques that can be used to prepare $S$ in a desired initial state, we will turn our attention to the method studied primarily in this paper: By turning on suitable external fields, etc., we attempt to tune the dynamics of $S$ so as to have the property that the state we want $S$ to prepare in, denoted $\Omega_S$, is the \textit{ground state} of the given dynamics; we then weakly couple $S$ to a dispersive environment, $E$, (e.g., the quantized lattice vibrations of a crystal, or the electromagnetic field) chosen in such a way that, in the vicinity of $S$, the composed system, $S\vee E$, relaxes to the ground state of $S\vee E$. By letting the strength of the interaction between $S$ and $E$ tend to zero sufficiently slowly in time, we can manage to asymptotically decouple $S$ from $E$ and have $S$ approach its own ground state, which is the desired state $\Omega_S$, as time $t$ tends to $\infty$. The method for preparing a quantum-mechanical system in a specific state sketched here has the advantage that it is very robust: It has the attractive property that $S$ approaches the desired state $\Omega_S$ \textit{with probability $1$}, as time tends to $\infty$. Moreover, the speed of approach of the state of $S$ to $\Omega_S$ can be estimated quite explicitly. (In the following, $\omega$ denotes the expectation with respect to a state $\Omega$, i.e., $\omega(\cdot) = 
\langle \Omega, (\cdot) \Omega \rangle$, and we will also use the expression ``state'' for $\omega$.)

Instead of engaging in a general abstract discussion of the problem of preparation of states in quantum mechanics, we explain our ideas and insights on the rather concrete example of a system $S$ with a finite-dimensional Hilbert space of pure state vectors, a simple caricature of a very heavy ``atom'', coupled to a free massless scalar quantum field (e.g., a quantized field of phonons or ``photons''). Mathematically, our analysis of this class of examples is based on methods introduced in \cite{FrDe} and further developed in \cite{DeR1} and \cite{DeR2}. Although the models discussed in this paper look very simple and idealized, the analysis of state preparation presented here is  somewhat intricate, mathematically. We believe that our analysis highlights many characteristic features of this particular method of state preparation.

Before  explaining our main results in more detail,  we propose to describe three alternative techniques that can be used to prepare a quantum-mechanical system $S$ in a specific initial state $\omega_S$.

\subsection{Three different methods for state preparation in quantum mechanics}

The first method described below is the one that we will subsequently explore in detail. We then sketch a technique based on the use of time-dependent Hamiltonians controlled from the outside; see, e.g., \cite{Bruder}. Finally, we describe the frequently used method to produce many essentially identical copies of the system $S$ and performing state selection with the help of projective measurements of some physical quantity, $A=A^{*}$, of $S$ and subsequently keeping only those copies of $S$ that correspond to a specific eigenstate of $A$ we want $S$ to prepare in; see \cite{FS} and references given there.

\subsubsection{ Quantum state preparation via weak interaction with a dispersive environment }
Let $\overline{S} := S \vee E$ denote the composition of $S$ with an ``environment'' $E$, let 
$\mathcal{H}_{\overline{S}} = \mathcal{H}_S \otimes \mathcal{H}_E$ denote the Hilbert space of pure state vectors of $\overline{S}$, and let $\mathcal{B}(\mathcal{H})$ denote the algebra of all bounded operators on a Hilbert space $\mathcal{H}$. We assume that $E$ is chosen in such a way that, to a good approximation, 
$\overline{S}$ can be considered to be a ``closed system''; see \cite{FFS}. Let $\mathcal{A}_{\overline{S}} \subseteq \mathcal{B}(\mathcal{H}_{\overline{S}})$ denote the kinematical algebra of operators (including all operators representing physical quantities or ``observables'' of $\overline{S}$) used to describe $\overline{S}$. To say that $\overline{S}$ is a closed system amounts to assuming that the time evolution of all operators in the algebra $\mathcal{A}_{\overline{S}}$ in the Heisenberg picture is given by  invertible linear maps, $\alpha_{t,s}$, from $\mathcal{A}_{\overline{S}}$ onto $\mathcal{A}_{\overline{S}}$ with the properties that 
$$
\alpha_{t,s} \circ \alpha_{s,r} = \alpha_{t,r},
$$
$$
\alpha_{t,s}(A\cdot B) = \alpha_{t,s}(A)\cdot \alpha_{t,s}(B),
$$
and
$$
\alpha_{t,s}(A^{*}) = (\alpha_{t,s}(A))^{*}.
$$
If $A=A^{*} \in \mathcal{A}_{\overline{S}}$ represents a physical quantity of $\overline{S}$ at some time $s$ then 
$\alpha_{t,s}(A)$ represents the same physical quantity at a different time $t$. Maps $\lbrace \alpha_{t,s} \rbrace_{t,s \in \mathbb{R}}$ with the properties specified above are called $^{*}$automorphisms of $\mathcal{A}_{\overline{S}}$.
Let $\mathcal{A}_S$ be the algebra of operators generated by all physical quantities or ``observables'' of the system $S$. In the examples of systems $S$ studied in this paper, one can take $\mathcal{A}_S$ to be given by $\mathcal{B}(\mathcal{H}_S)$. (Clearly $S$ is usually \textit{not} a closed system, and the time evolution automorphsims $\alpha_{t,s}$ do \textit{not} map $\mathcal{A}_S$ into itself.)  

Let $\mathcal{S}_{0}$ be a suitably large subset of  states of $\overline{S}$; (i.e., of positive normalized linear functionals on $\mathcal{A}_{\overline{S}}$, which, for the purpose of this discussion, can be thought of as density matrices on $\mathcal{H}_{\overline{S}}$). Let $\omega_S$ be some state of the system $S$, i.e., a density matrix on $\mathcal{B}(\mathcal{H}_S)$. We say that $S$ can be prepared in the state $\omega_S$ with \textit{certainty} with respect to $\mathcal{S}_{0}$ if and only if, the time evolution 
$\alpha_{t,s}, t,s \text{   } \text{in} \text{   } \mathbb{R},$ of operators in $\mathcal{A}_{\overline{S}}$ 
can be tuned in such a way that, for all operators 
$A = a\otimes \mathds{1}$, with $a \in \mathcal{A}_S$,
\begin{equation} \label{limdisp}
\lim_{t \rightarrow \infty} \omega(\alpha_{t,0}(A)) =\omega_S(a), \qquad  \qquad  \forall \omega
 \in\mathcal{S}_0.
\end{equation}
Our main aim, in the present work, is to prove property \eqref{limdisp}  for a class of models of systems $S$ with a finite-dimensional state space $\mathcal{H}_S$ (``atoms'') weakly coupled to an environment $E$ that consists of a massless free scalar quantum field (``photons''); such models are sometimes called ``generalized spin-boson models''. We will assume that the coupling strength, $\lambda(t)$, between $S$ and $E$ tends to $0$ slowly in time $t$, with $\lambda(0)$ small enough. The Hamiltonian $H_S$ generating the time evolution of the system $S$ before it is coupled to $E$ is chosen in such a way that the state $\omega_S$ 
of $S$ to be approached, as time $t$ becomes large, in the sense of \eqref{limdisp}, is the eigenstate of  $H_S$ corresponding to the lowest eigenvalue (the ground-state energy) of $H_S$. We will describe conditions that guarantee that Eq. (\ref{limdisp}) holds if $E$ is prepared in states corresponding to zero temperature.
If, initially, the environment $E$ is in a thermal equilibrium state at some temperature $T>0$ then $\omega_S$ turns out to be the canonical equilibrium state of $S$ at the same temperature $T>0$ corresponding to the Hamiltonian $H_S$. The  method developed in this paper is powerful enough to cover a broad class of time evolutions and apply to a very large set, $\mathcal{S}_{0}$, of initial states of $\overline{S}$. Our method can also be applied to \textit{autonomous} systems $\overline{S}=S \vee E$, with $\lambda(t)= \lambda_0$, $\forall t$, where $\lambda_0$ is assumed to be small. In this case, the limiting state in \eqref{limdisp} is a small perturbation of the ground state of $H_S$. Using an implicit function theorem argument, one can  then choose $H_S$ such that the limiting state  in \eqref{limdisp} is exactly the state we wanted to reach. 

\subsubsection{Preparation of states via adiabatic evolution}
We consider a quantum-mechanical system $S$ with a Hilbert space $\mathcal{H}_S$ of pure state vectors, and we assume that we know how to prepare the system in an initial state $\Omega \equiv \Omega(0) \in 
\mathcal{H}_S$, where $\Omega$ is the ground state of a Hamiltonian $H(0)$ generating the time evolution of states of $S$ at times $t\leq 0$. (For example, one may make use of the state preparation procedure outlined in the previous subsection to prepare $S$ in the state $\Omega$ at an early time). We would like to find out how, 
at a later time, one might manage to prepare $S$ in a state, $\Omega_{S} \equiv \Omega(1) \neq \Omega(0)$, 
of interest for the purpose of some observations or experiments. The idea explored in this subsection is to make use of \textit{adiabatic evolution} to transform the initial state $\Omega$ into the desired state $\Omega_{S}$.
By turning on suitable time-dependent external fields one may be able to tune the time evolution of $S$ to be
given by a family of time-dependent Hamiltonians $\lbrace H(s) \rbrace_{s \in\mathbb{R}}$ with the property that 
$\Omega_{S} = \Omega(1)$ is the ground state of the operator $H(1)$.

Given the family of Hamiltonians $\lbrace H(s) \rbrace_{s\in\mathbb{R}}$ and a time-scale parameter $\tau > 0$, the time evolution of state vectors in $\mathcal{H}_S$ from an initial time $t_0$ to time $t$ is given by a unitary propagator $U(t,t_0)$ that solves the equation
\begin{equation}
\label{adia}
\frac{d}{dt}U(t,t_0) = -iH(t/\tau) U(t,t_0), \qquad U(t,t)=\mathds{1},
\end{equation}
for all $t_0, t \in \mathbb{R}$. 
The so-called \textit{adiabatic limit} is the limit where $\tau$ tends to $\infty$.

In order to investigate the adiabatic limit mathematically, one has to require some assumptions on the Hamiltonians $H(s)$, $s\in \mathbb{R}$, (see, e.g., \cite{Avron-Elg}, \cite{Teufel} and  \cite{AbouSalFro}): We assume that all the operators $H(s)$ are self-adjoint on a common dense domain $\mathcal{D} \subset \mathcal{H}_S$, that the resolvents $R(s,i):=(H(s)-i)^{-1}$ are differentiable in $s$, with norm-bounded derivatives, and that the operators $H(s) \frac{d}{ds}R(s,i)$ are bounded uniformly in $s\in\mathbb{R}$. We assume that all the Hamiltonians $H(s), s \in \mathbb{R},$ have a non-degenerate ground state energy, 
$e(s)$, corresponding to a ground state eigenvector $\Omega(s)\in\mathcal{H}_S$, with 
$\Omega(0) = \Omega$. The projections $P(s):= \vert\Omega(s)><\Omega(s)\vert$ are assumed to be twice continuously differentiable in $s$, with norm-bounded first and second derivatives.
After rescaling the time $t$ by setting $s=t/\tau$, with $s_0=t_{0}/\tau$, Eq. \eqref{adia} takes the form 
\begin{equation}
\frac{d}{ds} U_{\tau}(s,s_0)=- i \tau H(s)U_{\tau}(s,s_0), 
\end{equation}
where $U_{\tau}(s,s_0)=U(\tau s, \tau s_0)$, and one can prove that
  \begin{equation}
  \label{adia2}
\underset{s \in [0,1]}{\sup} \| U_{\tau}(s,0) \Omega - \Omega(s) \| \underset{\tau \rightarrow \infty}{\longrightarrow} 0,
\end{equation}
see \cite{Avron-Elg}, \cite{Teufel}.

Eq. \eqref{adia2} tells us that it is possible to drive  $S$  from an initial state $\Omega(0)=\Omega$ to the desired state $\Omega(1)=\Omega_{S}$  adiabatically. Thus, $S$ can be prepared in the state $\Omega_{S}$. The drawback of this method is that it presupposes our ability to initially prepare $S$ in the ground state 
$\Omega$
of the Hamiltonian $H(0)$ and that suitable external fields must be turned on that lead to a family of time-dependent Hamiltonians slowly driving $\Omega(0)$ to the desired state $\Omega_S$.

There are variants of this method of state preparation that are \textit{not} based on very slow evolution (i.e., do not involve an adiabatic limit) but require some kind of ``optimal control'' used to construct a family of time-dependent Hamiltonians that determine a propagator driving $S$ from its initial state $\Omega(0)$ to the desired state 
$\Omega_{S}$ in as short a time as possible; see, e.g., \cite{Ruschhaupt}.
  
  \subsubsection{State preparation via duplication of systems and state selection}
We next sketch a method for state preparation that is presumably most often used in practice: One attempts to create a large number, $n$, of independent copies of the system $S$, which we denote by $S_i$, $i=1,...,n$. The closed system  
$\overline{S}=S_1 \vee S_2... \vee S_n \vee E$ 
is the union of $n$ copies of $S$ all of which are successively coupled to a measuring device $E$. 
The purpose of coupling the systems $S_1, ..., S_n$ to the device $E$ is to perform projective measurements of a physical quantity, represented by a self-adjoint operator $A=A^{*}$, common to $S_1, ...,S_n$, one of whose eigenvectors is the state, $\Omega_{S}$, 
in which we want to prepare the system $S$. After a projective measurement of $A$, the system $S_{i}$ is in an eigenstate, $\Omega_{k_{i}}$, of the operator $A$. Whenever $\Omega_{k_{i}} \neq \Omega_{S}$, the system $S_{i}$ is thrown into the waste basket. However, if, in the $i^{th}$ measurement of $A$, the measured (eigen-)value corresponds to the (eigen-)vector $\Omega_{S}$ of $A$ then the system $S_i$ is kept and has been successfully prepared in the desired state $\Omega_{S}$.

A typical  example of this method for state preparation is a Stern-Gerlach spin measurement: In this example, the system $S$ consists of a spin 1/2-particle, (e.g., a silver atom). The experimentalist successively sends a large number, $n$, of such particles through a very slightly inhomogeneous magnetic field essentially parallel to the z-axis, which is perpendicular to the initial direction of motion of the particles (parallel to the x-axis). Particles with spin up 
($S^{(z)}=+\hbar/2$) are deflected towards the positive z-direction, whereas  particles with spin down 
($S^{(z)} =-\hbar/2$) are deflected towards the negative z-direction. Thus, after traversing the magnetic field, 
the particle beam is split into two sub-beams that point into slightly distinct directions. One of these two sub-beams is then targeted towards a screen that destroys it. The remaining beam consists of particles prepared in a fixed eigenstate of $S^{(z)}$ and can be used for further experimentation.  

The method of state preparation discussed here demands creating many  essentially identical copies of a system $S$ of interest and cannot be applied if the system $S$ cannot be duplicated; (e.g., if $S$ is the sun).  
For the theorist, this method obviously poses the problem of first understanding what ``projective measurements'' are; (see, e.g, \cite{FS} and refs. given there).
 
\subsubsection{Plan of the paper} 
In concrete applications, the methods of state preparation discussed in this section are often combined with one another (as already indicated in subsection 1.1.2). The methods discussed in subsections 1.1.2 and 1.1.3 are reasonably well understood (disregarding from problems concerning a theoretical understanding of projective measurements relevant for the method discussed in subsection 1.1.3). They will therefore not be discussed any further in the bulk of this paper.

The goal of this paper is to make a solid mathematical contribution to our understanding of the first method, which has been described  in subsection 1.1.1. For this purpose, we will study  a specific (idealized) model introduced in Section  \ref{model}, below.  Our main results and the underlying hypotheses are described and explained in Subsections \ref{res} and \ref{List}, respectively. The proof of our main result is outlined in Subsection 1.4. All technical matters are treated in four rather tedious sections, Sections 2 through 5, and four appendices.

 \subsection{ The model} \label{model}
The model underlying our analysis is the so-called generalized spin-boson model, which describes an idealized very heavy atom coupled to a quantized free scalar field. (It is straightforward to replace the scalar field considered in this paper by the quantized electromagnetic field. We prefer to consider a scalar field purely for reasons of notational simplicity. But we will call the field quanta ``photons''.) The atom represents a quantum (sub)system henceforth denoted by  $S$, while the field represents a system denoted by $E$.  By   
$\mathcal{H}_S$ we denote the $n$-dimensional Hilbert space describing the internal states of the atom before it is coupled to the field; (the center-of-mass motion of the atom is neglected, because it is assumed to be very heavy). By $\mathcal{F}_{+}(L^{2}(\mathbb{R}^{3}) )$ we denote the symmetric Fock space over the one-particle Hilbert space $L^{2}(\mathbb{R}^3)$. As usual, physical quantities or ``observables'' of  $S \vee E$ are represented by certain bounded self-adjoint operators acting on the Hilbert space $\mathcal{H}:=\mathcal{H}_S \otimes  \mathcal{F}_{+}(L^{2}(\mathbb{R}^{3}) )$.  Bosonic annihilation- and creation operators on $\mathcal{F}_{+}(L^{2}(\mathbb{R}^{3}))$ are operator-valued distributions, $a(k)$ and $a^{*}(k)$, where $k\in \mathbb{R}^{3}$ denotes a wave vector, satisfying the usual canonical commutation relations 
$$
\left[a(k), a^{*}(k')\right]=\delta(k-k'), \text{   } \left[a^{\sharp}(k), a^{\sharp}(k')\right]=0, \qquad k,k' \in \mathbb{R}^{3}.
$$

\subsubsection{The Hamiltonian of the System}
Let $E_1 <...<E_n$ be real numbers and $(\varphi_i)_{i=1}^{n}$ an orthonormal basis of $\mathcal{H}_S$. We set  $P_i:= \vert \varphi_i \rangle \langle \varphi_i \vert$, $i=1,...,n$. We suppose that $\omega_{S}(\cdot) = 
\langle \varphi_1, (\cdot) \text{  }\varphi_1 \rangle$ is the state in which we want to prepare the system $S$. 
The time-dependent Hamiltonian of the  system $S \vee E$ is  given by 
\begin{equation}\label{Ham}
H(t):=H_0+ \lambda(t) H_I,
\end{equation}
where 
\begin{equation}
\label{H0}
H_0=\sum_{i=1}^{n} E_{i} P_{i} \otimes 1_{E} + 1_S \otimes \int_{\mathbb{R}^{3}}  d^{3} k  \text{ } \omega (k) a^{*}(k) a(k):=H_S \otimes 1_{E} + 1_S \otimes H_E,
\end{equation}
with $\omega(k) =\vert k \vert $, for all $k \in \mathbb{R}^{3}$. In (\ref{Ham}), $\lambda(t)$ is a positive, monotone-decreasing function of time $t$, and $H_I$ is the interaction Hamiltonian coupling $S$ to $E$, which we define next.
We assume that the coupling between the ``atom'' and the field is linear in creation- and annihilation operators. More precisely, the interaction Hamiltonian is given by 
\begin{equation}
\label{H_I}
H_I:=G \otimes \left( a (\phi)  + a^{*} (\phi) \right),
\end{equation}
where the ``form factor'' $\phi$ belongs to $L^{2} (\mathbb{R}^{3})$,  and
 \begin{equation}
 \label{a}
 a^*(\phi)=\int d^{3}k  \text{ } \phi(k) a^{*}(k), \qquad  a(\phi)=\int d^{3}k \text{ } \bar{\phi}(k) a(k),
 \end{equation}
 for arbitrary  $\phi \in L^{2}(\mathbb{R}^{3})$. 
 The physics of exchange of quanta of energy between the atom and the field is characterized by the property that it satisfies a Fermi-Golden-Rule  condition introduced in Section \ref{List}, below.

\subsubsection{Initial states and ``observables''}
\label{init}
We assume that the system is initially in a state of the form
 \begin{equation} \label{inistate}   \Psi = \varphi     \otimes \Omega,  \end{equation}
  where $ \varphi  \in \mathcal{H}_S$ and $ \Omega $ is the  vacuum Fock state.  This choice of a simple initial state is made merely to avoid cumbersome notations and lengthy formulae. Our results still hold true if the initial state of the atom, $\varphi$, is replaced by a density matrix on $\mathcal{H}_S$, and if the field is in a state where finitely many field modes are excited or in a coherent state; see Section \ref{S5} and \cite{DeR1}. Moreover, the initial state may entangle the atom $S$ with the field $E$. 
  
The situation where the field is initially prepared in an equilibrium state at positive temperature is discussed in Section \ref{S5}.

\subsubsection{Basic
  assumptions}
\label{List}
 
\begin{assumption}
 \label{Co} (\textit{Decay of  correlations})
We assume that the form factor $\phi$  in  \eqref{H_I}  is chosen  such  that  $\phi$ and $ \phi/\sqrt{\omega}$ belong to $L^{2}(\mathbb{R}^3)$. We define
\begin{equation} 
\label{decc}
f(t):=\int d^{3}k  \text{ }\vert \phi(k) \vert^{2} e^{-it \omega(k)},
\end{equation}
 $t \geq 0$. We assume that there exists a constant $\alpha>2$  such that
\begin{equation}
\label{requ}
\vert f(t) \vert \propto \frac{1}{(1+t)^{\alpha}}.
\end{equation}
\end{assumption}
\vspace{2mm}

\begin{assumption} \label{Fe} (\textit{Fermi-Golden-Rule Condition})
For all $i \in \lbrace 2,...,n \rbrace$,
\begin{equation}
\label{Fer}
\sum_{j=1}^{i-1} \int d^{3}k \text{ } \vert G_{ij} \vert^{2} \vert \phi(k) \vert^{2} \delta(E_j-E_i+\omega(k)) >0.
\end{equation}
\end{assumption}
\vspace{2mm}
\begin{assumption} \label{Ev} (\textit{Evolution of $\lambda(t)$})
 There exists a constant $\gamma$, with
 
$$ -1/2<\gamma<0,$$
 such that 
\begin{equation}\label{CC}
\lambda(t) = (\lambda(0)^{1/\gamma}+t)^{\gamma}.
\end{equation}
\end{assumption}
\vspace{2mm}

The first part of Assumption \ref{Co} ensures that $H_I$ is $H_E^{1/2}$-bounded. Therefore, by Kato's theorem (see \cite{Kato}), $H_E+ \lambda(t) H_I$ is self-adjoint on the domain of $H_E$, for all values of 
$\lambda(t)$. The second part of Assumption \ref{Co} specifies the minimal decay rate of the ``correlation function'' $f(t)$  in time $t$ needed to carry out our analysis. The behavior of $f(t)$, for large $t$, is determined by the infrared behavior of the form factor  $\phi$.  If $\phi$ is smooth in $k$, except at $k=0$, with 
$\phi(k) \simeq \vert k \vert^{\mu}, \text{as } \vert k \vert \rightarrow 0$, and if $\phi$ is invariant under rotation and has compact support,  the  theory of asymptotic expansions for Fourier integrals  shows that
\begin{equation*}
f(t) \propto t^{-3-2 \mu},
\end{equation*}
see \cite{Erd}. Therefore, $\mu$ must be strictly bigger than $-1/2$ for \eqref{requ} to be satisfied. Eq. \eqref{Fer} implies that  the exited states $  \varphi_i  $  ($i \neq 1$) of the atom  decay, i.e., correspond to resonances, when the coupling between the atom and the field is turned on.  Assumption  \ref{Ev} is an ``adiabatic'' condition: The coupling $ \lambda(t)$ must decrease  sufficiently slowly  in $t$  for a state, $\rho_{inv}(s)$, invariant under the  reduced dynamics of $S$ to exist on a time scale of $t-s \propto \lambda^{-2}(s)$, (Van Hove limit).
 
 \subsection{Main result} \label{res}
\begin{theorem}
\label{clu}
Suppose that Assumptions \ref{Co}, \ref{Fe} and  \ref{Ev} are satisfied.  Then, there exists a constant $\lambda_c>0$, such that, for any $0<\lambda(0)<\lambda_c$,
\begin{equation}
\label{resultat}
 \langle \Psi(t) \vert (O  \otimes 1 )  \Psi(t)  \rangle \underset{t \rightarrow \infty}{\longrightarrow}  \langle \varphi_1 \vert O  \varphi_1  \rangle \equiv \omega_{S}(O),
\end{equation}
for all initial states $\Psi $ of the form given in  \eqref{inistate} and  for all observables $O \in \mathcal{B}(\mathcal{H}_S)$. Here $  \varphi_1 $ is the ground state (unique up to a phase) of $H_S$ corresponding to the eigenvalue  $E_1$, and $  \Psi(t) $ is the state $\Psi$ evolves into, after time $t$, under the dynamics generated by the family of Hamiltonians $H(t)$.
\end{theorem}

\begin{remark}
In Section \ref{S5}, we will generalize Theorem \ref{clu} to a larger class of initial states (including ones with a non-zero, but finite number of occupied field modes and ones exhibiting entanglement). We also present a variant of Theorem \ref{clu} where the field modes are at some non-zero temperature $T>0$. We will show that  $S$  thermalizes at the same temperature $T$, as $t \rightarrow \infty$. 
\end{remark}

\subsection{Outline of the proof}
We will use expansion methods developed in  \cite{DeR1,DeR2} to prove convergence of the expectation values $\langle \Psi(t) \vert  (O \otimes 1)    \Psi(t)  \rangle$ to $ \langle \varphi_1   \vert O  \varphi_1   \rangle$, as $t \rightarrow \infty$.  In \cite{DeR1,DeR2}, the authors consider a coupling  constant $\lambda$ independent of time $t$. We  adapt the methods developed in these references to apply to the models considered in this paper, with $\lambda$ depending on time $t$ and decreasing to zero, as $t \rightarrow \infty$. We attempt to present a somewhat streamlined version of the arguments (in particular of the ``polymer expansion'') in \cite{DeR1,DeR2}. We employ the Heisenberg picture, and we only investigate the time evolution of observables, $O$, of the atom $S$.  \\

\subsubsection{Step 1.  Analysis of the reduced dynamics on the Van Hove time scale}  We introduce a linear  operator $\mathcal{Z}^{t,s}: \mathcal{B}(\mathcal{H}_S) \rightarrow \mathcal{B}(\mathcal{H}_S)$ describing the  effective dynamics of observables $O  \in \mathcal{B}(\mathcal{H}_S)$,  for times $= \mathcal{O}(\lambda^{-2})$ (van Hove time scale).  Let $U(t,s)$ be the unitary propagator generated by the family of time-dependent Hamiltonians $\lbrace{H(t)}\rbrace_{t\in \mathbb{R}_{+}}$. We define $\mathcal{Z}^{t,s}(O) \in \mathcal{B}(\mathcal{H}_S)$ by
\begin{equation}
\label{def_O}
\langle \varphi \vert \mathcal{Z}^{t,s} (O) \psi \rangle:=  \langle  \varphi  \otimes \Omega \vert  U^{*}(t,s) (O \otimes \mathds{1}) U(t,s)  (\psi \otimes \Omega) \rangle,
 \end{equation}
for all  $\varphi,\psi \in \mathcal{H}_S$. In defining $\mathcal{Z}^{t,s}$ we take an average in the vacuum vector, $\Omega$, of the field variables, because ``photons'' emitted by the atom escape towards infinity. On the Van Hove time scale, $\mathcal{O}(\lambda(s)^{-2})$, the atom decays to its ground state  with a probability very close to 1, and the photons have escaped from the vicinity of the atom and will never return to it. Thus, $\mathcal{Z}^{t,s}(\cdot)$ can be expected to describe the Heisenberg time evolution of atomic observables fairly accurately if $t-s = \mathcal{O}(\lambda(s)^{-2})$. If the function $\lambda(t)$ decays slowly in time, the effective time evolution $\mathcal{Z}^{t,s}$ is well approximated by a semi-group of completely positive maps on 
$\mathcal{B}(\mathcal{H}_S)$ generated by a Lindblad operator, for time differences, $t-s$, of order $\mathcal{O}(\lambda(s)^{-2})$; see   \eqref{Lind}-\eqref{difference}. (The error actually tends to zero in norm  as  $s \rightarrow \infty$.) We use this result to show that 
\begin{equation}
\label{deco}
\mathcal{Z}^{s+\tau \lambda^{-2}(s),s} =P(s)+R(s),
\end{equation}
where $P(s):  \mathcal{B}(\mathcal{H}_S) \rightarrow \mathcal{B}(\mathcal{H}_S)$ is a one-dimensional projection, and $R(s):  \mathcal{B}(\mathcal{H}_S) \rightarrow \mathcal{B}(\mathcal{H}_S)$ is a perturbation that can be made arbitrarily small by choosing the parameter $\tau>0$ large enough.  We show in Paragraph \ref{Peruu} that   $P(s)$ converges in norm to the one-dimensional projection $\vert 1_S \rangle \langle \Pi_{11} \vert$, as $s \rightarrow \infty$, with $\Pi_{11} = |\varphi_1\rangle \langle \varphi_1 |$. Here elements of $\mathcal{B}(\mathcal{H}_S)$ are written as vectors, 
$\vert \cdot \rangle$, (more precisely, as vectors in the Hilbert space of matrices).

Eq. \eqref{deco} captures the dissipative behavior of the effective dynamics of the  atomic system on the Van Hove time scale.

\subsubsection{Step 2.  Reduced dynamics at arbitrarily large times: the cluster expansion}
Eq.  \eqref{deco} is only valid on the Van Hove time scale $t-s=\mathcal{O}( \lambda^{-2}(s))$. However, we  intend  to prove   that
\begin{equation}
\label{limit}
\mathcal{Z}^{t,0} \underset{t \rightarrow \infty}{\longrightarrow} \vert 1_S \rangle \langle \Pi_{11}  \vert.
\end{equation}
The polymer expansion introduced in \cite{DeR1,DeR2} offers a way  to pass from the Van Hove time scale to arbitrarily large times, $t \rightarrow \infty$. It is based on the intuition that the dynamics of atomic observables is close to one given by a quantum Markov dynamics whose only invariant state is given by $\vert 1_S \rangle \langle \Pi_{11} \vert$, with errors that can be controlled with the help of a cluster expansion for a one-dimensional system of ``extended particles'', called ``polymers'', of ever smaller density of $\mathcal{O}(\lambda(s)^{2})$, as 
$s\rightarrow \infty$.

In somewhat more precise terms, our expansion is set up as follows: We start by labeling all terms in  the  Dyson expansion of $\mathcal{Z}^{t,s}(\cdot)$ by Feynman diagrams.  For each time interval $I_i = [t_i, t_{i+1})$, with
 $t_{i+1} = t_i + \tau\lambda (t_i)^{-2}$, we  sum all contributions labeled by diagrams with the property that any ``photon'' emitted at a time in the intervall $I_i$ is re-absorbed by the atom at another time in the \textit{same} interval $I_i$. This yields the contribution corresponding to the operator $\mathcal{Z}^{t_i+\tau \lambda^{-2}(t_i),t_i}$. It is at this point where the decomposition \eqref{deco} comes into play:  We use that $P(s)$ is  a one-dimensional projection  to  rewrite  the expectation value $\langle \Psi(t_N)  \vert (O \otimes 1)  \Psi(t_N) \rangle$ in the form of a cluster  expansion for a system of ``extended particles'' /``polymers'' in one dimension. We show in Section  \ref{graphph} that
\begin{equation}
\label{cpoli}
\langle \Psi(t_N)  \vert (O \otimes 1) \Psi(t_N) \rangle  =  \sum_{q=1}^{N} \frac{1}{q!} \underset{\text{dist}(\mathcal{X}_i,\mathcal{X}_j) \geq 2,\text{ diam}(\mathcal{X}_j) \leq N}{\underset{\mathcal{X}_1,...,\mathcal{X}_q}{\sum}}   p\left(\mathcal{X}_1\right)...\text{ }p\left(\mathcal{X}_q\right),
\end{equation}
where $p\left(\mathcal{X}\right) \in \mathbb{C}$ are the statistical weights of certain polymers, 
$\mathcal{X}$; see Paragraph \ref{secon}.

\subsubsection{Step 3.  The limit $t \rightarrow \infty$}
In  Section \ref{laco} we prove that our cluster expansion, see Eq. \eqref{cpoli}, converges uniformly in $N$, and, as a corollary, that $\langle \Psi(t)  \vert (O \otimes 1) \Psi(t) \rangle$ tends to  $\langle \varphi_1 \vert  O \varphi_1 \rangle$, as  $t \rightarrow + \infty$.  In our proof of convergence of the cluster expansion we have to require that  the correlation function $f(t)$ defined  in \eqref{decc} decay sufficiently fast in $t$; see Assumption \ref{Co}. In Paragraph \ref{bound}, we use the decay properties of $f(t)$ to derive an upper bound for the statistical weights $p(\mathcal{X})$ of polymers 
$\mathcal{X}$ appearing in  \eqref{cpoli} that implies the so-called Kotecky-Preiss criterion, 
\begin{equation}
\label{cricril}
\sum_{\mathcal{G'}, \text{ dist} (\mathcal{X}',\mathcal{X}) \leq 1} \vert p (\mathcal{X}') \vert \text{ } e^{a(\mathcal{X}')} \leq a(\mathcal{X}),
\end{equation}
for a suitably chosen positive function  $a$. Using standard results in the theory of cluster expansions (see Appendix \ref{clustersec} for a short recap), it is straightforward to show that \eqref{cricril} implies that \eqref{cpoli} converges to $\langle \varphi_1 \vert O \varphi_1 \rangle$, as $N \rightarrow \infty$. A crucial point in this analysis is that the right side of (\ref{cpoli}) can be written as an exponential of a convergent sum. Dividing this expression by a corresponding expression for $1=\langle \Psi(t_N) \vert \Psi(t_N) \rangle$, one observes that the number of  terms contributing in the limit  $N \rightarrow \infty$ is quite small;  see Section \ref{Proo}. 

\subsubsection{Generalization of Theorem \ref{clu} to initial states with finitely many ``photons'' and to thermal equilibrium states} Such generalizations of Theorem \ref{clu} are formulated in Section \ref{S5}. Sketches of the proofs are given in Appendix \ref{sketch}.
\vspace{2mm}

\section{ \large Analysis of $\mathcal{Z}^{t,s}$ for $t-s \propto \lambda^{-2}(s)$ }
\label{S2}
We begin this section with a list of notations and conventions that are used throughout the paper. In Subsections \ref{dy1} and \ref{dy2}, we   compare  the Dyson expansion for  $\mathcal{Z}^{t,s}(O)$  with the Dyson expansion for  $\mathcal{Z}_0^{t,s}(O) $, where  $\mathcal{Z}_0^{t,s}(O)\in \mathcal{B}(\mathcal{H}_S)$ is defined by
\begin{equation}
\label{nonev}
\langle \varphi \vert \mathcal{Z}_0^{t,s}(O) \psi \rangle :=  \langle \varphi \otimes \Omega \vert  e^{i(t-s)H(s)} (O \otimes 1) e^{-i(t-s)H(s)} (\psi \otimes \Omega)\rangle,
\end{equation}
for arbitrary $\phi,\psi \in \mathcal{H}_S$. We then compare  $\mathcal{Z}_0^{t,s}(\cdot) $  with  the semigroup generated by a Lindbladian, using results in  \cite{DeR1}. The calculation of the Lindbladian   is explicitly   carried out in Appendix B. Using estimates from perturbation theory, we will prove that, for large values of the parameter $\tau$ and small values of $\lambda(0)$,
 \begin{equation} \label{encore} \mathcal{Z}^{s+\tau \lambda^{-2}(s),s} =P(s)+R(s),
 \end{equation}
for all  $s \geq 0$, where $P(s): \mathcal{B}(\mathcal{H}_S) \rightarrow \mathcal{B}(\mathcal{H}_S)$ is a one-dimensional projection, and $R(s):\mathcal{B}(\mathcal{H}_S) \rightarrow \mathcal{B}(\mathcal{H}_S)$ is a small perturbation. Various straightforward but  lengthy calculations are deferred to Appendix B.\\

\subsection{Notations} \label{not}
 \subsubsection{Inner products and norms}
The norm on  $\mathcal{H}_S$ determined by the scalar product $\langle   \cdot , \cdot \rangle_S$ is 
denoted by $\vert \vert \cdot \vert \vert_S$.   On the tensor product  space $\mathcal{H}=\mathcal{H}_S \otimes  \mathcal{F}_{+}(L^{2}(\mathbb{R}^{3}) )$, the scalar product is  given by
$$\langle \phi \otimes \psi, \phi' \otimes \psi' \rangle:=\langle \phi , \phi'  \rangle_S \langle \psi , \psi'  \rangle_{\mathcal{F}_+}, $$
where $\langle   \cdot , \cdot \rangle_{\mathcal{F}_+}$ is the scalar product on $  \mathcal{F}_{+}(L^{2}(\mathbb{R}^{3}) )$ and is defined  by
\begin{equation}
  \langle \psi , \chi  \rangle_{\mathcal{F}_+}= \overline{\psi^{(0)}} \chi^{(0)} +\sum_{n \geq 1} \int d^3 k_1\text{ }... \text{ }d^3k_n \text{ }  \overline{\psi^{(n)}}(k_1,...,k_n) \chi^{(n)} (k_1,...,k_n),
\end{equation}
for all $\psi=\lbrace \psi^{(n)} \rbrace_{n\geq 0}$ and all $\chi=\lbrace \chi^{(n)} \rbrace_{n\geq 0}$.

The algebra of bounded operators on  $\mathcal{H}_S$ is denoted by $\mathcal{B}(\mathcal{H}_S)$.   
Since $\text{dim}\mathcal{H}_S =: n < \infty$, $\mathcal{B}(\mathcal{H}_S)$ is a Hilbert space equipped 
with the scalar product 
\begin{equation}
\label{scalar}
\langle  X, Y \rangle_{\mathcal{B}(\mathcal{H}_S)}:= \text{Tr}(X^{*} Y), \qquad \forall X,Y \in  \mathcal{B}(\mathcal{H}_S),
\end{equation}
and the Hilbert-Schmidt norm.
The operator norm on $\mathcal{B}(\mathcal{H}_S)$ is given by 
\begin{equation}
\vert \vert X \vert \vert= \underset{ \psi \neq 0, \text{ } \psi \in \mathcal{H}_S }{\sup} \frac{\vert \vert  X \psi \vert \vert_S}{\vert \vert \psi \vert \vert_S}
\end{equation}
for all $X \in \mathcal{B}(\mathcal{H}_S)$. The algebra of bounded linear operators on  $\mathcal{B}(\mathcal{H}_S)$  is  denoted by  $\mathcal{B}(\mathcal{B}(\mathcal{H}_S))$. It is isomorphic to the finite dimensional Hilbert space of $n^2 \times n^2$ complex matrices.  We will use two equivalent norms on
 $\mathcal{B}(\mathcal{B}(\mathcal{H}_S))$, which are defined by 
\begin{equation}
\label{norm}
\vert \vert B \vert \vert_{\infty}:= \underset{X \neq 0, \text{ }X \in \mathcal{B}(\mathcal{H}_S) }{\sup} \frac{\vert \vert BX \vert \vert}{\vert \vert X \vert \vert} \text{ }\qquad  \text{and} \qquad \vert \vert B \vert \vert_{2}^2:= \underset{X \neq 0, \text{ }X \in \mathcal{B}(\mathcal{H}_S) }{\sup} \frac{   \langle  BX, BX \rangle_{\mathcal{B}(\mathcal{H}_S)}}{  \langle  X, X \rangle_{\mathcal{B}(\mathcal{H}_S)}},
\end{equation}
for all $B \in \mathcal{B}(\mathcal{B}(\mathcal{H}_S))$.
\vspace{1mm}

\subsubsection{Dyson expansion}
The starting point of the methods  developed in \cite{DeR1,DeR2} and used in the present paper is the Dyson expansion. We need to  introduce some shorthand notations, in order to avoid cumbersome complications in the representation of the Dyson expansion. We define the quantities
\begin{align}
\label{PHII}
\Phi(\phi)&:=a(\phi)+a^{*}(\phi),\\
\label{Gt}
G(t)&:=e^{it H_S} G  e^{-it H_S},\\
\label{HH}
H_{I}(t)&:=e^{itH_{0}} H_{I} e^{-it H_{0}},\\
\label{phihi}
\phi(t)(k)&:=e^{it \omega(k)} \text{ }\phi(k),
\end{align}
for all $t \in \mathbb{R}$, where $H_0$ and $H_S$  are defined in Eq. \eqref{H0}. Time-integrations will usually extend over an $n$-dimensional simplex
\begin{equation}
\label{simpl}
\Delta^{n}\left[ t,s\right]:=\lbrace (t_1,...,t_n) \mid  s \leq t_1<...<t_n \leq t \rbrace, \qquad 0<s<t<\infty.
\end{equation}
For any  $n$-tuple $\underline{t}:=(t_1,...,t_n)$, ordered as in (\ref{simpl}), and for  any time-dependent operator-valued function $O(t)$, we set
\begin{eqnarray}
\label{ordon}
\lambda(\underline{t})&:=& \lambda(t_1) \text{ }...\text{ } \lambda(t_n),\\
O(\underline{t})&:=&O(t_1) \text{ }... \text{ }O(t_n).
\end{eqnarray}
It turns out to be useful to introduce operators $\textbf{R}(A)$ (right multiplication by $A$) and $\textbf{L}(A)$ (left multiplication by $A$), 
$A\in \mathcal{B}(\mathcal{H}_S)$, by setting
\begin{eqnarray}
\label{Mult}
\textbf{R}(A)(O) &:=&OA^{*},\\
\textbf{L}(A)(O) &:=&AO,
\end{eqnarray}
 for all $A,O \in \mathcal{B}(\mathcal{H}_S)$.
\vspace{1mm}

\subsection{Dyson expansions for $\mathcal{Z}^{t,s}$ and $\mathcal{Z}_0^{t,s}$ } \label{dy1}
We first recall the Dyson expansion for the propagator $U(t,s)$ of the system.
\begin{lemma}
\label{Dyson}
Let $t,s \in \mathbb{R}$. The Dyson  series 
\begin{equation}
\label{D1}
U(t,s) =e^{-i(t-s)H_0} +\sum_{k=1}^{\infty} (-i)^{k} \int_{s}^{t}d u_{k} ...\int_{s}^{u_{2}} du_{1} \lambda(u_1) ...\lambda(u_k) \text{ }   e^{-i t H_0}  H_{I}(u_k)... H_{I}(u_1)  e^{i s H_0},   
\end{equation}
and
\begin{equation}
\label{D2}
e^{-i(t-s)H(s) }=e^{-i(t-s)H_0} +\sum_{k=1}^{\infty} (-i)^{k} \int_{s}^{t}d u_{k} ...\int_{s}^{u_{2}} du_{1} \lambda^{k}(s) \text{ }   e^{-i t H_0} H_{I}(u_k)... H_{I}(u_1)    e^{i s H_0}   
\end{equation}
converge strongly on $\mathcal{H}_S \otimes F(L^{2}(\mathbb{R}^3))$, where  $ F(L^{2}(\mathbb{R}^3)) \subset \mathcal{F}_+(L^{2}(\mathbb{R}^3))$  is the dense set of vectors in Fock space describing configurations of \textit{finitely many} ``photons''.
\end{lemma}
\vspace{1mm}

  Lemma \ref{Dyson} is standard, (but see Appendix \ref{A1}). It can be used to deriving the Dyson expansion of the effective atom propagator $\mathcal{Z}^{t,s}$. We remind the reader that $\mathcal{Z}^{t,s}$ is defined by 
  \begin{equation}
\label{def_OO}
\langle \varphi \vert  \mathcal{Z}^{t,s}(O) \psi \rangle:=  \langle  \varphi  \otimes \Omega \vert  U^{*}(t,s) (O \otimes \mathds{1}) U(t,s)  (\psi \otimes \Omega) \rangle, \qquad \forall \varphi,\psi \in \mathcal{H}_S,
\end{equation}
for arbitrary $O \in \mathcal{B}(\mathcal{H}_S)$. We  compute the vacuum expectation values of products of interaction Hamiltonians $H_I(u)$ in the Dyson expansion of the propagators on the right side of \eqref{def_OO} using Wick's theorem. 
\vspace{2mm}

\begin{theorem}[Wick's Theorem]
If $ \Omega $ denotes the Fock vacuum then
\begin{align}
\label{qfree}
\langle \Omega \vert \Phi(\phi(v_{1}))...  \Phi(\phi(v_{2k}))   \Omega \rangle  &=\sum_{\text{pairings } \pi} \prod_{(i,j)\in \pi}  \langle \Omega \vert  \Phi(\phi(v_i)) \Phi(\phi(v_j))  \Omega \rangle,\\
\langle \Omega \vert \Phi(\phi(v_{1}))...  \Phi(\phi(v_{2k+1}))  \Omega \rangle  &=0,
\end{align}
 for arbitrary times  $v_1,...,v_{2k}, v_{2k+1} \in \mathbb{R}$, where pairings, $\pi$, are sets of pairs $(i,j)$, with $i<j$, whose union is the set  $\lbrace 1,...,2k \rbrace$.
\end{theorem}
\vspace{1mm}

The ``two-point functions'' $\langle \Omega \vert  \Phi(\phi(v_i)) \Phi(\phi(v_j))  \Omega \rangle$ can all be expressed in terms of the correlation function $f(t)$, see (\ref{decc}), which is given by  
 \begin{equation}
 \label{cooco}
f(t) = \langle \Omega \vert \Phi(\phi(t)) \Phi(\phi) \vert \Omega \rangle.
\end{equation} 
for all $t \in \mathbb{R}$. 
We note that $f(-t)=\overline{f(t)}$, for all $t \in \mathbb{R}$. 

Next, we introduce an index $r \in \lbrace 0,1 \rbrace$ attached to each time $u$ labeling an interaction Hamiltonian that appears in the Dyson expansion of the right side of \eqref{def_OO}: $r$ takes the value $0$ if $H_{I}(u)$ in \eqref{D1} appears on the left of $O$, and $r=1$ if $H_{I}(u)$ appears on the right of $O$. 
We then write $(u,r)$, instead of $u$.  We also  define
\begin{equation}
\label{MM}
A(u,r)=\left\{ \begin{array}{ll}
 \textbf{L}(A(u)) &\mbox{ if } r=0, \\[4pt]
 \textbf{R}(A(u))  &\mbox{ if } r=1,\\
 \end{array} \right.
\end{equation}
for all $(u,r)$ and for any time dependent family of  operators $A(u)$ on $\mathcal{H})$. Remark that $A(u,r)$ is an operator on $\mathcal{B}(\mathcal{H})$ for all $(u,r)$.

We introduce some convenient short-hand notations for two-point functions as follows.
\begin{definition}{ (The functions $F$ and ${\bf{F}}$)}
\begin{equation}\label{F3}
 \text{F}(u, r; v, r'):= \langle\Omega | \big( \Phi(\phi(u,r))  \Phi(\phi(v,r'))\big) ( 1_{\mathcal{F}_+} ) \Omega \rangle,
\end{equation}
and 
\begin{equation}
\label{F2}
{\bf{F}}(u, r; v, r'): =  \text{F}(u, r; v, r')  (iG)(u,r) (iG)(v,r').
\end{equation}
\end{definition}
\vspace{1mm}

 We also introduce a time-ordering operator acting on products of operators in $\mathcal{B}(\mathcal{B}(\mathcal{H}_S))$. We will use it  to order products of operators $(iG)(u,r)$.
\vspace{2mm}

 \begin{definition}[Time-ordering]
 Let $0<u_1<...<u_n$ be an ordered n-tuple of times, and let $A(u_1),...,A(u_n) \in \mathcal{B}(\mathcal{B}(\mathcal{H}_S)) $ be a family of operators. We  define an operator $ \mathcal{T}_S:\mathcal{B}(\mathcal{B}(\mathcal{H}_S))  \rightarrow \mathcal{B}(\mathcal{B}(\mathcal{H}_S)) $ by
 \begin{equation}
 \label{time}
 \mathcal{T}_S ( A(\pi(u_1)) \text{ }... \text{ }A(\pi(u_n))) :=  A(u_1) \text{ }...\text{ }A(u_n)
 \end{equation}
for all  permutations $\pi$ of $\lbrace1,...,n\rbrace$.
 \end{definition}
\vspace{2mm}

 We denote by $w :=(u, r; v, r')$   a pair of times, $u<v$, decorated by indices $r$ and $r'$, and by 
  $\underline{w}=(w_1,...,w_k)$ a $k$-tuple of such pairs.  A similar underlined notations is used for $k$-tuples of times $u$, denoted by $\underline{u}=(u_1,...,u_k)$, and $k$-tuples of indices $r$, denoted by $\underline{r}=(r_1,...,r_k)$.
  
 Next, we introduce a measure on $k$-tuples of pairs, $k=1,2,3,...$
\begin{equation}
\label{mumu}
d\mu_k(\underline{w})  :=  \sum_{\underline{r},\underline{r}'  \in \lbrace 0,1 \rbrace^{k}}  \chi(u_1<...<u_k) \prod_{i=1}^{k} \chi(u_i<v_i)  \text{ } du_1 \text{ }... \text{ }du_k \text{ } dv_1 \text{ }... \text{ }dv_k \text{ }.
\end{equation}
Our next lemma describes the Dyson expansions of $\mathcal{Z}^{t,s}(O)$ and $\mathcal{Z}_{0}^{t,s}(O) $. We  make use of the notations introduced above. 

\begin{lemma}
\label{lem3}
\begin{equation}
\label{DD1'}
e^{is H_S}\mathcal{Z}^{t,s}(O) e^{-is H_S} = \sum_{k=0}^{\infty} \underset{[s,t]^{2k}}{\int} d\mu_k(\underline{w})  \lambda( \underline{w}) \mathcal{T}_S \left[ \prod_{i=1}^{k}   {\bf{F}}(u_{i}, r_{i}; v_{i}, r'_{i})  \right] \left[ O(t) \right],
\end{equation}
where $\lambda( \underline{w}) :=\lambda(\underline{u}) \lambda(\underline{v})$; and
 
\begin{equation}
\label{DD2'}
e^{is H_S}\mathcal{Z}_{0}^{t,s}(O) e^{-is H_S}=\sum_{k=0}^{\infty} \underset{[s,t]^{2k}}{\int} d\mu_k(\underline{w}) \lambda^{2k}(s)   \mathcal{T}_S \left[ \prod_{i=1}^{k}   \textbf{F}(u_{i}, r_{i}; v_{i}, r'_{i})  \right] \left[ O(t) \right]
 \end{equation}
for all $s,t \in \mathbb{R}_+$ with $t \geq s$. 

The series in \eqref{DD1'} and \eqref{DD2'} converge in norm, for all $O \in \mathcal{B}(\mathcal{H}_S)$, and are  bounded by 
\begin{equation}
\label{labound}
 e^{4 \vert t-s \vert \text{ } \vert \vert f \vert \vert_{L^1} \vert \vert G \vert \vert^{2} \lambda^{2}(s)} \vert \vert O \vert \vert.
\end{equation}
 \end{lemma}
 
 A  proof of  Lemma \ref{lem3}  is given in Appendix \ref{A22}.

\vspace{3mm}

\subsection{Comparison of the effective propagator $\mathcal{Z}^{t,s}$ with the semigroup generated by a Lindbladian } \label{dy2}
In Lemma \ref{lem4}, below, we present an estimate  on the norm of the  difference  
$$\mathcal{Z}^{t,s}-\mathcal{Z}_0^{t,s},$$
 for $t-s = \tau \lambda(s)^{-2}$.  We will then use  a result from  \cite{DeR1,DeR2} to compare $\mathcal{Z}_0^{t,s}$ with the semigroup generated by a Linbladian. These findings will enable us to represent the effective propagator $\mathcal{Z}^{t,s}$, with $t= s + \tau \lambda(s)^{-2}$, as the sum of a one-dimensional projection $P(s)$ 
  and a ``small perturbation'' $R(s)$.  We will show in subsection \ref{Peruu} that the parameter $\tau$ can be chosen in such a way that, for any given 
$\varepsilon_0>0$,  $\| R(s) \|_{\infty}< \varepsilon_0$.

\begin{lemma}
\label{lem4}
  Let $t>s \geq 0$. Then
\begin{equation}
\label{compi}
\vert \vert \mathcal{Z}^{t,s}-\mathcal{Z}_0^{t,s} \vert \vert_{\infty} \leq e^{ 4 (t-s) \vert  \vert f   \vert  \vert_{L^1}  \vert \vert G \vert \vert^{2} \lambda^{2}(s) }-e^{ 4 (t-s) \vert  \vert f   \vert  \vert_{L^1}  \vert \vert G \vert \vert^{2} \lambda^{2}(t) }.
\end{equation}
If Assumption \ref{Ev}, see Eq. (\ref{CC}), is satisfied then
\begin{equation}
\label{compi}
\vert \vert \mathcal{Z}^{s+ \tau \lambda^{-2}(s),s}- \mathcal{Z}_0^{s+ \tau \lambda^{-2}(s),s} \vert \vert_{\infty}  \underset{ s \rightarrow \infty}{\longrightarrow}0.
\end{equation}
Furthermore, given any $\varepsilon>0$, there exists $\lambda_{\varepsilon}>0$ such that, for any $0<\lambda(0)<\lambda_{\varepsilon}$,
\begin{equation}
\label{compi}
\vert \vert \mathcal{Z}^{s+ \tau \lambda^{-2}(s),s}- \mathcal{Z}_0^{s+ \tau \lambda^{-2}(s),s} \vert \vert_{\infty}  \leq \varepsilon,
\end{equation}
for all $s \geq 0$.
\end{lemma}
\vspace{2mm}

\begin{proof}
 Let $O \in \mathcal{B}(\mathcal{H}_S)$. Eq. \eqref{DD1'} implies that 
 \begin{equation*}
 \label{cki}
 \begin{split}
 e^{is H_S} &\left( \mathcal{Z}^{t,s}(O)-\mathcal{Z}_{0}^{t,s}(O) \right) e^{-is H_S}\\
=&   \sum_{k=0}^{\infty} \underset{[s,t]^{2k}}{\int} d\mu_k(\underline{w}) \text{ }  (\lambda(\underline{w}) -\lambda^{2k}(s))  \mathcal{T}_S \left[ \prod_{i=1}^{k}  \textbf{F}(u_{i}, r_{i}; v_{i}, r'_{i})  \right] \left[ O(t) \right]. 
\end{split}
\end{equation*}
Using that $\lambda(t)$ decreases in time $t$ and that $\| G(u) G(v) \| \leq \|G \|^2$ (see (2.8) for the definition of $G_u$), we find that
 \begin{equation*}
 \begin{split}
 \vert \lambda(\underline{w}) -\lambda^{2k}(s) \vert    \vert \vert \mathcal{T}_S& \left[ \prod_{i=1}^{k}  \textbf{F}(u_{i}, r_{i}; v_{i}, r'_{i})  \right] \left[ O(t) \right]  \vert \vert \\
&  \leq   \vert    \vert O \vert \vert  \left( \lambda^{2k}(s)-\lambda^{2k}(t) \right)  \vert    \vert G \vert \vert^{2k}   \prod_{i=1}^{k} \vert  f(v_i-u_i)\vert.
 \end{split}
 \end{equation*}
We thus conclude that
\begin{eqnarray*}
\vert \vert \mathcal{Z}^{t,s}(O) -\mathcal{Z}_{0}^{t,s}(O)\vert \vert & \leq &  \vert    \vert O \vert \vert  \sum_{k=1}^{\infty}   \underset{[s,t]^{2k}}{\int} d\mu_k(\underline{w}) \text{ }        \left( \lambda^{2k}(s)-\lambda^{2k}(t) \right) \vert \vert G \vert \vert^{2k} \text{ }  \prod_{i=1}^{k} \vert  f(v_i-u_i)  \vert \\
&\leq&    \vert    \vert O \vert \vert  \sum_{k=1}^{\infty}  \frac{4^k (t-s)^{k}}{k!}   \text{ } \left( \lambda^{2k}(s)-\lambda^{2k}(t) \right) \text{ }  \vert \vert  G \vert \vert^{2k} \text{ }   \vert   \vert  f   \vert \vert^{k}_{L^1}
\end{eqnarray*}
by integrating  first over all the $v$- variables and subsequently over  all the $u$- variables. The  factor 
$\frac{(t-s)^k}{k!}$  comes from integrating over the $k$-dimensional simplex $\Delta^{k}[t,s]$. Hence
\begin{equation*}
\vert \vert \mathcal{Z}^{s+ \tau \lambda^{-2}(s),s} -\mathcal{Z}_0^{s+ \tau \lambda^{-2}(s),s} \vert \vert_{\infty} \leq e^{4\tau \vert   \vert  f   \vert \vert_{L^1}   \vert \vert G \vert \vert^{2} }-e^{ 4\tau \lambda^{-2}(s)  \vert   \vert  f   \vert \vert_{L^1}    \text{ } \vert \vert   G \vert \vert^{2} \lambda^{2}(t) },
\end{equation*}
with $t=s+ \tau \lambda^{-2}(s)$. By (\ref{CC}), the ratio $\lambda(t)/\lambda(s)$ is given by 
\begin{equation} \label{rara}
\frac{\lambda(t)}{\lambda(s)}=\left(\frac{\lambda(0)^{1/\gamma}+t}{\lambda(0)^{1/\gamma}+s} \right)^{\gamma}=\left(1+\tau  (\lambda(0)^{1/\gamma}+s)^{-2 \gamma-1}\right)^{\gamma}.
\end{equation}
Thus $\left(1+\tau  (\lambda(0)^{1/\gamma}+s)^{-2 \gamma-1}\right)^{\gamma} \rightarrow 1$, as 
$s \rightarrow \infty$, because $-2 \gamma -1<0$. Furthermore, the maximum in \eqref{rara} is reached at $s=0$; it is equal to 
$$ (1+  \tau  \lambda(0)^{(-2 \gamma-1)/\gamma})^{\gamma}.$$ 
Given any $\tau$, we can choose the coupling $\lambda(0)$ in such a way that this term is as close to $1$ as we wish, because $-1/2<\gamma<0$.
\end{proof}
\vspace{2mm}
 
 \noindent We define the Liouvillian  $\mathcal{L}_S \in \mathcal{B}(\mathcal{B}(\mathcal{H}_S)) $ by
\begin{equation}
\label{LS}
\mathcal{L}_S :=\textbf{L}(H_S)-\textbf{R}(H_S).
 \end{equation}
 
\noindent The eigenvalues of $\mathcal{L}_S$ are energy differences $\epsilon_{ij}:=E_{i}-E_{j}$, $i,j \in \lbrace 1,...,n \rbrace$. The associated eigenvectors are the $n\times n$ matrices $ \Pi_{ij} =\vert \varphi_{i} \rangle \langle \varphi_{j} \vert  \in  \mathcal{B}(\mathcal{H}_S)$, where  $\varphi_{i}$ is the eigenvector of $H_S$ corresponding to the eigenvalue $E_{i}$, $i=1,...,n$. The eigenvalues $E_{i}$ are assumed to be non-degenerate, and we  may assume that  the eigenvalues $\epsilon_{ij}$ are non-degenerate, too, for $i  \neq j$. The eigenvalue $0$ is $n$-fold degenerate, and the corresponding eigenvectors  are given by $\Pi_{11},...,  \Pi_{nn}$. We denote by  $P_{\epsilon} \in \mathcal{B}(\mathcal{B}(\mathcal{H}_S)) $ the orthogonal projection onto the eigenspace of $\mathcal{L}_S$ corresponding to the eigenvalue $\epsilon (= \epsilon_{ij}$, for some 
$i$ and $j$). This projection is one-dimensional if $\epsilon \neq 0$ and $n$-dimensional if  $\epsilon=0$. The spectrum of   $\mathcal{L}_S$ is denoted by $ \sigma(\mathcal{L}_S)$. Following  \cite{DeR1,DeR2,Da}, we define the Lindbladian  $\mathcal{M} \in \mathcal{B}(\mathcal{B}(\mathcal{H}_S)) $.

\begin{definition} (Lindbladian) 
A Lindblad generator  $\mathcal{M} \in \mathcal{B}(\mathcal{B}(\mathcal{H}_S)) $ is defined  by
\begin{equation}
\label{Lind}
\mathcal{M}:=\sum_{\epsilon \in \sigma(\mathcal{L}_S)} \int_{0}^{\infty} e^{-is \epsilon } P_{\epsilon } \mathcal{K}_s P_{\epsilon } \text{ }ds,
\end{equation}
where
\begin{equation}
\label{KK}
\mathcal{K}_{u_2-u_1} :=e^{-i u_1 \mathcal{L}_S} \sum_{(r_1,r_2) \in \lbrace 0,1 \rbrace^2}   {\bf{F}}(u_{1},r_1;u_{2},r_2)  \text{ } e^{i u_2 \mathcal{L}_S}.
\end{equation}
\end{definition}

An easy calculation shows that the right side of \eqref{KK}  only depends on $u_2-u_1$ and that the operator $\mathcal{M}$ is well-defined.  

\begin{lemma}
\label{compa}
There exists a constant $C>0$ independent of $\lambda(\cdot)$ such that 
\begin{equation} \label{difference}
 \vert \vert \mathcal{Z}_0^{t,s} - e^{i(t-s) \mathcal{L}_S +(t-s) \lambda^{2}(s)\mathcal{M}} \vert \vert_{\infty} \leq C  \lambda^{2}(s) e^{C \lambda^{2}(s) (t-s)}  \vert \ln(\lambda(s) ) \vert ,
\end{equation}
 for all $t,s \geq 0$.
\end{lemma}
The proof of Lemma \ref{compa} is similar to Proposition 3.3 in \cite{DeR1}. 

\subsection{Properties of  $\mathcal{Z}^{s+\tau  \lambda^{-2}(s),s}$ }
\label{2.4.2}
We calculate the Lindblad operator $\mathcal{M}$ quite explicitly in Appendix B.3.  Using those calculations, it is easy to deduce that $\Pi_{11}$ is  a left-eigenvector of $\mathcal{M}$ with eigenvalue $0$. Indeed, $\langle \Pi_{11} \vert \mathcal{M} X \rangle =\text{Tr}(\Pi_{11} \mathcal{M} X)=0$, for all  $X \in \mathcal{B}(\mathcal{H}_S)$.   Using  formulae  \eqref{M} and \eqref{M'}, one verifies that $\mathcal{M} \vert 1_S \rangle=0$, where 
$$   1_S  =\sum_{i=1}^{n}  \Pi_{ii} $$
is the identity matrix  in  $\mathcal{B}(\mathcal{H}_S)$. The one-dimensional projection
  \begin{equation}
  \label{P}
  P:=\vert 1_S \rangle \langle \Pi_{11} \vert \in \mathcal{B}(\mathcal{B}(\mathcal{H}_S)) 
  \end{equation}
satisfies 
$$P \vert X   \rangle=\vert 1_S \rangle \langle     \Pi_{11} , X \rangle_{\mathcal{B}(\mathcal{H}_S)}=\vert 1_S \rangle \text{Tr}(  \Pi_{11}  X)$$
for all $X \in \mathcal{B}(\mathcal{H}_S)$. Moreover, $P$ commutes with $\mathcal{M}$:
 \begin{equation}
 P\mathcal{M}=\vert 1_S \rangle \langle \Pi_{11} \vert \mathcal{M}=0=\mathcal{M}\vert 1_S \rangle \langle \Pi_{11} \vert=\mathcal{M}P.
 \end{equation}

\begin{lemma} 
\label{spectrum}
If the Fermi-Golden-Rule conditions \eqref{Fer} are satisfied then $0$ is a non-degenerate eigenvalue of the Lindbladian $\mathcal{M}$.  The other eigenvalues of   $\mathcal{M}$  have a strictly negative real part. Furthermore, the projection $P=\vert 1_S \rangle \langle \Pi_{11} \vert$ satisfies 
$P\mathcal{M}=\mathcal{M}P=0$. 
\end{lemma}

That the non-zero eigenvalues of $\mathcal{M}$ have a strictly negative real  part is  a consequence of the Fermi-Golden-Rule conditions \eqref{Fer}. Since $\mathcal{L}_S$ and $\mathcal{M}$ commute, Lemma \ref{spectrum} yields the following corollary.
\begin{corollary}
\label{spec2}
The operator $e^{i(t-s)\mathcal{L}_S+(t-s)  \lambda^{2}(s) \mathcal{M}}$ has a non-degenerate eigenvalue $1$ corresponding to the eigenvector $ 1_S$. The projection $P$ given in  \eqref{P} commutes with the operator $e^{i(t-s)\mathcal{L}_S+(t-s)  \lambda^{2}(s) \mathcal{M}}$ and 
$P e^{i(t-s)\mathcal{L}_S+(t-s)  \lambda^{2}(s) \mathcal{M}}=P$.
\end{corollary}
\vspace{2mm}

\subsubsection{Spectrum of $\mathcal{Z}^{s+\tau  \lambda^{-2}(s),s}$ }
\label{Peruu}

\begin{lemma} 
\label{spectrum2}
Suppose that  Assumption \ref{Co}, Eq.(\ref{requ}), Assumption \ref{Fe} and Assumption \ref{Ev}, Eq. (\ref{CC}), are satisfied.  Let $0<\varepsilon_0<1$. There  are positive constants  $\tau_{\varepsilon_{0}}$ and  
$\lambda_{\varepsilon_{0}, \tau}>0$ such that, for any $\tau> \tau_{\varepsilon_{0}}$ and for any $\lambda(0) <\lambda_{\varepsilon_{0}, \tau}$,
\begin{equation}
\label{dec}
\mathcal{Z}^{s+\tau  \lambda^{-2}(s),s}= P(s)+ R(s),
\end{equation}
 for all  $s \geq 0$. The operators $P(s)$ and $R(s)$ have the following properties:
 \begin{itemize}
 \item  R(s) is a small perturbation, with $\vert \vert R(s) \vert \vert_{\infty} <\varepsilon_0$;
 
 \item  $P(s)$ has the form 
 $\vert 1_S  \rangle \langle \Pi(s) \vert$, where $\Pi(s)$ is a rank-1 projection, with $\Pi(s) \simeq \Pi_{11}$. 
 More precisely, $\Pi(s)$ converges to $\Pi_{11}$ in norm, as $s \rightarrow \infty$.
 
 \noindent The operator $P(s)$  projects onto  the (subspace spanned by the) eigenvector 
 
\noindent  $ 1_S $ of $\mathcal{Z}^{s+\tau  \lambda^{-2}(s),s}$. 
 \item  $P(s)$ commutes with $\mathcal{Z}^{s+\tau  \lambda^{-2}(s),s}$  and  \begin{equation} \label{proji} P(s)R(s)=R(s) P(s)=0. \end{equation}
 \end{itemize}

\end{lemma}
\vspace{2mm}

\begin{proof}
We first remark  that $  1_S $ is  an eigenvector of the operator $ \mathcal{Z}^{s+\tau  \lambda^{-2}(s),s}$ with associated eigenvalue $1$.  Indeed,
\begin{equation*}
\mathcal{Z}^{s+\tau  \lambda^{-2}(s),s}(1_S) P_{\Omega}=P_{\Omega} U^{*}(s+\tau  \lambda^{-2}(s),s) U(s+\tau \lambda^{-2}(s),s) P_{\Omega} = P_{\Omega}.
\end{equation*}
Let  $0<\varepsilon_0 < 1$ and  $s \geq 0$. We consider the disk $D_r$ of radius $1>r>0$ centered at the eigenvalue $1$ of $e^{   i \tau \lambda^{-2}(s) \mathcal{L}_S+ \tau \mathcal{M}}$ in the complex plane, and we choose $r$ such that $D_r \cap \sigma(e^{   i \tau \lambda^{-2}(s) \mathcal{L}_S+ \tau \mathcal{M}})=\{1\}$.   We  introduce 
$$m_{\sigma}:= \max \{ \Re (z ) \setminus \{0\} \vert z \in \sigma(\mathcal{M})\}.$$
 The finite set $\sigma(e^{   i \tau \lambda^{-2}(s) \mathcal{L}_S+  \tau \mathcal{M}}) \setminus \lbrace 1 \rbrace$ lies to the left of the vertical line given by the equation $\Re (z)=e^{\tau m_{\sigma}}$.  Since $m_{\sigma}<0$,  the radius 
 $r$ of the disk $D_r$ can be  set  to $1/2$, for sufficiently large $\tau$.

Corollary \ref{spec2} shows  that $P=\vert 1_S \rangle \langle \Pi_{11} \vert$ commutes with $e^{   i \tau \lambda^{-2}(s) \mathcal{L}_S+  \tau \mathcal{M}}$, and that $$Pe^{  i \tau \lambda(s)^{-2} \mathcal{L}_S+ \tau \mathcal{M}}=e^{  i \tau \lambda(s)^{-2} \mathcal{L}_S+ \tau \mathcal{M}}P=P.$$ The projection $\Pi_{11}$ is the only operator  of trace $=1$ such that $\langle \Pi_{11} \vert e^{  i \tau \lambda(s)^{-2} \mathcal{L}_S+ \tau \mathcal{M}}=\langle \Pi_{11} \vert$, and $P$ coincides  with  the Riesz projection
$$P=-\frac{1}{2 i\pi} \int_{\partial D_{1/2}}  \frac{1}{ e^{  i \tau \lambda(s)^{-2} \mathcal{L}_S+ \tau \mathcal{M}}-z} dz.$$ 
\noindent 
The  sequence $\vert \vert (e^{\mathcal { M}}-P)^n \vert \vert_{\infty}^{1/n}$ tends to $e^{m_{\sigma}}$, as $n$ tends to infinity, (a consequence of the spectral radius formula). Therefore there exists   $\tau_{\varepsilon_0} >0$ such that, for any $\tau >\tau_{\varepsilon_0}$, 
\begin{equation}
\label{choix1}
 \vert \vert  e^{ i  \tau \lambda(s)^{-2} \mathcal{L}_S+ \tau \mathcal{M}} -P \vert \vert_{\infty} \leq \vert \vert (e^{\mathcal{M}} -P)^{\tau} \vert \vert_{\infty}  < e^{\tau m_{\sigma}/2} <\varepsilon_0/2,
\end{equation}
for all $s \geq 0$. 

We  now choose $\tau>\tau_{\varepsilon_0}$ and  compare the spectra of $\mathcal{Z}^{s+\tau \lambda^{-2}(s),s}$  and $e^{  i \tau \lambda(s)^{-2} \mathcal{L}_S+ \tau \mathcal{M}}$. The second resolvent formula yields the formal Neumann series 
\begin{equation}
\begin{split}
\label{encorune}
\frac{1}{\mathcal{Z}^{s+\tau \lambda^{-2}(s),s}-z}=&\frac{1}{e^{  i \tau \lambda(s)^{-2} \mathcal{L}_S+ \tau \mathcal{M}}-z}  \\
&\qquad  \cdot \sum_{k=0}^{\infty}\left[  (e^{  i \tau \lambda(s)^{-2} \mathcal{L}_S+ \tau \mathcal{M}} -\mathcal{Z}^{s+\tau \lambda^{-2}(s),s})  \frac{1}{e^{  i \tau \lambda(s)^{-2} \mathcal{L}_S+ \tau \mathcal{M}}-z} \right]^{k}.
\end{split}
\end{equation}
 The resolvent  $  (e^{  i \tau \lambda(s)^{-2} \mathcal{L}_S+ \tau \mathcal{M}}-z)^{-1} $ is bounded in norm  on the circle $\partial D_{1/2}$ by  a constant   $C(\tau)>1$  that  depends on $\tau$, but not on  $\lambda(\cdot)$ or $\gamma$, because 
 $\sigma(\mathcal{L}_S )\subset \mathbb{R}$.
 
  Let $0< \varepsilon \ll 1$.  Lemmas \ref{lem4} and  \ref{compa} show that there exists  a constant $\lambda_{\varepsilon}$ depending on $\gamma$ and $\tau$ such that, for any 
 $\lambda(0)<\lambda_{\varepsilon}$,
\begin{equation}
\label{pertutu}
\vert \vert  e^{  i \tau \lambda(s)^{-2} \mathcal{L}_S+ \tau \mathcal{M}}  - \mathcal{Z}^{s+\tau \lambda^{-2}(s),s}\vert \vert_{\infty}<  \frac{\varepsilon}{  C^{2}(\tau)},
\end{equation}
for all $s \geq 0$.
If \eqref{pertutu} holds then the Neumann series in  \eqref{encorune} converges in norm 
$\vert \vert \cdot \vert \vert_{\infty}$ ,  uniformly in $s \geq 0$, and there exists a  bounded operator $A(s,z)$, such that
\begin{eqnarray*} 
\frac{1}{\mathcal{Z}^{s+\tau \lambda^{-2}(s),s}-z}=\frac{1}{e^{  i \tau \lambda(s)^{-2} \mathcal{L}_S+ \tau \mathcal{M}}-z}+ A(s,z),
\end{eqnarray*}
for all $s \geq 0$ and all $z \in \partial D_{1/2}$. Furthermore, $ \vert \vert A(s,z) \vert \vert_{\infty} < C \varepsilon$,   for all $s \geq 0$ and all $z \in \partial D_{1/2}$. Here  $C$ is a positive  constant  independent of  $\lambda$ and $\tau$. The Riesz projection $P(s)$, defined by
\begin{eqnarray*}
P(s):=-\frac{1}{2 i\pi} \int_{ \partial D_{1/2}} \frac{1}{\mathcal{Z}^{s+\tau \lambda^{-2}(s),s}-z} dz, 
\end{eqnarray*}
is one-dimensional,  and $\|P - P(s)\|_{\infty}=\mathcal{O}(\varepsilon)$,  for all $s \geq 0$.   That $P(s)$ is rank-one follows from the property that two projections $P$ and $Q$ with $\dim(\text{Ran} P) \neq \dim(\text{Ran} Q)$ must satisfy $\vert \vert P-Q \vert \vert_{2} \geq 1$, where the norm $\vert \vert \cdot  \vert \vert_{2}$ has been defined in \eqref{norm}.   For small $\varepsilon>0$, $P(s)$ must be rank-one, because the norms $\vert \vert \cdot \vert \vert_{2}$ and  $\vert \vert \cdot \vert \vert_{\infty}$ are equivalent.   The identity matrix, $  1_S $,  is an eigenvector  of $\mathcal{Z}^{s+\tau \lambda^{-2}(s),s}$ corresponding to the eigenvalue $1$, for all $s \geq 0$,  and   the Riesz projection $P(s)$ must project onto the subspace spanned by $ 1_S$.  Therefore
$$P(s)=\vert 1_S \rangle \langle \Pi(s) \vert,$$
where $\Pi(s) \in \mathcal{B}(\mathcal{H}_S)$ is an $n \times n$ matrix with trace one. 

Furthermore, 
$\|\Pi(s)-\Pi_{11}\|=\mathcal{O}(\varepsilon ),$ because 
$\vert \vert (P-P(s))(\vert \Pi_{11} \rangle-\vert \Pi(s) \rangle) \vert \vert = \langle  \Pi_{11} -\Pi(s) , \Pi_{11} -\Pi(s) \rangle_{\mathcal{B}(\mathcal{H}_S)}$.  To complete our proof we note that
 \begin{align*}
\vert \vert R(s) \vert \vert_{\infty}&=\vert \vert  \mathcal{Z}^{s+\tau \lambda^{-2}(s),s}-P(s) \vert \vert_{\infty}\\
&=  \vert \vert  \mathcal{Z}^{s+\tau \lambda^{-2}(s),s}-e^{  i \tau \lambda(s)^{-2} \mathcal{L}_S+ \tau \mathcal{M}}+ e^{  i \tau \lambda(s)^{-2} \mathcal{L}_S+ \tau \mathcal{M}}-P + P-P(s)\vert \vert_{\infty}\\
& \leq   \vert \vert  \mathcal{Z}^{s+\tau \lambda^{-2}(s),s}-e^{  i \tau \lambda(s)^{-2} \mathcal{L}_S+ \tau \mathcal{M}} \vert \vert_{\infty} +   \vert \vert  e^{  i \tau \lambda(s)^{-2} \mathcal{L}_S+ \tau \mathcal{M}}-P  \vert \vert_{\infty} +  \vert \vert   P-P(s)\vert \vert_{\infty},
\end{align*}
and  $\vert \vert  e^{  i \tau \lambda(s)^{-2} \mathcal{L}_S+ \tau \mathcal{M}}-P  \vert \vert_{\infty}$ is bounded by $\varepsilon_0/2$, thanks to our choice of  $\tau$;  see \eqref{choix1}.  The other terms  are bounded by a constant of order $\varepsilon$. Choosing $\varepsilon$ small enough, we can make sure that $\vert \vert R(s) \vert \vert_{\infty}<\varepsilon_0$. Furthermore, $\varepsilon$ can be made arbitrary small by  an appropriate choice of  $\lambda(0)$.
 \end{proof}

\section{ \large Rewriting $\mathcal{Z}^{t,0}$ as a sum of terms labelled by graphs}
\label{graphph}
 We  rewrite the Dyson expansion for  $\mathcal{Z}^{t,0}$  using graphs to label terms arising from partial re-summations of the Dyson series.  Our analysis involves three steps.\\
\vspace{1mm}

\noindent \textit{First step.} We discretize time on the Van Hove time scale and consider intervals $I_i:=[t_{i}, t_{i+1})$ with $t_{i+1}=t_{i} +\tau \lambda^{-2}(t_{i})$, $i=0,...,N-1$,  and $t_0=0$.   We  introduce  four Feynman rules that correspond to the four possible contraction schemes  in the Dyson expansion \eqref{DD1'} and we associate a Feynman diagram to each pairing  appearing under the integrals in  \eqref{DD1'}; see Paragraph \ref{Par1}.\\

\noindent \textit{Second step.}   We re-sum the contributions to the Dyson series corresponding to all  diagrams  that have the property that any correlation line starting in an interval $I_i$  is ending in the \textit{same} interval $I_i$, for any $i=0,...,N-1$.   We observe that this re-summation just yields the contribution of the operator 
$\mathcal{Z}^{t_{i+1},t_{i}}$ to the Dyson series for $\mathcal{Z}^{t_N,0}$; see subsection \ref{suba}. \\

 \noindent \textit{Third step.} We use the decomposition \eqref{dec} in Lemma \ref{spectrum2}, in order to  express  $ \langle \Psi(t) \vert O  \Psi(t) \rangle$ in the form of a convergent cluster expansion. The fact that the range of the projections $P(t_i)$ is one-dimensional plays an important role. Indeed, a product of operators 
 $A_1,..., A_n \in \mathcal{B} (\mathcal{B}(\mathcal{H}_S))$, when sandwiched  between two  projections $P(s_1)=\vert 1_S \rangle \langle \Pi(s_1) \vert $ and $P(s_2)=\vert 1_S \rangle \langle \Pi(s_2) \vert$, is equal to  
 \begin{equation*}
 P(s_1)A_1... A_n P(s_2)= P(s_2)  \text{ } \langle \Pi(s_2) \vert A_1... A_n \vert 1_s \rangle. 
\end{equation*}
Using this identity, we are able to assign a  scalar weight  to each element, $\mathcal{X}$, of a set, denoted by $\mathbb{P}_{N}$, of  ``polymers''. The set  $\mathbb{P}_{N}$ is constructed from the possible pairings (correlation lines) appearing in Wick's theorem;   see subsection \ref{secon}.

\subsection{Discretization of time, and Feynman rules }
\label{Par1}
We introduce a sequence $(t_i)_{i \geq 0}$ of times with the help of a recursion formula
\begin{equation}
\label{discr}
t_{i+1}=t_i +\tau \lambda^{-2}(t_i), \qquad t_0=0.
\end{equation}
We set
\begin{align}
\label{Ii}
I_{i}&:=\left[ t_{i}, t_{i+1}\right).
\end{align}
\small
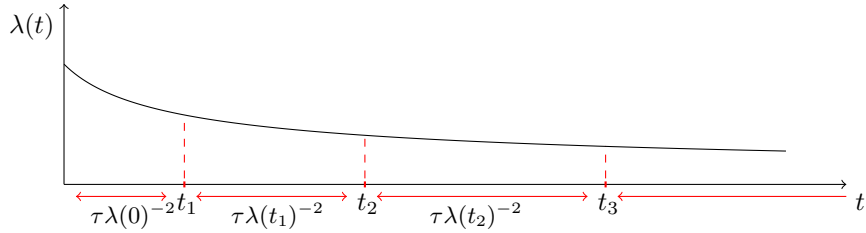
\begin{center}
\begin{tikzpicture}[scale=0.8]
\draw[->] (0,0) -- (13.0,0);
\draw[->] (0,0) -- (0,3.0);
\draw[-,color=red,thick] (2.0,-0.05) -- (2.0,0.05);
\draw[-,color=red,thick] (5.0,-0.05) -- (5.0,0.05);
\draw[-,color=red,thick] (9.0,-0.05) -- (9.0,0.05);

\draw[-,color=red,dashed] (2.0,0.1) -- (2.0,1.14);
\draw[-,color=red,dashed] (5.0,0.1) -- (5.0,0.8);
\draw[-,color=red,dashed] (9.0,0.1) -- (9.0,0.64);

 \draw (2.05,0) node[below] {$t_1$};
 \draw (5.05,0) node[below] {$t_2$};
 \draw (9.05,0) node[below] {$t_3$};

 \draw (1.15,-0.15) node[below] {\footnotesize $ \tau \lambda(0)^{-2}$};
 \draw (3.55,-0.15) node[below] { \footnotesize $ \tau \lambda(t_1)^{-2}$};
 \draw (6.85,-0.15) node[below] { \footnotesize $ \tau \lambda(t_2)^{-2}$};

\draw[<->,color=red] (0.2,-0.2) -- (1.7,-0.2);
\draw[<->,color=red] (2.2,-0.2) -- (4.7,-0.2);
\draw[<->,color=red] (5.2,-0.2) -- (8.7,-0.2);
\draw[<-,color=red] (9.2,-0.2) -- (13,-0.2);

 \draw (0,3.0) node[below left] {$\lambda(t)$};
 \draw (13.0,0) node[below right] {$t$};
\draw plot [domain=0:12.0,samples=200] (\x,{2.0*sqrt((1+\x)^(-1))});
\end{tikzpicture}
\end{center}
\normalsize

\label{ff}
We introduce four  Feynman rules  corresponding to the four possible contractions displayed in \eqref{F2}:  
\begin{center}
\begin{tikzpicture}[scale=0.7]

\draw[-] (0,0) -- (2.0,0);
\draw[-] (4.0,0) -- (6.0,0);
\draw[-,] (4,-3) -- (6,-3);
\draw[-] (8,-3) -- (10,-3);
\draw[-] (8,0) -- (10,0);
\draw[-] (12,-3) -- (14,-3.0);

 \draw (-0.5,0) node[left] {$  (r=0)$};
 \draw (-0.5,-3) node[left] {$(r=1)$};
 
 \draw[decorate, decoration={snake, segment length=1mm, amplitude=0.4mm}] (0.2,0) to[bend left=90](1.5,0);
 \draw[decorate, decoration={snake, segment length=1mm, amplitude=0.4mm}] (4.2,0) to[bend left=10](5.5,-3);
  \draw[decorate, decoration={snake, segment length=1mm, amplitude=0.4mm}] (8.2,-3) to[bend left=10](9.5,0);
   \draw[decorate, decoration={snake, segment length=1mm, amplitude=0.4mm}] (12.2,-3) to[bend right=90](13.5,-3);

 \draw (0.2,0) node[below] {$u_i$};
 \draw (1.5,0) node[below] {$v_i$};

  \draw (4.2,0) node[above] {$u_i$};
 \draw (5.5,-3) node[below] {$v_i$};
 
  \draw (8.2,-3) node[below] {$u_i$};
 \draw (9.5,0) node[above] {$v_i$};
 
   \draw (12.2,-3) node[above] {$u_i$};
 \draw (13.5,-3) node[above] {$v_i$};
 
  \draw (1,-4) node[below] {\textbf{(a)}};
    \draw (5,-4) node[below] {\textbf{(b)}};
      \draw (9,-4) node[below] {\textbf{(c)}};
        \draw (13,-4) node[below] {\textbf{(d)}};

\end{tikzpicture}
\end{center}
corresponding to the ``operator-valued amplitudes''
\begin{align}
\label{fey1}
\textbf{(a)}&:=\lambda(u_i) \lambda(v_i)  f(u_i-v_i) \textbf{L}(iG(u_i)) \textbf{L}(iG(v_i)),  \\
\textbf{(b)}&:=\lambda(u_i) \lambda(v_i) f(u_i-v_i)  \textbf{L}(iG(u_i)) \textbf{R}(iG(v_i)),   \\
\textbf{(c)}&:=\lambda(u_i) \lambda(v_i) f(v_i-u_i)  \textbf{R}(iG(u_i)) \textbf{L}(iG(v_i)),   \\
\label{fey4}
\textbf{(d)}&:=\lambda(u_i) \lambda(v_i)  f(v_i-u_i) \textbf{R}(iG(u_i)) \textbf{R}(iG(v_i)).
\end{align}
The expressions (\ref{fey1})-(\ref{fey4}) can be read off directly from (\ref{F2}). To any operator 
$$ \lambda( \underline{u})  \lambda( \underline{v} )   \mathcal{T}_S \left[ \prod_{i=1}^{k}   \textbf{F}(u_{i},r_i ;v_{i},r'_i)  \right] $$
 in \eqref{DD1'} there corresponds a unique Feynman diagram constructed according to the rules (\ref{fey1})-(\ref{fey4}), above. (Time ordering has to be carried out to determine the corresponding contribution to \eqref{DD1'}).  We are led to considering  diagrams of a kind indicated in Figure 1.
  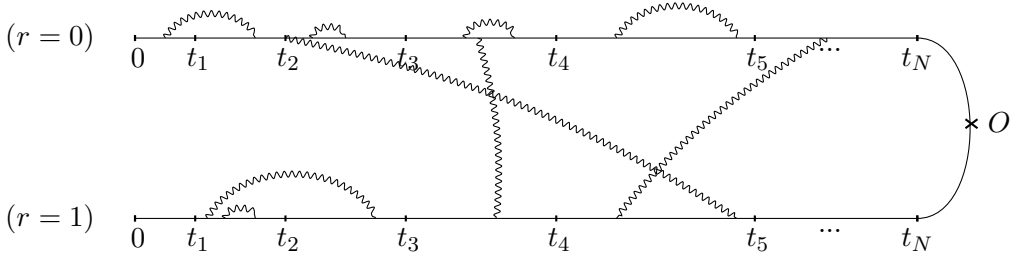
\begin{figure}[h]
\begin{center}
\begin{tikzpicture}[scale=0.8]
\draw[-] (0,0) -- (7.0,0);
\draw[-] (10,0) -- (13.0,0);
\draw[-,] (0,-3) -- (1,-3);
\draw[-] (1,-3) -- (4.5,-3);
\draw[-] (10,-3) -- (13.,-3.0);
\draw[-] (13,0.0) to[bend left=90] (13.0,-3);

\draw[-](7,0) -- (10.0,0);
\draw[-](4.5,-3) -- (10.0,-3);

\draw[-,thick] (0.0,-0.05) -- (0.0,0.05);
\draw[-,thick]  (13,-0.05) -- (13,0.05);

\draw[-,thick] (0.0,-3.05) -- (0.0,-2.95);
\draw[-,thick]  (13,-3.05) -- (13,-2.95);

\draw[-,thick]  (13.8,-1.5) -- (14,-1.35);
\draw[-,thick]  (13.8,-1.35) -- (14,-1.5);
\draw(14,-1.4) node[right] {$O$};

 \draw (-0.5,0) node[left] {$  (r=0)$};
 \draw (-0.5,-3) node[left] {$(r=1)$};

\draw[decorate, decoration={snake, segment length=1mm, amplitude=0.4mm}]  (0.5,0.0) to[bend left=60] (2,0.0);
\draw[decorate, decoration={snake, segment length=1mm, amplitude=0.4mm}] (2.95,0.0) to[bend left=90] (3.5,0.0);
\draw[decorate, decoration={snake, segment length=1mm, amplitude=0.4mm}] (5.7,0.0) to[bend left=10] (6.0,-3.0); 
\draw[decorate, decoration={snake, segment length=1mm, amplitude=0.4mm}] (5.5,0.0) to[bend left=90] (6.3,0.0);  

\draw[decorate, decoration={snake, segment length=1mm, amplitude=0.4mm}]  (1.5,-3.0) to[bend left=90] (2,-3.0);
\draw[decorate, decoration={snake, segment length=1mm, amplitude=0.4mm}] (1.2,-3.0) to[bend left=60] (4.0,-3.0);   
\draw[decorate, decoration={snake, segment length=1mm, amplitude=0.4mm}] (2.5,0.0) to[bend left=10] (10.0,-3.0); 
\draw[decorate, decoration={snake, segment length=1mm, amplitude=0.4mm}] (11.5,0.0) to[bend right=10] (8.0,-3.0);  
\draw[decorate, decoration={snake, segment length=1mm, amplitude=0.4mm}](8,0) to[bend left=60] (10,0.0);

\draw[-,thick] (0.0,-0.05) -- (0.0,0.05);
\draw[-,thick]  (1,-0.05) -- (1,0.05);
\draw[-,thick]  (2.5,-0.05) -- (2.5,0.05);
\draw[-,thick]  (4.5,-0.05) -- (4.5,0.05);
\draw[-,thick]  (7,-0.05) -- (7.0,0.05);
\draw[-,thick]  (10.3,-0.05) -- (10.3,0.05);
\draw[-,thick]  (13,-0.05) -- (13,0.05);

\draw[-,thick] (0.0,-3.05) -- (0.0,-2.95);
\draw[-,thick]  (1,-3.05) -- (1,-2.95);
\draw[-,thick]  (2.5,-3.05) -- (2.5,-2.95);
\draw[-,thick]  (4.5,-3.05) -- (4.5,-2.95);
\draw[-,thick]  (7,-3.05) -- (7.0,-2.95);
\draw[-,thick]  (10.3,-3.05) -- (10.3,-2.95);
\draw[-,thick]  (13,-3.05) -- (13,-2.95);

 \draw (0.05,0) node[below] {$0$};
 \draw (1.05,0) node[below] {$t_1$};
 \draw (2.55,0) node[below] {$t_2$};
 \draw (4.55,0) node[below] {$t_3$};
 \draw (7.05,0) node[below] {$t_4$};
 \draw (10.35,0) node[below] {$t_5$};
 \draw (11.55,0) node[below] {$...$};
 \draw (13.0,0) node[below] {$t_N$};

 \draw (0.05,-3) node[below] {$0$};
 \draw (1.05,-3) node[below] {$t_1$};
 \draw (2.55,-3) node[below] {$t_2$};
 \draw (4.55,-3) node[below] {$t_3$};
 \draw (7.05,-3) node[below] {$t_4$};
 \draw (10.35,-3) node[below] {$t_5$};
 \draw (11.55,-3) node[below] {$...$};
 \draw (13.0,-3) node[below] {$t_N$};
\end{tikzpicture}
\end{center}
\caption{\footnotesize A Feynman diagram.} \label{M1}
\end{figure}
\vspace{2mm}

\subsection{Resummation}
\label{presum}

\subsubsection{Re-summing  the Dyson expansion inside isolated double intervals }
We fix a natural number $N \in \mathbb{N}$.  To each  interval $I_i$, as defined in (\ref{Ii}), we associate an index $r$ that takes the values $0$ or $1$. We set $\textbf{I}_i:=\left(I_{i},0;I_{i},1\right)$  and call $\textbf{I}_i$ a  double interval.  It corresponds to the picture
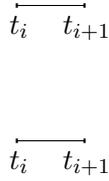
\begin{center}
\begin{tikzpicture}[scale=0.6]

\draw[-] (1,0) -- (2.5,0);
\draw[-,] (1,-3) -- (2.5,-3);

\draw[-,thick]  (1,-0.05) -- (1,0.05);
\draw[-,thick]  (2.5,-0.05) -- (2.5,0.05);

\draw[-,thick]  (1,-3.05) -- (1,-2.95);
\draw[-,thick]  (2.5,-3.05) -- (2.5,-2.95);

 \draw (1.05,0) node[below] {$t_{i}$};
 \draw (2.55,0) node[below] {$t_{i+1}$};
 \draw (1.05,-3) node[below] {$t_{i}$};
 \draw (2.55,-3) node[below] {$t_{i+1}$};
\end{tikzpicture}
\end{center}

\noindent The curved part from $(t_N,r=0)$ to $(t_N,r=1)$ in Figure \ref{M1}, which contains a point representing the observable $O$, is  also considered to be a double interval and is denoted by  $\textbf{I}_{N}$.  There are two types of  double intervals in Figure \ref{M1}: either $\textbf{I}_i$  is  connected to  some  $\textbf{I}_j$ ($j \neq i$) by a wavy (correlation) line; or all wavy (correlation) lines starting  in  $\textbf{I}_i$ end  in  $\textbf{I}_i$. In this latter case, we say that $\textbf{I}_i$ is isolated. Let  
$\textbf{I}_{i_0}$, with $i_0 \neq N$, be an isolated interval. We re-sum the Dyson expansion inside this interval  . Let $A \subseteq \{0,...,N -1\}$. A function 
$\mathds{1}_{A }$  is defined on the set of all possible pairings (Wick contraction schemes) as follows:
\begin{equation}
\mathds{1}_{A}(\underline{w}):=\left\{ \begin{array}{cl}  1 & \text{ if  }  \exists i \in A \text{ such that } \textbf{I}_{i}  \text{ is isolated, }\\0 & \text{ otherwise}. \end{array} \right.
\end{equation}
In the example where $A=\{i_0\}$ we rewrite
\label{suba}
\begin{equation}
\label{DD13}
 \mathcal{Z}^{t_N,0}(O)    = \mathcal{Z}_{\{i_0\}}^{t_N,0}(O) +( \mathcal{Z}^{t_N,0}(O) -\mathcal{Z}_{\{i_0\}}^{t_N,0}(O))  \end{equation}
where
\begin{align}
\label{DD13}
 \mathcal{Z}_{\{i_0\}}^{t_N,0}(O)   & = \sum_{k=0}^{\infty}  \underset{[0,t_N]^{2k}}{\int} d\mu_k(\underline{w})  \lambda( \underline{w})  \text{ } \mathds{1}_{\{i_0\}}(\underline{w}) \mathcal{T}_S \left[    
  \textbf{F}( \underline{w})  \right] \left[ O_{t_N} \right].
\end{align}
To shorten our formulae, we have introduced the notation
\begin{equation}
\label{FF}
 \textbf{F}( \underline{w}) := \prod_{i=1}^{k}   {\bf{F}}(u_{i}, r_{i}; v_{i}, r'_{i}),
 \end{equation}
 and
 \begin{equation}
 \label{F}
 F( \underline{w}) := \prod_{i=1}^{k} F(u_{i}, r_{i}; v_{i}, r'_{i}),
 \end{equation}
 where ${\bf{F}}(u, r; v, r')$ has been defined in \eqref{F2} and $F(u_i,r_i;v_i,r'_i)$ in \eqref{F3}. We now explain how to split the integrations in the formula for $ \mathcal{Z}_{\{i_0\}}^{t_N,0}(O)$. Interchanging summations and integrations will be carried out without further mention, because the Dyson series converges in norm;  see $\eqref{labound}$. For every pairing $\underline{w}$ consisting of $k$ pairs, it is convenient to write
\begin{equation}
\label{1G}
\mathds{1}_{\{i_0\}}(\underline{w})=  \sum_{m=0}^{k} \mathds{1}_{\{i_0\},m} (\underline{w}),
  \end{equation}
where $ \mathds{1}_{\{i_0\},m} (\underline{w})=1$ if $\textbf{I}_{i_0}$ is isolated and contains exactly $m$ pairs, and is equal to zero otherwise. We plug (\ref{1G})  into (\ref{DD13}). This yields
\begin{equation}
\label{DD14}
\begin{split}
 \mathcal{Z}_{\{i_0\}}^{t_N,0}(O)=\sum_{k=0}^{\infty}  \sum_{m=0}^{k}  \underset{[0,t_N]^{2k}}{\int} d\mu_k (\underline{w})  \lambda( \underline{w})    \text{ } \mathds{1}_{\{i_0\} ,m}(\underline{w}) \mathcal{T}_S \big[   {\bf{F}}(\underline{w})  \big] \left[ O_{t_N} \right].
\end{split}
\end{equation}
Given some pairing $\underline{w}$, with $ \mathds{1}_{\{i_0\} ,m}(\underline{w}) =1,$ for some $m \in \mathbb{N}$, there are two unique pairings $\underline{\tilde{w}}$ and $\underline{\hat{w}}$,  with  $\underline{w}=\underline{\tilde{w}} \cup \underline{\hat{w}}$,  such  that \textit{all} times in $\underline{\hat{w}}$  lie in $\textbf{I}_{i_0}$, whereas  \textit{no} time in   $\underline{\tilde{w}}$ lies in $\textbf{I}_{i_0}$. After a change of variables we can factorize the integral in \eqref{DD14} into two distinct integrals. Exchanging summation over $k$ with summation over $m$, we  get that 
\begin{equation*}
\begin{split}
\mathcal{Z}_{\{i_0\}}^{t_N,0}(O)=\sum_{k=0}^{\infty}  & \underset{[0,t_N]^{2k}}{\int} d\mu_{k}  (\underline{\tilde{w}})  \lambda( \underline{\tilde{w}})  \text{ }  \chi_{\{0,...N-1\} \setminus \{i_0\}}( \underline{\tilde{w}})\\
 & \mathcal{T}_S \Big(    \textbf{F}( \underline{\tilde{w}})  \sum_{m=0}^{\infty}  \underset{ I_{i_0}}{\int} d\mu_m (\underline{\hat{w}})  \lambda( \underline{\hat{w}})    \mathcal{T}_S \left[   \textbf{F}( \underline{\hat{w}})  \right] \Big) \left[ O_{t_N} \right].
 \end{split}
\end{equation*}
We recognize the expansion of the effective propagator $e^{i t_{i_{0}} \mathcal{L}_S}\mathcal{Z}^{t_{i_0+1},t_{i_0}}e^{-i t_{i_0+1} \mathcal{L}_S}$  inside the parenthesis on the right side of this equation; see \eqref{DD1'}.  Note that $ \chi_{\{0,...N-1\} \setminus \{i_0\}}$  only selects pairings  with no pairs inside $\textbf{I}_{i_0}$ or linked to $\textbf{I}_{i_0}$. Repeating this procedure for each interval,  we obtain the following lemma.

\begin{lemma} 
\label{coiso}
\begin{equation}
\label{DDtt}
\begin{split}
 \mathcal{Z}^{t_N,0}(O)  = \sum_{A \subseteq \{0,...,N-1 \} } & \sum_{k=0}^{\infty}      \underset{[0,t_N]^{2k}}{\int} d\mu_{k} (\underline{w})  \lambda( \underline{w})    \text{ }   \big(1- \mathds{1}_{\{0,...,N-1 \} \setminus A}   (\underline{w}) \big)  \chi_{\{0,...,N-1\} \setminus A}(\underline{w}) \\
 &\mathcal{T}_S \Big(      \textbf{F}( \underline{w} ) \text{ }  \prod_{j \in A} \big( e^{i t_{j} \mathcal{L}_S}\mathcal{Z}^{t_{j+1},t_{j}}e^{-i t_{j+1} \mathcal{L}_S} \big) \Big) \left[ O_{t_N} \right] ,
 \end{split}
\end{equation}
for all $O \in \mathcal{B}(\mathcal{H}_S)$.
\end{lemma}
\vspace{2mm}

\subsection{The cluster expansion}
\label{secon}
We plan to use the Dyson series for $\mathcal{Z}^{t,s}$ and Lemma \ref{spectrum2}, namely the identity
\begin{equation}
 \label{dec2}
 \mathcal{Z}^{t_{i+1}, t_i} = P(t_i)+R(t_i),
 \end{equation}
to  rewrite $\langle \Psi(t_N)  \vert (O  \otimes 1)   \Psi(t_N) \rangle$  in the form of a cluster- or polymer expansion. Observing that
\begin{equation*}
e^{i t_{i_{0}} \mathcal{L}_S}\mathcal{Z}^{t_{i_0+1},t_{i_0}}e^{-i t_{i_0+1} \mathcal{L}_S} e^{i t_{i_{0}+1} \mathcal{L}_S}\mathcal{Z}^{t_{i_0+2},t_{i_0+1}} e^{-i t_{i_0+2} \mathcal{L}_S}
=e^{i t_{i_{0}} \mathcal{L}_S}\mathcal{Z}^{t_{i_0+1},t_{i_0}} \mathcal{Z}^{t_{i_0+2},t_{i_0+1}} e^{-i t_{i_0+2}  \mathcal{L}_S}
\end{equation*}
and that
\begin{equation*}
e^{i t_{N-1} \mathcal{L}_S} \mathcal{Z}^{t_{N},t_{N-1}} e^{-i t_{N} \mathcal{L}_S} O_{t_N}=e^{i t_{N-1} \mathcal{L}_S} \mathcal{Z}^{t_{N},t_{N-1}} O,
\end{equation*}
we see that the operators $e^{it_j \mathcal{L}_S}$  cancel each other in the product inside the parentheses in \eqref{DDtt} unless they are located at the endpoints of  a union of adjacent isolated  intervals. We   extend   the definition of the time-ordering operator $\mathcal{T}_S$   in such a way that  $\mathcal{T}_S$ places 
$e^{i t_{l} \mathcal{L}_S}$ on the right of $e^{-i t_{l} \mathcal{L}_S}$  and $e^{-i t_{l+1} \mathcal{L}_S}$ on the right of  $\mathcal{Z}^{t_{l+1},t_{l}}$. We may then write
\begin{equation}
\label{DDt}
\begin{split}
 \mathcal{Z}^{t_N,0}(O)=\sum_{A  \subseteq \{0,...,N-1 \} }\sum_{k=0}^{\infty} &    \underset{[0,t_N]^{2k} }{\int} d\mu_{k} (\underline{w})  \lambda( \underline{w})    \text{ }   \big(1- \mathds{1}_{\{0,...,N-1 \} \setminus A}   (\underline{w}) \big) \chi_{\{0,...,N-1\} \setminus A}(\underline{w} )   \\
 &\mathcal{T}_S \Big(    \big( \prod_{l \in A}  e^{-i t_{l+1} \mathcal{L}_S}  \big)    \textbf{F}( \underline{w} )   \big( \prod_{l \in A}  e^{i t_{l} \mathcal{L}_S}  \big)  \big( \prod_{j \in A} \mathcal{Z}^{t_{j+1},t_{j}} \big)  \Big) [ O_{t_N} ].
 \end{split}
\end{equation} 

\noindent Next, we  insert (\ref{dec2}) into (\ref{DDt})  and expand the resulting expression as a sum of products of $P's$ and $R's$. This yields
\begin{equation}
\label{DDt2}
\begin{split}
 \mathcal{Z}^{t_N,0}(O)&=\sum_{A \subseteq \{0,...,N-1 \}} \sum_{C \subseteq A } \sum_{k=0}^{\infty}     \underset{[0,t_N]^{2k} }{\int} d\mu_{k} (\underline{w})  \lambda( \underline{w})    \text{ }  \big(1- \mathds{1}_{\{0,...,N-1 \} \setminus A}   (\underline{w}) \big) \chi_{\{0,...N-1\} \setminus A}(\underline{w} )   \\
 &\mathcal{T}_S \Big(    \big( \prod_{l \in A}  e^{-i t_{l+1} \mathcal{L}_S}  \big)    \textbf{F}( \underline{w} )   \big( \prod_{l \in A}  e^{i t_{l} \mathcal{L}_S}  \big)  \big( \prod_{j \in C} R(t_j) \big)  \big( \prod_{m \in A \setminus C} P(t_m) \big)  \Big) [ O_{t_N} ].
 \end{split}
\end{equation} 
\vspace{2mm}

\noindent

\subsubsection{Construction of the set $\mathbb{P}_{N}$}\label{PN}
We now elucidate the structure of the polymer set $\mathbb{P}_N$  that we use to re-organize the sums and integrals in \eqref{DDt2}.\\
\begin{itemize}[leftmargin=*]
  \item  We introduce ``decorated" vertices  by associating capital letters $R$ (for red), $P$ (for purple), or $B$ (for blue) to every integer $i \in \{0,...,N-1 \}$. An integer $i$ labeled with an $R$, $(i,R)$, corresponds to the perturbation $R(t_i)$.  An integer $i$ labelled with a $P$, $(i,P)$, corresponds to the projection $P(t_i)$.  An integer $i$ labelled with a $B$, $(i,B)$, corresponds to an interval $\textbf{I}_i$ that is not isolated. In formula \eqref{DDt2} above, integers in the sets $C$ are labelled with an $R$, integers in the sets $A \setminus C$ are labeled with a $P$, and integers in the sets $\{0,...,N-1 \} \setminus A$ are labelled with a $B$. The integer $N$ (corresponding to the observable $O$) is labelled with an $R$ and is considered to be an $R$-vertex. \\ 

 \item  We introduce \textit{decorated graphs} on the the set $\{0,...,N \}$. A decorated graph $\mathcal{G}$ is a pair $(\mathcal{V}(\mathcal{G}),\mathcal{E}(\mathcal{G}))$. The vertex set  $\mathcal{V}(\mathcal{G})$  consists of $B$ and/or $R$-vertices. The edge set  $\mathcal{E}(\mathcal{G})$ consists of  edges $((i,B);(j,B))$ joining two distinct $B$-vertices.  The distance between two decorated graphs is defined by 
 \begin{equation}
\mbox{dist} \left(\mathcal{G}_1, \mathcal{G}_2\right):=\underset{i \in \mathcal{G}_1, \text{ } j \in \mathcal{G}_2}{\min}   \text{ }\vert i-j  \vert. 
\end{equation}

\item A  \textit{connected} graph is an $R$-vertex or a  decorated graph $\mathcal{G}=(\mathcal{V}(\mathcal{G}),\mathcal{E}(\mathcal{G}))$ such that  $\mathcal{V}(\mathcal{G})$ only contains $B$-vertices and such that the graph $\mathcal{G}$ is connected (in the usual sense of graph theory).\\

 \item A polymer $\mathcal{X} \in \mathbb{P}_N$ is  a union of \textit{disjoint connected} graphs  $\mathcal{G}_1,....,\mathcal{G}_n$ (for some $n \in \mathbb{N}$)  such that $\mbox{dist}\left( \cup_{j \in J}\mathcal{G}_j, \mathcal{X} \setminus (\cup_{j \in J}\mathcal{G}_j) \right) =1$, for all  $J \subsetneq \{1,...,n \}$. The vertices of $\mathcal{X}$ are denoted by $\mathcal{V}(\mathcal{X})$, and the edges  by $\mathcal{E}(\mathcal{X})$.\\
 \end{itemize}

\subsubsection{The cluster expansion}
\noindent We use the  polymer set $\mathbb{P}_N$ to  rewrite the Dyson series in \eqref{DDt2} in a more convenient form. Starting from \eqref{DDt2}, with contributions corresponding to $P$- and $R$-vertices 
re-summed, we remark that the intervals $\textbf{I}_i$ corresponding to B-vertices  (where $i \in \{0,...,N-1 \} \setminus A$) may be connected by pairings, $\underline{w}$, with  $ \mathds{1}_{\{0,...,N-1 \} \setminus A}   (\underline{w})=0$, in many different ways.   We associate  a  decorated  graph $\mathcal{G}$  to every subset $A \subseteq \{0,...,N-1 \}$ and to every pairing $\underline{w}$  with $ \mathds{1}_{\{0,...,N-1 \} \setminus A}   (\underline{w})=0$ by labeling elements of $A$ by an  index $R$ or  $P$  and by drawing an edge between $(i,B)$ and  $(j,B)$  if  there is a correlation line starting in $\textbf{I}_i$  and ending in  $\textbf{I}_j$.  The vertex $(N,R)$, corresponding to the observable $O$, is added to $\mathcal{G}$. The  decorated graph $\mathcal{G}$ can be  rewritten as a disjoint union of connected components. We fuse  the adjacent connected components of $\mathcal{G}$ and obtain a collection of non-adjacent  polymers $\mathcal{X}_1,...,\mathcal{X}_n \in  \mathbb{P}_N$, for some $n \leq N$. 
\begin{figure}[h]
\begin{center}
\begin{tikzpicture}[scale=0.7]
\draw[-] (0,0) -- (7.0,0);
\draw[-] (10,0) -- (13.0,0);
\draw[-,] (0,-3) -- (1,-3);
\draw[-] (1,-3) -- (4.5,-3);
\draw[-] (10,-3) -- (13.,-3.0);
\draw[-] (13,0.0) to[bend left=90] (13.0,-3);

\draw[-](7,0) -- (10.0,0);
\draw[-](4.5,-3) -- (10.0,-3);

\draw[-,thick] (0.0,-0.05) -- (0.0,0.05);
\draw[-,thick]  (13,-0.05) -- (13,0.05);

\draw[-,thick] (0.0,-3.05) -- (0.0,-2.95);
\draw[-,thick]  (13,-3.05) -- (13,-2.95);

\draw[-,thick]  (13.8,-1.5) -- (14,-1.35);
\draw[-,thick]  (13.8,-1.35) -- (14,-1.5);
\draw(14,-1.4) node[right] {$O$};

\draw[decorate, decoration={snake, segment length=1mm, amplitude=0.4mm}]  (0.5,0.0) to[bend left=20] (9,0.0);
\draw[decorate, decoration={snake, segment length=1mm, amplitude=0.4mm}] (2.95,0.0) to[bend left=90] (3.5,0.0);
\draw[decorate, decoration={snake, segment length=1mm, amplitude=0.4mm}] (5.7,0.0) to[bend left=10] (6.0,-3.0); 
\draw[decorate, decoration={snake, segment length=1mm, amplitude=0.4mm}] (5.5,0.0) to[bend left=90] (6.3,0.0);  

\draw[decorate, decoration={snake, segment length=1mm, amplitude=0.4mm}]  (1.5,-3.0) to[bend left=90] (2,-3.0);
\draw[decorate, decoration={snake, segment length=1mm, amplitude=0.4mm}]  (0.4,-3.0) to[bend left=90] (0.9,-3.0);
\draw[decorate, decoration={snake, segment length=1mm, amplitude=0.4mm}] (0.8,0.0) to[bend left=10] (10.0,-3.0); 
\draw[decorate, decoration={snake, segment length=1mm, amplitude=0.4mm}] (11.5,0.0) to[bend right=10] (8.0,-3.0);  
\draw[decorate, decoration={snake, segment length=1mm, amplitude=0.4mm}](8,0) to[bend left=60] (10,0.0);

\draw[-,thick] (0.0,-0.05) -- (0.0,0.05);
\draw[-,thick]  (1,-0.05) -- (1,0.05);
\draw[-,thick]  (2.5,-0.05) -- (2.5,0.05);
\draw[-,thick]  (4.5,-0.05) -- (4.5,0.05);
\draw[-,thick]  (7,-0.05) -- (7.0,0.05);
\draw[-,thick]  (10.3,-0.05) -- (10.3,0.05);
\draw[-,thick]  (13,-0.05) -- (13,0.05);

\draw[-,thick] (0.0,-3.05) -- (0.0,-2.95);
\draw[-,thick]  (1,-3.05) -- (1,-2.95);
\draw[-,thick]  (2.5,-3.05) -- (2.5,-2.95);
\draw[-,thick]  (4.5,-3.05) -- (4.5,-2.95);
\draw[-,thick]  (7,-3.05) -- (7.0,-2.95);
\draw[-,thick]  (10.3,-3.05) -- (10.3,-2.95);
\draw[-,thick]  (13,-3.05) -- (13,-2.95);

 \draw (0.05,0) node[below] {$0$};
 \draw (1.05,0) node[below] {$t_1$};
 \draw (2.55,0) node[below] {$t_2$};
 \draw (4.55,0) node[below] {$t_3$};
 \draw (7.05,0) node[below] {$t_4$};
 \draw (10.35,0) node[below] {$t_5$};
 \draw (13.0,0) node[below] {$t_6$};

 \draw (0.05,-3) node[below] {$0$};
 \draw (1.05,-3) node[below] {$t_1$};
 \draw (2.55,-3) node[below] {$t_2$};
 \draw (4.55,-3) node[below] {$t_3$};
 \draw (7.05,-3) node[below] {$t_4$};
 \draw (10.35,-3) node[below] {$t_5$};
 \draw (13.0,-3) node[below] {$t_6$};

 
\fill (0.5,-6) circle (0.1cm);
\fill[color=black!30] (1.75,-6) circle (0.1cm);
\fill[color=black!70] (3.5,-6) circle (0.1cm);
\fill[color=black!30](5.5,-6) circle (0.1cm);
\fill (8.5,-6) circle (0.1cm);
\fill (11.5,-6) circle (0.1cm);
\fill (13.5,-6) circle (0.1cm);

 \draw (0.5,-6) node[below] {  \footnotesize $ (0,B)$};
  \draw (1.75,-6) node[below] {  \color{black!30} \footnotesize $  (1,P)$};
  \draw (3.5,-6) node[below] { \color{black!60} \footnotesize$  (2,R)$};
  \draw (5.5,-6) node[below] {  \color{black!30} \footnotesize $  (3,P)$};
  \draw (8.5,-6) node[below] {\footnotesize $(4,B)$};
  \draw (11.5,-6) node[below] {\footnotesize $  (5,B)$};
    \draw (13.5,-6) node[below] { \footnotesize $   (6,R)$};
  
\draw[-] (0.5,-6) to[bend left=40] (8.5, -6.0); 
\draw[-] (8.5,-6) to[bend left=60] (11.5, -6.0);

\end{tikzpicture}
\end{center}
\footnotesize \caption{ \footnotesize A  pairing $\underline{w}$ is represented by a Feynman diagram.   An example is given in the upper drawing of this figure. We  generate graphs on $\{0,....,N\}$  by drawing  edges between intervals that are paired at least by one correlation line and by  assigning an index  $P$ or $R$ to isolated intervals.  The pairing we chose here  generates 8  decorated graphs on the vertex set $\{0,...,6\}$  because  there are  $8$ possible  combinations of  decorations (R or P) associated to the isolated vertices $(1, \cdot)$, $(2, \cdot)$ and $(3, \cdot)$.   One of them is drawn on the lower picture. This decorated graph has three connected components: $\{(2,R)\}$, $\{(6,R)\}$, and the connected graph $\mathcal{G}$ with vertex set $\mathcal{V}(\mathcal{G})=\{(0,B),(4,B),(5,B)\}$ and edge set  $\mathcal{E}(\mathcal{G})=\{ ((0,B);(4,B)), ((4,B);(5,B))\}$. Fusing the adjacent connected components, this generates two-non adjacent polymer:  $\mathcal{X}_1=\{(2,R)\}$ and $\mathcal{X}_2=\mathcal{G} \cup \{(6,R)\}$.  }
\end{figure}
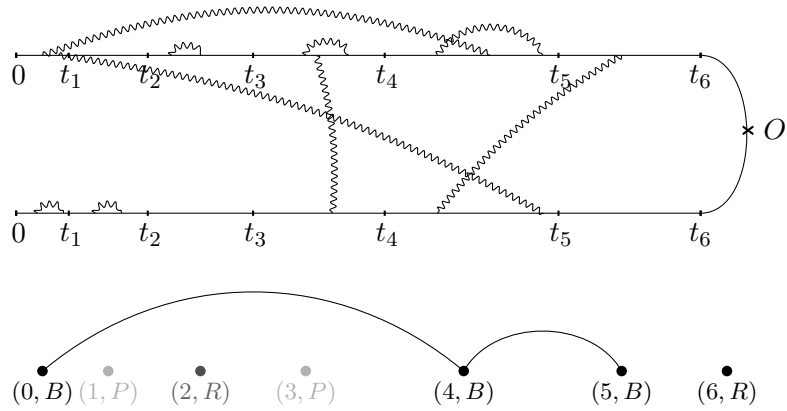

 \noindent Summing terms in the Dyson series labelled by subsets  $A \subseteq \{0,...,N -1 \}$,  after  integrating over all  pairings $\underline{w}$  such that  
 $\mathds{1}_{\{0,...,N-1 \} \setminus A}   (\underline{w})=0$  in \eqref{DDt2}, amounts to the same as summing terms labelled by arbitrary collections of  non-adjacent polymers, after integrating over all pairings compatible with these polymers (pairings whose correlation lines follow the edges of the polymers). At their boundaries, polymers are surrounded by $P$-vertices (corresponding to one-dimensional projections), and the contributions corresponding to non-adjacent polymers factorize. This observation implies that the expectation value $\langle \Psi(t_N) \vert  O    \Psi(t_N) \rangle$ can be represented in the form of a cluster expansion for a one-dimensional gas of polymers, with polymers corresponding to the elements of the set $\mathbb{P}_N$. In what follows, the cardinality of a set $X$ is denoted by $\vert X  \vert$.
\vspace{2mm}

\begin{proposition}
\label{clustt}
\leavevmode
\begin{enumerate} 
\item There is a  complex-valued function $p: \mathbb{P}_N \rightarrow \mathbb{C}$  such that 
\begin{align}
\label{eqclu1}
\langle \Psi (t_N) \vert (O  \otimes 1)   \Psi(t_N) \rangle&=  \sum_{n=1}^{N} \frac{1}{n!} \underset{\text{dist}(\mathcal{X}_i,\mathcal{X}_j) \geq 2, \text{ } (N,R) \in  \cup_{i=1}^{n}\mathcal{V}( \mathcal{X}_{i})}{\underset{\mathcal{X}_1,...,\mathcal{X}_n \in \mathbb{P}_{N}}{\sum}}   p(\mathcal{X}_1) \text{ }... \text{ } p(\mathcal{X}_n). 
\end{align}
The weight $p(\mathcal{X})$ depends  on the observable $O$ only if $(N,R)$ is a vertex of the polymer $\mathcal{X}$.
\vspace{2mm}

\item If $\lambda(0)$  is sufficiently small,  then 
\begin{equation} \label{kokori}
\sum_{\mathcal{X}', \text{ dist}(\mathcal{X},\mathcal{X}') \leq 1} \vert  p(\mathcal{X}') \vert  \text{ }e^{ \vert  \mathcal{V}(\mathcal{X}') \vert } \leq \vert  \mathcal{V}(\mathcal{X}) \vert , \qquad \forall \chi \in \mathbb{P}_N, 
\end{equation}
where   $\mathcal{V}(\mathcal{X})$ is the vertex set of the polymer $\mathcal{X}$. The  critical value of  $\lambda(0)$ such that  \eqref{kokori} is satisfied can be chosen  uniformly in $N$.
\end{enumerate}

\end{proposition}
\vspace{4mm}

\begin{remark}  The factors $1/n!$ on the right side of  \eqref{eqclu1} account for over-counting that originates in summing over all permutations of the polymers $\mathcal{X}_1,...,\mathcal{X}_n$ in $\mathbb{P}_{N}$ . Clearly, the maximal number, $n$, of polymers  appearing in the sums on the right side of Eq. \eqref{eqclu1}
  is finite, with $n<N$.
\end{remark}

\begin{remark} \label{332}
Let $\mathcal{X} \in \mathbb{P}_N$ and  let  $ \beta \in (0,\alpha-2)$; (we remind the reader that $\alpha$ is the decay rate of the two point correlation function $f$  with time: $\vert f(s) \vert \propto (1+s)^{-\alpha}$).  If $\lambda(\cdot)$ were constant, we would have to  replace the positive function $a(\mathcal{X}')= \vert \mathcal{V}(\mathcal{X}') \vert$ in the Kotecky-Preiss criterion \eqref{kokori} by  $$a(\mathcal{X}')=\vert \mathcal{V}(\mathcal{X}') \vert+ \beta  \ln(d(\mathcal{X}')) $$  in order to control the limit $N \rightarrow +\infty$, where $$d(\mathcal{X}'):=1+ \max( \{i \mid (i , \cdot )\in \mathcal{V}(\mathcal{X}')\})-\min( \{i \mid (i , \cdot )\in \mathcal{V}(\mathcal{X}')\})$$ is the diameter of the polymer $\mathcal{X}'$.  We refer the reader to  \cite{DeR1}  for more details, and to remarks  \ref{411} and \ref{412} in Section \ref{laco}. We can omit the term $\beta \ln(d(\mathcal{X}'))$  here, because the coupling  $\lambda(t)$ tends to $0$ as $t$ tends to $+\infty$.
\end{remark}

\begin{remark} \label{1.1}
Part (1) of Proposition \ref{clustt} is proven in Appendix \ref{AppC}. The proof of the Kotecky-Preiss criterion \eqref{kokori} needs some amount of work and is  given in Section \ref{sectionko}. The cluster expansion \eqref{eqclu1} can also be applied when $O$ happens to be the identity, $1_S$. Then
 \begin{align}
\label{eqclu11}
\langle \Psi (t_N) \vert (1_S  \otimes 1)   \Psi(t_N) \rangle&=  1+\sum_{n=1}^{N} \frac{1}{n!} \underset{\text{dist}(\mathcal{X}_i,\mathcal{X}_j) \geq 2, \text{ } (N,R) \notin \cup_{i=1}^{n}\mathcal{V}( \mathcal{X}_{i})}{\underset{\mathcal{X}_1,...,\mathcal{X}_n \in \mathbb{P}_{N}}{\sum}}   p(\mathcal{X}_1) \text{ }... \text{ } p(\mathcal{X}_n), 
\end{align}
and the weights $p(\mathcal{X})$ are then all independent of an ``observable''.  The right side of \eqref{eqclu11} is then equal to 1, because the initial state $\Psi $ is assumed to be normalized. 
\end{remark}

\begin{remark}  The cluster expansion converges in the limit where $N \rightarrow \infty$, because the coupling function $\lambda(t)$ is small, for all  times $t$, and because the two-point correlation function $f(t)$ is ``twice integrable''. We investigate this limit in Section \ref{conv}.
\end{remark}

We conclude this paragraph by introducing some useful notions and  notations that enable us to write the exact expressions for  the weights $p(\cdot)$. Let $\mathcal{X}$ be a polymer. We say that a  set $U=\{(i,.),(i+1,.),...,(i+k,.)\} \subseteq \mathcal{V}(\mathcal{X})$  is  a \textit{maximal block} of neighboring  vertices  if there is no  set $V \subseteq \mathcal{V}(\mathcal{X})$ of  neighboring  vertices such that $ U \subsetneq V$. We denote by  
$\mathcal{U}(\mathcal{X})$ the set of maximal blocks of neighboring vertices  of  $\mathcal{X}$.  Maximal blocks of neighboring vertices are  surrounded at their boundaries (extreme points) by vertices corresponding to double intervals where a one-dimensional projection $P$ is chosen. The notion of ``maximal blocks of neighboring vertices'' is therefore helpful in the formulation of the cluster expansion \eqref{eqclu1} and the weights $p(\mathcal{X)}$. We prove in Appendix \ref{AppC} that 
\begin{equation}
\label{p(G)}
p(\mathcal{X}):=\sum_{k=0}^{\infty }  \underset{[0,t_N]^{2k}} {\int} d\mu_k  (\underline{w})  \lambda( \underline{w})    \text{ }   \chi_{\mathcal{X} } (\underline{w})  \text{ F}(\underline{w})   \prod_{U \in \mathcal{U}(\mathcal{X})} h_U(\underline{w}),
\end{equation}
with
\begin{equation}
\label{hU}
 h_U(\underline{w}):=\langle \Pi (t_{m(U)}) \vert \mathcal{T}_S \Big( \underset{(l,B) \in U  }{\prod} \Big[ e^{-it_{l} \mathcal{L}_S} \big( \prod_{(t,r) \in  \underline{w} \cap \textbf{I}_l } (i G)(t,r) \big) e^{it_{l+1} \mathcal{L}_S} \Big]   \underset{(j,R) \in U  }{\prod} R(t_j) \Big) 1_S \rangle,
\end{equation}
 and where we have set  
 \begin{equation}
 \label{mmm}
 \Pi(t_{-1}):= \vert \varphi  \rangle \langle \varphi \vert, \qquad  R(t_{N}):=\textbf{L}(O), \qquad \text{and} \qquad m(U):= \min \{  i \mid (i, \cdot) \in U \}-1.
 \end{equation}
The function $ \chi_{\mathcal{X}} (\underline{w})$ is equal to $1$ only  for pairings $\underline{w}$ that are compatible with the edges of  $\mathcal{X}$ (in the sense that the correlation lines of the pairing $\underline{w}$ ``follow" the edges of $\mathcal{X}$, and that there is at least one correlation line in $\underline{w}$  for each edge of $\mathcal{X}$). It takes the value zero otherwise. 
\vspace{2mm}

\subsubsection{The exponentiated form of the cluster expansion} \label{expf}
 In Appendix \ref{clustersec}, some important results concerning convergence criteria for cluster expansions are summarized; (see \cite{Ue,FPr} for more details). We now  use Proposition \ref{Ue} (see Appendix A) and Proposition \ref{clustt} to rewrite the cluster expansions given in  \eqref{eqclu1} and  \eqref{eqclu11} as exponentials of convergent series.  In Section \ref{proo} below, we  use the  exponentiated form of the cluster expansion to  prove our main result, Theorem \ref{clu}. We introduce a function 
$\xi: \mathbb{P}_{N} \times  \mathbb{P}_{N} \rightarrow \lbrace -1,0 \rbrace$ by setting

\begin{equation} \label{xixi}
\xi(\mathcal{X}, \mathcal{X}'):=\left\{
    \begin{array}{ll}
        -1& \mbox{if } \text{ dist}(\mathcal{X},\mathcal{X}') \leq 1,\\[5pt]
       \text{ } \text{ }0& \mbox{otherwise},
    \end{array}
\right.
\end{equation}
for all $\mathcal{X},\mathcal{X}' \in \mathbb{P}_{N}$. The ``Ursell functions'', $\varphi^{T}$, are then defined by
\begin{equation} 
\label{tronc}
\varphi^{T} \left(\mathcal{X}_1,...,\mathcal{X}_k \right):= \left\{ \begin{array}{ll} \text{ } \text{ }  \text{ }  1& k=1,\\[6pt]  \underset{g \in \textbf{C}(\mathbb{N}_{k})}{\sum}  \text { }\underset{(i, j) \in \mathscr{E}(g)}{\prod} \xi(\mathcal{X}_i, \mathcal{X}_j) & k\geq 2, \end{array} \right.
\end{equation}
using the same notations as in Appendix A, where $\textbf{C}(\mathbb{N}_{k})$ is the set of all connected graphs with vertex set $\mathbb{N}_{k}:=\{1,...,k\}$.  To exponentiate the cluster expansion for the expectation value of the observable $O$, we start from  \eqref{eqclu1} and single out the polymer $\mathcal{X}$ that contains the vertex $(N,R)$ in the sum on the right side of  \eqref{eqclu1}. This yields
\begin{equation*}
\langle \Psi(t_N) \vert  (O \otimes 1)    \Psi(t_N) \rangle =  \sum_{ \tiny \begin{array}{c} \mathcal{X} \in \mathbb{P}_N\\ (N,R) \in \mathcal{V}(\mathcal{X}) \end{array}} p(\mathcal{X})  \Big( 1 + \sum_{n=1}^{N} \frac{1}{n!} \underset{\text{dist}(\mathcal{X}_i,\mathcal{X}_j) \geq 2,  \text{ dist}(\mathcal{X}_i,\mathcal{X}) \geq 2}{\underset{\mathcal{X}_1,...,\mathcal{X}_n \in \mathbb{P}_{N}}{\sum}}   p(\mathcal{X}_1) \text{ }... \text{ } p(\mathcal{X}_n) \Big).
\end{equation*}
The polymers $\mathcal{X}_1,...,\mathcal{X}_n$  in the  sum  above are separated from the polymer $\mathcal{X}$  by a distance greater or equal to $2$.  The weights $p(\cdot)$ satisfy the Kotecky-Preiss criterion \eqref{kokori}, and  Proposition \ref{Ue} shows that we can exponentiate the term inside the  parenthesis. We get that 
\begin{equation}\label{e1}
\langle \Psi(t_N) \vert  (O \otimes 1)    \Psi(t_N) \rangle=  \sum_{\mathcal{X} \in \mathbb{P}_N, \text{ } (N,R) \in \mathcal{V}(\mathcal{X})} p(\mathcal{X})  \text{ } \tilde{z}(\mathcal{X}),  
\end{equation}
where 
\begin{equation}
\label{e2}
\tilde{z}(\mathcal{X}):= \exp \Big( \sum_{k \geq 1} \frac{1}{k!}  \text{ }\sum_{ \tiny \begin{array}{c} \mathcal{X}_1,...,\mathcal{X}_k \in \mathbb{P}_{N}\\ \textrm{dist}(\mathcal{X}_1 \cup... \cup \mathcal{X}_k ,\mathcal{X}) \geq 2  \end{array}} \text{ } p(\mathcal{X}_1)...p(\mathcal{X}_k)  \text{ }\varphi^{T}(\mathcal{X}_1,...,\mathcal{X}_k ) \Big).
\end{equation}

The weights $p(\mathcal{X}_i)$ on the right side of  \eqref{e2}  \textit{do not} depend  on the observable $O$. Next, we exponentiate the cluster expansion for the expectation value of the identity operator. If  $O=1_S$, we start from \eqref{eqclu11}.  We exponentiate the right side of \eqref{eqclu11} and obtain that 
\begin{equation} \label{e3}
1=\exp \Big( \sum_{k \geq 1}  \frac{1}{k!}  \text{ }\sum_{ \tiny \begin{array}{c} \mathcal{X}_1,...,\mathcal{X}_k \in \mathbb{P}_{N} \\  (N,R) \notin \mathcal{X}_i  \end{array}} \text{ } p(\mathcal{X}_1)...p(\mathcal{X}_k) \text{ }\varphi^{T}(\mathcal{X}_1,...,\mathcal{X}_k ) \Big)
\end{equation}
We now divide the right side of \eqref{e1} by \eqref{e3}. Many terms in the exponent cancel, and we are left with
\begin{equation}
\label{eqf}
\langle \Psi(t_N) \vert  (O \otimes 1)    \Psi(t_N) \rangle=  \sum_{\mathcal{X} \in \mathbb{P}_N, \text{ } (N,R) \in \mathcal{V}(\mathcal{X})} p(\mathcal{X})  \text{ } z(\mathcal{X}),  
\end{equation}
where 
\begin{equation}
\label{eqft}
z(\mathcal{X}):= \exp \Big( \sum_{k \geq 1}  \frac{1}{k!}  \text{ }\sum_{ \tiny \begin{array}{c} \mathcal{X}_1,...,\mathcal{X}_k \in \mathbb{P}_{N }, \text{ }(N,R) \notin \mathcal{X}_i   \\ \textrm{dist}(\mathcal{X}_1 \cup ... \cup \mathcal{X}_k, \mathcal{X}) \leq 1\end{array}} \text{ } p(\mathcal{X}_1)...p(\mathcal{X}_k)  \text{ }\varphi^{T}(\mathcal{X}_1,...,\mathcal{X}_k ) \Big),
\end{equation}
for all  $N \in \mathbb{N}$.  We  show in Section  \ref{proo}  that the main contribution to the right side of \eqref{eqf} comes from the polymer $\mathcal{X}=\{(N,R)\}$ when  $N$  is  large.
\vspace{2mm}

 \section{ \large Proof of the Kotecky-Preiss criterion and convergence of the Cluster expansion  when  $N \rightarrow \infty$}
\label{laco}
 To simplify our exposition, we assume that $\|O\| \leq \varepsilon_0$ (see Lemma \ref{spectrum2}).  This amounts to rescaling the weights $p(\mathcal{X})$ by a factor $ \varepsilon_0/\| O \|$ for all  polymers $\mathcal{X}$ such that $(N,R) \in  \mathcal{V}(\mathcal{X})$. We propose to verify the  Kotecky-Preiss convergence criterion $(2)$ in Proposition \ref{clustt} for the weights $p$.  We establish first an upper bound on 
$\vert p(\mathcal{X}) \vert$; (see subsection \ref{bound}).  We show that an edge $\mathscr{E}=((i,B);(j,B)) \in \mathcal{X}$ contributes  to this upper bound by a factor  $$C^{2} \int_{t_{i}}^{t_{i+1}} du \int_{t_{j}}^{t_{j+1}} dv \vert f(v-u) \vert \lambda(u) \lambda(v),$$ where $C$ is some  positive constant larger than 1.  This last  term has the  important property to be summable in $j$, for every fixed $i$, and satisfies 
\begin{equation}
\label{upup}
\sum_{j \neq i} \int_{t_{i}}^{t_{i+1}} du \int_{t_{j}}^{t_{j+1}} dv \vert f(v-u) \vert \lambda(u) \lambda(v) = \mathcal{O}(\lambda(0)^2),
\end{equation}
 uniformly in $i$ and in the parameter $N$. This is a direct consequence of the hypothesis  that $f(t)  \propto (1+t)^{-\alpha}$, with $\alpha>2$; see \eqref{requ}. Since every  polymer $\mathcal{X}'$ is a collection of $R$-vertices and edges, we can estimate 
 \begin{equation}\label{eq111}
\sum_{\mathcal{X}', \text{dist}(\mathcal{X},\mathcal{X}') \leq 1} \vert  p(\mathcal{X}') \vert e^{\vert \mathcal{V}(\mathcal{X}')\vert } 
\end{equation}
using  \eqref{upup} to sum over all possible polymers $\mathcal{X}'$ with $\textrm{dist}(\mathcal{X'},\mathcal{X}) \leq 1$. The perturbations $R(\cdot)$ are norm-bounded by $\varepsilon_0 \in (0,1)$, and it is not difficult to  control   \eqref{eq111}   by a term of order $\vert V(\mathcal{X})\vert (\varepsilon_0 + \lambda(0)^2)$, using \eqref{upup}; see Section \ref{proo} for a detailed proof. 

 In Section \ref{Proo}, we show that the expansion in Eq. \eqref{eqft} converges, as $N \rightarrow \infty $,  using  the  Kotecky-Preiss criterion for the polymer weights $p(\mathcal{X})$. For large values of $N$, we show  that the main contribution to the cluster expansion  on the right side of  \eqref{eqf} comes from the polymer  $\mathcal{X}=\{(N,R)\}$. Polymers of larger size and containing $\{(N,R)\}$  make  a negligible contribution, for large $N$, because  the coupling $\lambda(t)$  tends to zero,  as $t$ tends to infinity. A rigorous proof is given in Section \ref{conv}.
\vspace{2mm}

\subsection{ The  Kotecky-Preiss criterion for $p(\mathcal{X})$} \label{sectionko}
\subsubsection{Upper bound on $ \vert p(\mathcal{X}) \vert$ and summability of weights}
\label{bound}
Let  $\mathcal{X} \in \mathbb{P}_N$.
For any edge $\mathscr{E}=((i,B);(j,B)) \in \mathcal{E}(\mathcal{X})$, we define
\begin{equation}
\label{gL}
\eta(\mathscr{E}):= 4 \| G \|^{2} \int_{t_{i}}^{t_{i+1}} du \int_{t_{j}}^{t_{j+1}} dv \vert f(v-u) \vert \lambda(u) \lambda(v),
\end{equation}
where $G$ is the form factor introduced in \eqref{H_I}.  We remind the reader that the cardinality of a set $X$ is denoted by $\vert X  \vert$. 

\vspace{1mm}

\begin{lemma}  \label{4.111}
Let $\mathcal{X}$ belong to $\mathbb{P}_N$. Then
\begin{equation}
\label{borne2}
\vert   p(\mathcal{X}) \vert   \leq    2^{ \vert  \mathcal{U}(\mathcal{X}) \vert}   e^{4 \tau   \vert    \vert  f \vert   \vert_{L^1}  \text{ }  \| G \|^{2} \vert \mathcal{B}(\mathcal{X}) \vert}   \Big(\prod_{\mathscr{E} \in \mathcal{E}(\mathcal{X})} \eta(\mathscr{E}) \Big)   \varepsilon_{0}^{\vert \mathcal{R}(\mathcal{X}) \vert},
\end{equation}
where $\mathcal{B}(\mathcal{X})$ and $\mathcal{R}(\mathcal{X}) $ are respectively the sets of  $B$- and  $R$-vertices of $\mathcal{X}$, and  $ \mathcal{U}(\mathcal{X})$ is defined to be the set of maximal blocks of neighboring vertices  of  $\mathcal{X}$, as described at the end of subsection 3.3.2.
\end{lemma}
\vspace{1mm}
 
Before proving Lemma \ref{4.111}, we state an important (though easy) lemma which claims that, for an arbitrary but fixed $i$, the weights 
 $\eta((i,B);(j,B))$ are summable in $j$, uniformly in $i$. The proof of this Lemma follows by inspection; it is a direct consequence of the definition of the weights $\eta((i,B);(j,B))$ given in \eqref{gL}, using that the decay rate $\alpha$ of the function $f(t)$,  ($\vert f(t) \vert \propto \frac{1}{(1+t)^{\alpha}}$, as $t \rightarrow \infty$), satisfies $\alpha>2$.
  \begin{lemma}
\label{4.112}
If $\alpha > 2$, as assumed, there is  a constant $C(\alpha) < \infty$ independent of $N$ such that

\begin{equation}
\label{ineq11}
\sum_{j \neq i}   \lambda(t_{i})^{-1} \lambda(t_{j})^{-1}  \text{ } \eta((i,B);(j,B)) \leq C(\alpha),
\end{equation}
 for all $i \in \mathbb{N}$.
\end{lemma}
 \vspace{2mm}
\begin{remark} \label{411}
A  better estimate is actually satisfied by the weights $\eta(\cdot)$. It is fairly easy to show that, given any $\beta \in (0, \alpha-2)$,  there is  a constant $C(\alpha,\beta) < \infty$ independent of $N$ such that
\begin{equation}
\label{ineq1111}
\sum_{j \neq i}   (1 +\vert i-j\vert)^{\beta} \lambda(t_{i})^{-1} \lambda(t_{j})^{-1}  \text{ } \eta((i,B);(j,B)) \leq C(\alpha,\beta),
\end{equation}
 for all $i \in \mathbb{N}$. Inequality \eqref{ineq1111} is useful if the coupling $\lambda(\cdot)$ stays constant, since it implies that  polymers $\mathcal{X}$ with  $d(\mathcal{X}) \gg1$   have  a negligible  weight $p(\mathcal{X})$; see also Remark \ref{332} in Section \ref{graphph}. Eq. \eqref{ineq1111} shows that the weight associated to an edge  $\mathscr{E}=((i,B);(j,B))$ tends to zero, as $\vert j-i \vert$ tends to infinity. Eq. \eqref{ineq11} is sufficient for our purpose since the coupling $\lambda(t)$ tends to zero as $t$ tends to infinity, and  the factor $ \lambda(t_{i})^{-1} \lambda(t_{j})^{-1} $ in \eqref{ineq11} plays the same role as $(1+\vert i-j\vert)^{\beta}$ in \eqref{ineq1111}.
\end{remark}
\vspace{2mm}

\subsubsection{Proof of Lemma \ref{4.111}}
In Lemma \ref{spectrum2} it is shown that the operators $P(s)$ converge in norm to the projection  
$P=\vert 1_S \rangle \langle \Pi_{11} \vert,$
 as  the parameter $\lambda(0)$ (see Lemma \ref{spectrum2}) tends to $0$. Since  $\|P\|_{\infty}=1$,  it follows that  $\|P(s)\|_{\infty} \leq 2$, for all 
 $s \geq 0$, provided that  $\lambda(0)$ is small enough.
Let  $\underline{w}$ be a pairing. We rewrite the amplitude $h_{U}(\underline{w})$ (see Eq. \eqref{hU})  as 
\begin{equation} \label{hU2}
 \vert  1_S \rangle  h_U (\underline{w}) = P (t_{m(U)})  \mathcal{T}_S  \Big( \underset{(l,B) \in U  }{\prod} \Big[ e^{-it_{l} \mathcal{L}_S} \big( \prod_{(t,r) \in  \underline{w} \cap \textbf{I}_l } (i G)(t,r) \big) e^{it_{l+1} \mathcal{L}_S} \Big]  
 \underset{(j,R)\in U}{\prod} R(t_j) \Big) \vert 1_S \rangle,
\end{equation}
for all $U \in \mathcal{U}(\mathcal{X})$ with $m(U) >0$. It follows that 
\begin{equation*}
\vert h_U (\underline{w} ) \vert   \leq    2\text{ } \varepsilon_{0}^{\vert \mathcal{R}(\mathcal{X}) \cap U \vert } \text{ } \underset{(l,B)  \in U  }{\prod} \| G \|^{2  \vert \underline{w}  \cap \textbf{I}_l  \vert} 
\end{equation*}
If $m(U)=-1$,    $ 1_S $  has to be replaced by $P_{\varphi}=\vert \varphi \rangle \langle \varphi \vert$  (see \eqref{inistate}) on the left side of \eqref{hU2}, but  the upper bound remains unchanged, because   $\| P_{\varphi} \| =1$. Thus,
\begin{equation}
\label{ctnous}
\prod_{U \in\mathcal{U}(\mathcal{X})}  \vert h_U (\underline{w} ) \vert  \leq     2^{ \vert  \mathcal{U}(\mathcal{X}) \vert} \text{ } \varepsilon_{0}^{\vert \mathcal{R}(\mathcal{X}) \vert } \text{ } \| G \|^{2 k }, 
\end{equation}
for all pairings $\underline{w} $ with precisely $k$ pairs. These bounds are used to estimate the weights $p(\mathcal{X})$; (see right side of \eqref{p(G)}).
The characteristic function $\chi_{\mathcal{X}}$ in \eqref{p(G)} selects the pairings compatible with the polymer $\mathcal{X}$; (the correlation lines of the pairings $\underline{w}$  must ``follow" the edges of $\mathcal{X}$, and that there must be  at least one correlation line in $\underline{w}$  for each edge of $\mathcal{X}$). For each edge $\mathscr{E}=((i,B);(j,B)) \in \mathcal{E}(\mathcal{X})$, there exists at least one pair 
$(u,r;v,r') \in \underline{w} $ with $u \in I_i$ and $v \in I_j$. We choose such a pair and then estimate the integrals over $u$ and $v$, factoring out the rest. Carrying out this procedure for each  line $\mathscr{E}\in \mathcal{E}(\mathcal{X})$,  we conclude that
 \begin{equation}
\label{out}
\begin{split}
  \underset{[0,t_N]^{2k} } {\int} d\mu_{k} (\underline{w})  \lambda( \underline{w})   \text{ } &\chi_{\mathcal{X}} (\underline{w} )  \text{ } \vert   F(\underline{w} )   \vert \leq  \| G \|^{-2 \vert \mathcal{E}(\mathcal{X}) \vert}  \Big(\prod_{\mathscr{E} \in \mathcal{E}(\mathcal{X})} \eta(\mathscr{E}) \Big)  \\
  &  \cdot \int_{[0,t_N]^{2(k-\vert \mathcal{E}(\mathcal{X}) \vert)}} d\mu_{k-\vert \mathcal{E}(\mathcal{X}) \vert} (\underline{\hat{w}})  \lambda(\underline{\hat{w}})    \text{ }    \vert  F(\underline{\hat{w}})   \vert  \text{ } \prod_{i=1}^{k-\vert \mathcal{E}(\mathcal{X}) \vert} \chi  ( u_i,v_i \in  \mathcal{B}(\mathcal{X}) ),
\end{split}
\end{equation}
where $F(\underline{w})$ is defined in \eqref{F}, and $\underline{\hat{w}}$ contains $m:=k-\vert \mathcal{E}(\mathcal{X}) \vert$ pairs. If $m=0$ the integral is replaced by $1$. To estimate the integral on the right side of \eqref{out},
we first integrate over the $v_i$'s and use that $\lambda(u_i)<\lambda(v_i)$. This yields
\begin{equation}
\label{ctkiki}
\begin{split}
\int_{[0,t_N]^{2m}}  d\mu_{m} (\underline{w})  \lambda( \underline{w})  &   \text{ } \vert    F(\underline{w} )   \vert   \prod_{i=1}^{m} \chi  ( u_i,v_i \in  \mathcal{B}(\mathcal{X}) ) \leq 4^{m} \int_{[0,t_N]^{m}}  d\underline{u}  \text{ } \chi(u_1<...<u_{m})  \text{ }  \text{ }   \| f \|_{L^1}^{m} \\
&\qquad  \qquad \qquad \prod_{i=1}^{m} \chi(u_i \in \mathcal{B}(\mathcal{X}))  \lambda^{2}( u_i),
\end{split}
\end{equation} 
for all $m \in \mathbb{N}$.
We now set $q=\vert \mathcal{B}(\mathcal{X}) \vert$ and assume that $ \mathcal{B}(\mathcal{X})= \lbrace (i_1,B),...,(i_q,B) \rbrace$; (the case where $\mathcal{B}(\mathcal{X})=\emptyset$ yields a factor $1$). For  each tuple  $\underline{u}=(u_1,...,u_m) $ under the integral on the right side of \eqref{ctkiki}, there are  $q$ integers, $n_1,...,n_q$, given by $n_j= \vert \lbrace u_1,...,u_m \rbrace  \cap I_{i_j} \vert$, with $j=1,...,q$. The times $u_i$  are ordered, $u_1<...<u_{m}$, and  
$$\int_{t_{i_{j}}<u_1<...<u_{n_j}<t_{i_j +1}} [\lambda(u_1)....\lambda(u_{n_j})]^{2} du_1...du_{n_j} \leq  \frac{\tau^{n_j}}{n_j !},$$
with $\tau$ as in \eqref{deco}. Hence
\begin{eqnarray*}
\int_{[0,t_N]^{m}}  d\underline{u}  \text{ } \chi(u_1<...<u_{m})  \prod_{i=1}^{m} \chi(u_i \in \mathcal{B}(\mathcal{X}))  \lambda^{2}( u_i) &\leq & \tau^{m}  \text{ }\sum_{n_1+...+n_{q }=m} \frac{1}{n_1!} ... \frac{1}{n_{q}!},
\end{eqnarray*}
and it follows that
\begin{align*}
\vert p(\mathcal{X})  \vert &\leq   2^{ \vert  \mathcal{U}(\mathcal{X}) \vert} \text{ } \varepsilon_{0}^{\vert \mathcal{R}(\mathcal{X}) \vert } \text{ }  \Big(\prod_{\mathscr{E} \in \mathcal{E}(\mathcal{X})} \eta(\mathscr{E}) \Big)   \sum_{m=0}^{\infty} 4^{m}  \tau^{m} \vert   \vert   f \|_{L^1}^{m}   \| G \|^{2m} \sum_{n_1+...+n_{q}=m} \frac{1}{n_1!} ... \frac{1}{n_{q }!}\\
&\leq  2^{ \vert  \mathcal{U}(\mathcal{X}) \vert} \text{ } \varepsilon_{0}^{\vert \mathcal{R}(\mathcal{X}) \vert } \text{ }  \Big(\prod_{\mathscr{E} \in \mathcal{E}(\mathcal{X})} \eta(\mathscr{E}) \Big)    \exp \left(4 \tau  \vert   \vert   f \|_{L^1}  \text{ }\| G \|^{2} q \right).
\end{align*}
\vspace{2mm}

Next, we prove the  Kotecky-Preiss criterion for the weight $p$ and the function $a(\mathcal{X})=\vert\mathcal{V}( \mathcal{X}) \vert$ introduced above.

\subsubsection{Proof of the Kotecky-Preiss criterion} \label{proo}
  We first  decompose the set $\mathcal{R}(\mathcal{X}')$ of  vertices of $\mathcal{X}'$ decorated with a perturbation $R$ into a disjoint union of maximal blocks of neighboring vertices. For each polymer $\mathcal{X}'$ in $\mathbb{P}_{N}$, there exists an $m>0$ such that  $\mathcal{R}(\mathcal{X}')=A_1\cup ... \cup A_m$, where the sets $A_i$ are \textit{maximal blocks of neighboring $R$-vertices}.  The set of maximal  blocks of neighboring $R$-vertices  in $\mathcal{X}'$ is denoted by  $\mathcal{R}_{max}(\mathcal{X}')$. Every polymer 
 $\mathcal{X}' \in \mathbb{P}_{N}$ is entirely characterized by the collection of maximal  blocks of neighboring $R$-vertices in
  $\mathcal{X}'$ and by its set of edges $\mathcal{E}(\mathcal{X}')$.  The  polymer  drawn below, for instance, has  two maximal blocks of neighboring $R$-vertices,  $\{(0,R), (1,R)\}$ and  $\{(5,R)\}$,  and three edges, $\mathscr{E}_1= ((2,B); (4,B))$, $\mathscr{E}_2=((4,B); (9,B))$ and  $\mathscr{E}_3=  ((6,B); (8,B))$. 
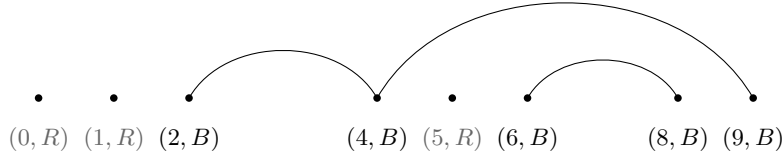
 \begin{center}
\begin{tikzpicture}[scale=0.5]

\fill (-2,0) circle (0.1cm);
\fill (-4,0) circle (0.1cm);
\fill (0,0) circle (0.1cm);
\fill (5,0) circle (0.1cm);
\fill (9,0) circle (0.1cm);
\fill (13,0) circle (0.1cm);
\fill (7,0) circle (0.1cm);
\fill (15,0) circle (0.1cm);

\draw (-2,-0.5) node[below] { \color{black!60} \footnotesize $(1,R)$};
\draw (-4,-0.5) node[below] { \color{black!60} \footnotesize $(0,R)$};
\draw (7,-0.5) node[below] {\color{black!60} \footnotesize $(5,R)$};
\draw (0,-0.5) node[below] {\footnotesize $(2,B)$};
\draw (5,-0.5) node[below] {\footnotesize $(4,B)$};
\draw (15,-0.5) node[below] {\footnotesize $(9,B)$};
\draw (13,-0.5) node[below] {\footnotesize $(8,B)$};
\draw (9,-0.5) node[below] {\footnotesize $(6,B)$};

\draw[-] (0,0)  to[bend left=60] (5, 0.0); 
\draw[-] (9,0) to[bend left=60](13,0); 
\draw[-] (5,0)  to[bend left=60](15,0);

 \draw (6.5,-2) node[below] {\footnotesize A polymer is a union of maximal blocks of neighboring $R$-vertices  and edges.};
 
\end{tikzpicture}
\end{center}
We  use the decomposition of polymers into blocks of neighboring $R$-vertices and edges to prove the Kotecky-Preiss criterion.  

We start from the bound  \eqref{borne2}.  Multiplying both sides by  $e^{\vert \mathcal{V}(\mathcal{X}) \vert}$, we get that 
\begin{equation*}
  \vert  p(\mathcal{X}) \vert  \text{ } e^{\vert \mathcal{V}(\mathcal{X}) \vert} \leq     2^{ \vert  \mathcal{U}(\mathcal{X}) \vert} (e \varepsilon_{0})^{\vert \mathcal{R}(\mathcal{X}) \vert}  \Big(\prod_{\mathscr{E} \in \mathcal{E}(\mathcal{X})} \eta(\mathscr{E}) \Big)  e^{ (4 \tau   \vert    \vert  f \vert   \vert_{L^1}  \text{ }  \| G \|^{2} +1) \vert \mathcal{B}(\mathcal{X}) \vert },
\end{equation*}
  where the function $\eta$ has been defined in \eqref{gL}. The integer  $\vert \mathcal{U}(\mathcal{X}) \vert$ is bounded by $ \vert \mathcal{B}(\mathcal{X}) \vert$  if  $\vert \mathcal{B}(\mathcal{X}) \vert \neq 0$, because $\mathcal{X}$ is a union of adjacent  connected graphs.  If $\vert \mathcal{B}(\mathcal{X}) \vert = 0$ then   $\vert \mathcal{U}(\mathcal{X}) \vert=1$.  Defining  $K:=2e^{ (4 \tau   \vert    \vert  f \vert   \vert_{L^1}  \text{ }  \| G \|^{2} +1)}$, we introduce
\begin{equation}
\label{weig}
\eta_{K}(\mathscr{E}): =K^{ 2}   \eta(\mathscr{E})
\end{equation} 
for every edge $\mathscr{E}$ and 
\begin{equation} \eta_{K}(A):= (e\varepsilon_{0})^{\vert A \vert } \end{equation}
for every block $A$ of neighboring $R$-vertices. Denoting by $\mathcal{A}d(\mathcal{X})$ the  set of vertices adjacent to the polymer $\mathcal{X}$, we obtain that
\begin{equation*}
\label{splspl}
     \sum_{ \tiny \begin{array}{c}\mathcal{X}' \in \mathbb{P}_{N} \\  \text{ dist} (\mathcal{X},\mathcal{X}') \leq 1 \end{array}}   \vert  p(\mathcal{X}') \vert  \text{ } e^{\vert \mathcal{V}(\mathcal{X}') \vert} \leq 6  \vert \mathcal{V}(\mathcal{X}) \vert  \underset{(j, \cdot) \in \mathcal{A}d(\mathcal{X}) \vee \mathcal{V}(\mathcal{X})}{\sup} \sum_{\underset{ (j,\cdot) \in \mathcal{V}(\mathcal{X}')}{\mathcal{X}' \in \mathbb{P}_{N} } }  \text{ }  \prod_{A \in \mathcal{R}_{\max}(\mathcal{X}') \vee \mathcal{E}(\mathcal{X}') }  \eta_{K}(A).
\end{equation*}
In order to avoid too much over-counting, we must carefully estimate  the right side of this inequality. We use that  every polymer $\mathcal{X}' \in \mathbb{P}_N$ is entirely characterized by its set of edges, $ \mathcal{E}(\mathcal{X}')$, and its set of maximal  blocks of neighboring $R$-vertices,  $\mathcal{R}_{\max}(\mathcal{X}')$. We  denote   by $\mathbb{A}_N$ the set  of all possibles edges and blocks of neighboring $R$-vertices.  It is sometimes useful to explicitly distinguish edges and block of $R$-vertices, and we denote the set of all possible edges by $\mathbb{E}_N$ and the set of all possible blocks of neighboring $R$-vertices by $\mathbb{R}_N$.  We can estimate the right side of the last inequality by summing over collections of elements in the set $\mathbb{A}_N=\mathbb{E}_N \vee \mathbb{R}_N$-- but not over all  of them!  To carry out this sum without intolerable over-counting, we introduce graphs: We denote by $g(A_1,...,A_n)$ the graph on $\mathbb{N}_n:=\{1,...,n \}$ that has an edge $(i,j)$  between $i$ and $j$ if and only if one of the following properties is satisfied. Namely, $A_i$ and/or  $A_j$   belong to the set $\mathbb{E}_N$ and are adjacent next to each other, or $A_i,$ and $A_j$ both belong to  $\mathbb{E}_N$ and share a common vertex.  If $\mathcal{X}$ is a polymer, and if $A_1,...,A_n$ consist of its set of edges and maximal blocks of neighboring $R$-vertices, then $g(A_1,...,A_n)$ is connected. Singling out  (\textit{one of}) the decorated edge(s) - or \textit{the} block of  neighboring $R$-vertices - that contains the vertex $(j,\cdot)$  and belongs to  $\mathcal{X}'$, we get the estimate
\begin{equation*}
\label{jojo}
  \sum_{\underset{ (j, \cdot) \in \mathcal{V}(\mathcal{X}')}{\mathcal{X}' \in \mathbb{P}_{N} } }  \text{ }   \prod_{A \in \mathcal{R}_{\max}(\mathcal{X}') \vee \mathcal{E}(\mathcal{X}') }  \eta_{K}(A) \leq      \sum_{\underset{(j,\cdot) \in A}{A \in \mathbb{A}_N}}  \eta_{K}(A) \text{ } \Big(1+ \sum_{n=1}^{\infty} \frac{1}{n!} \underset{ \underset{g(A,A_1,...,A_n) \in \textbf{C}(\mathbb{N}_{n+1})}{A_1,...,A_n  \in \mathbb{A}_N}}{\sum}   \prod_{i=1}^{n} \eta_{K}(A_i) \Big),
\end{equation*}
where $\textbf{C}(\mathbb{N}_{n+1})$ is the set of connected graphs with vertex set  $\mathbb{N}_{n+1}=\{1,...,n+1\}$.  We now follow ideas from \cite{Ka}   and sum over spanning trees to bound  the right side of  the last inequality. We denote by $ \textbf{T}(\mathbb{N}_{n+1})$ the set of labelled trees with vertex set $\mathbb{N}_{n+1}=\lbrace 1,...,n+1 \rbrace$, and we write $t \subset g$ if $\mathcal{V}(t)=\mathcal{V}(g) = \mathbb{N}_{n+1}$ and $\mathcal{E}(t) \subset \mathcal{E}(g)$.  Then
\begin{align*}
 \sum_{\underset{g(A,A_1,...,A_n) \in \textbf{C}(\mathbb{N}_{n+1})}{A_1,...,A_n \in \mathbb{A}_N}}    &\prod_{i=1}^{n} \eta_{K}(A_i)   = \sum_{g \in \textbf{C}(\mathbb{N}_{n+1})}    \sum_{\underset{g(A,A_1,...,A_n)=g}{A_1,...,A_n \in \mathbb{A}_N}}   \prod_{i=1}^{n} \eta_{K}(A_i)  \\
 & \leq   \sum_{t \in \textbf{T}(\mathbb{N}_{n+1})}   \sum_{\underset{g(A,A_1,...,A_n)  \supset t, \text{ } g \in \textbf{C}(\mathbb{N}_{n+1}) }{A_1,...,A_n \in \mathbb{A}_N}}  \prod_{i=1}^{n} \eta_{K}(A_i),
\end{align*}
for all $A \in \mathbb{A}_N$. Every tree $t \in \textbf{T}(\mathbb{N}_{n+1})$ has  $n$ edges  and each vertex  of $t$  is  linked to at least  one other vertex by an edge.  Using \eqref{weig}, we find that  
\begin{equation}
\begin{split}
\label{ineq22}
 \sum_{\underset{g(A,A_1,...,A_n)  \supset t }{A_1,...,A_n \in \mathbb{A}_N}} \prod_{i=1}^{n} \eta_{K}(A_i)  &\leq  \underset{ (i,j) \in \mathcal{E}(t)}{\prod} \Big(  \underset{A_i \in \mathbb{A}_N}{\sup} \underset{\underset{ A_j \sim A_i }{A_j \in \mathbb{A}_N}}{\sum} \eta_{K}(A_j) \Big),
 \end{split}
\end{equation}
where $A_j \sim A_i $ if $A_i$ \textit{and/or} $A_j$  belong to $\mathbb{E}_N$ and they are adjacent to one another, or if $A_i$ \textit{and} $A_j$  belong both to  $\mathbb{E}_N$   and share a common vertex. If $\varepsilon_0<e^{-1}$, Lemma \ref{4.112} and an easy calculation imply  that  
\begin{equation}\label{cbien}
 \underset{\underset{ A_j \sim A_i }{A_j \in \mathbb{A}_N}}{\sum} \eta_{K}(A_j) \leq 4 \frac{e \varepsilon_0}{1-e \varepsilon_0} + C(\tau)  \lambda(t_{m(A_i)})  \lambda(0), 
 \end{equation}
for all $A_i \in \mathbb{A}_N$, uniformly in  $N$. Here $C(\tau)$ is a positive constant that depends on $\tau$ through the constant $K$ appearing in \eqref{weig}. The number of labelled trees in $ \textbf{T}(\mathbb{N}_{n+1})$ is equal to $(n+1)^{n-1}$, and we deduce that 
\begin{equation*}
  \frac{1}{n!} \sum_{t \in \textbf{T}(\mathbb{N}_{n+1})}   \sum_{\underset{g(A,A_1,...,A_n)  \supset t }{A_1,...,A_n \in \mathbb{A}_N}}  \prod_{i=1}^{n} \eta_{K}(A_i)   \leq \frac{ (n+1)^{n-1}}{n!}  \left( C(\tau) \lambda(0)^{2} + 4 \frac{ e \varepsilon_0}{1- e \varepsilon_0}  \right)^{n}
\end{equation*}
by plugging \eqref{cbien} into the right side of  \eqref{ineq22}. Using  Stirling formula, it is easy to see that we can sum the right side over $n$ if $\varepsilon_0$ and $\lambda(0)$  are sufficiently small. This yields the upper bound
\begin{equation}
\label{Rpset}
 \sum_{\underset{ (j,\cdot) \in \mathcal{V}(\mathcal{X}')}{\mathcal{X}' \in \mathbb{P}_{N} } }  \text{ }  \prod_{A \in \mathcal{R}_{\max}(\mathcal{X}') \vee \mathcal{E}(\mathcal{X}') }  \eta_{K}(A) \leq  C_1(\tau,\lambda(0))  \sum_{\underset{(j,\cdot) \in A}{A \in \mathbb{A}_N}}   \eta_{K}(A), 
 \end{equation}
 for some constant $C_1(\tau,\lambda(0))$ of order $1$.   Using \eqref{weig} and  \eqref{cbien}, it is easy to see that  there exist constants $\tau_c>0$ and $\lambda_{\tau}>0$ such that the Kotecky-Preiss criterion  \eqref{kokori} is satisfied for all $\tau>\tau_c$ and all $0<\lambda(0) <\lambda_{\tau}$.
 \vspace{2mm}
 
 \begin{remark} \label{412}
Following the lines of the proof we just carried out above, we can use  \eqref{ineq1111} instead of \eqref{ineq11}  to show that  
\begin{equation}
\label{kokori2}
 \sum_{ \tiny \begin{array}{c}\mathcal{X}' \in \mathbb{P}_{N} \\  \text{ dist} (\mathcal{X},\mathcal{X}') \leq 1 \end{array}}   \vert  p(\mathcal{X}') \vert  \text{ } e^{\vert \mathcal{V}(\mathcal{X}')\vert  + \beta \ln(d(\mathcal{X}'))}  \leq \vert \mathcal{V}(\mathcal{X}) \vert
\end{equation}
  for all polymers $\mathcal{X} \in \mathbb{P}_N$, where $\beta \in (0, \alpha-2)$.  Indeed, if $\mathcal{X}'$ is a polymer, and if $A_1, ... , A_n$ consists of its set of  edges and maximal block of $R-$vertices, then
 \begin{equation} \label{dia}
 d(\mathcal{X}') \leq d(A_1) + ... +  d(A_n) \leq (d(A_1)+1)...(d(A_n)+1),
 \end{equation}
 and we can replace $\eta_{K}(A)$ in the equations above by $(d(A)+1)^{\beta}\eta_{K}(A)$. The first inequality in \eqref{dia} holds true because a polymer is a fusion of adjacent connected graphs. The second inequality follows  from the positivity of the diameter. As we already mentioned, \eqref{kokori2} is useful to investigate the limit $N \rightarrow \infty$ if the coupling $\lambda(\cdot)$ is constant.
\end{remark}
\vspace{2mm}

 \subsection{ Convergence of the cluster expansion as $N \rightarrow \infty$}
\label{Proo}
 We use the exponentiated form of the cluster expansion derived in Section \ref{expf}; see \eqref{eqf} and \eqref{eqft}.
We show  that the main contribution  to the right side of \eqref{eqf}  comes from the polymer $\mathcal{X}=\{(N,R)\}$ if  $N$ is large. We remark that if $\mathcal{X} \in \mathbb{P}_N$, $\mathcal{X} \neq   \{(N,R)\} $ and $(N,R) \in \mathcal{V}(\mathcal{X})$, then necessarily $(N-1,\cdot) \in \mathcal{V}(\mathcal{X})$; see Section \ref{PN}. We use this remark, and Lemma \ref{presquefini} below, to  prove Theorem \ref{clu}.

\subsubsection{Convergence to  the ground state}
\label{conv}
 \vspace{2mm}

\begin{lemma}
\label{presquefini}
We introduce
\begin{equation}
\label{eqf2}
{Z}_{N}(O):=   \sum_{\mathcal{X} \in \mathbb{P}_N, \text{ } (N-1,\cdot), (N,R)  \in \mathcal{V}(\mathcal{X})} p(\mathcal{X})  \text{ } z(\mathcal{X}).
\end{equation}
Then 
\begin{equation}
\label{Z_N}
Z_{N}(O) \underset{ N \rightarrow \infty}{\longrightarrow} 0.
\end{equation}
\end{lemma}
\vspace{2mm}

We postpone the proof of Lemma  \ref{presquefini} to the next  paragraph and turn to the proof of our main Result, namely Theorem \ref{clu}. 
\begin{proof} (Theorem  \ref{clu})
If $\mathcal{X} =  \{(N,R)\}$,  then
\begin{equation}
\label{pI1}
   p( \{(N,R)\})  \text{ }z( \{(N,R)\} ) = \langle \Pi(t_{N-1}) \vert O \rangle  \text{ }  z( \{(N,R)\} ).
\end{equation}
 Lemma \ref{spectrum2} shows that  $ \langle \Pi(t_{N-1}) \vert O \rangle$ converges to $ \langle \varphi_1 \vert O  \varphi_1 \rangle$.  We note that $z( \{(N,R)\} )$ does not depend   on the choice of the observable $O$.  It is therefore sufficient to exhibit an observable $O$ for which we can show that $ z( \{(N,R)\}) \rightarrow 1$, as $N \rightarrow \infty$, in Eq. \eqref{eqf}: Choosing $O=1_S$, we find that 
 \begin{equation}
\langle \Psi(t_N) \vert \Psi(t_N)  \rangle = Z_N(1_S)+  \langle \Pi(t_{N-1}) \vert 1_S \rangle  z( \{(N,R)\}  )=Z_N(1_S)+  z(  \{(N,R)\} )=1.
\end{equation} 
As $ Z_N(1_S) \rightarrow 0$, as $N \rightarrow \infty$ (see \eqref{Z_N}), we deduce that $ z( \{(N,R)\} )$ converges to $1$. This completes the proof of Theorem \ref{clu}.
\end{proof}

\subsubsection{Proof of Lemma \ref{presquefini}}
We first establish an upper bound on the weights  $z(\mathcal{X})$  introduced  in \eqref{eqft}. Clearly\begin{equation}
\label{xixix}
\vert  \prod_{i=1}^{k} \left(1+ \xi(\mathcal{X}_i,\mathcal{X}) \right)-1  \vert  \leq \sum_{i=1}^{k}  \vert \xi(\mathcal{X}_i,\mathcal{X}) \vert.
\end{equation}
Moreover, the argument  of the exponential in \eqref{eqft} is bounded by 
\begin{equation} \label{4.42}
\sum_{k \geq 1}  \frac{1}{k!}  \text{ }\sum_{ \tiny \begin{array}{c} \mathcal{X}_1,...,\mathcal{X}_k \in \mathbb{P}_{N } \\  (N,R) \notin \mathcal{X}_i  \end{array}} \text{ } \vert p(\mathcal{X}_1)...p(\mathcal{X}_k) \vert \sum_{i=1}^{k}  \vert \xi(\mathcal{X}_i,\mathcal{X}) \vert  \text{ } \vert \varphi^{T}(\mathcal{X}_1,...,\mathcal{X}_k ) \vert  \leq \vert \mathcal{V}(\mathcal{X}) \vert .
\end{equation}
Inequality \eqref{4.42}  follows from the inequality 
\begin{equation}
\label{cmbb}
1+ \sum_{k \geq 1}  \frac{1}{k!}  \text{ }\sum_{\tiny \begin{array}{c} \mathcal{X}_1,...,\mathcal{X}_k \in \mathbb{P}_{N } \\  (N,R) \notin \mathcal{X}_i  \end{array}} \text{ } \vert  p(\mathcal{X}_1)...p(\mathcal{X}_k)  \vert  \text{ }  \vert  \varphi^{T}(\mathcal{X}',\mathcal{X}_1,...,\mathcal{X}_k )  \vert  \leq  e^{\vert \mathcal{V}(\mathcal{X}') \vert},
\end{equation}
 by multiplying both sides of  \eqref{cmbb} by $\vert  \xi(\mathcal{X},\mathcal{X}') \vert \text{ } \vert p(\mathcal{X}') \vert$ and by  summing over $\mathcal{X}'  \in \mathbb{P}_{N }$; see also Appendix A and  \cite{Ue} for more details. We deduce that 
\begin{equation}
\label{222}
\vert  Z_{N}(O) \vert  \leq   \sum_{\mathcal{X} \in \mathbb{P}_N, \text{ } (N-1,\cdot), \text{ }(N,R) \in \mathcal{V}(\mathcal{X})}   \vert  p(\mathcal{X}) \vert   e^{ \vert \mathcal{V}(\mathcal{X}) \vert}.
\end{equation}
The  upper  bound  for $\vert p(\mathcal{X})\vert$ given in  \eqref{borne2}  is not sharp enough to show that the right side of  \eqref{222} tends to zero as $N$ tends to infinity. We derive a slightly refined upper bound by exploiting the particular structure of the polymer set $\mathbb{P}_N$. We start again  from the definition of $p(\mathcal{X})$ given in \eqref{p(G)} and use that
\begin{equation}
\label{217}
P(t_i)R(t_i)=R(t_i)P(t_i)=0,
\end{equation}
for all $i=0,...,N$. Eq. \eqref{217} follows from the  relations $[P(t_i),\mathcal{Z}^{t_{i+1},t_i}]=0$,  $R(t_i)=\mathcal{Z}^{t_{i+1},t_i}-P(t_i)$, and   $\mathcal{Z}^{t_{i+1},t_i} ( 1_S )=  1_S$. The product $P(t_i)R(t_i)$ does not appear in  our expansion. However, terms of the form $P(t_{i-1})R(t_i)$  $do$ arise. We also remark that $R(t_i) P(t_{i+1})=0$, but we will not use this fact.  We define  
\begin{equation}
\label{u_i}
u_n:= \| P(t_{n-1})R(t_{n})  \|_{\infty}, \qquad n \in \mathbb{N}.
\end{equation}
The sequence  $(u_n)_{n=1}^{\infty}$ tends to zero, as  $n \rightarrow + \infty$, because
 \begin{equation*}
 \| P(t_{n-1})R(t_{n})  \|_{\infty}= \| (P(t_{n-1}) -P(t_{n}))  R(t_{n})  \|_{\infty} \leq \varepsilon_{0}  \| P(t_{n-1}) -P(t_{n})  \|_{\infty}.
\end{equation*}
\vspace{1mm}

The main idea of our proof is to use the sequence $(u_n)_{n=1}^{\infty}$ and the decay of the coupling $\lambda(\cdot)$  towards zero to prove that the right side of \eqref{222} tends to zero as $N$ tends to $+\infty$. To do so, it is  useful to distinguish two  classes of polymers $\mathcal{X}$ in the sum on the right side of \eqref{222}.
\vspace{3mm}

\noindent \textbf{Class 1}:  $\vert \mathcal{B}(\mathcal{X}) \vert =0$.\\
Every vertex in $ \mathcal{V}(\mathcal{X})$ carries a  perturbation  $R(\cdot)$. There are only $N$ polymers $\mathcal{X} \in \mathbb{P}_N$ with $(N-1,R),(N,R) \in \mathcal{V}(\mathcal{X})$ and  $\vert \mathcal{B}(\mathcal{X}) \vert =0$. Using Formula \eqref{p(G)},  we deduce that
\begin{equation}
\label{mgg}
 p(\mathcal{X})= \langle \Pi(t_{N-\vert \mathcal{V}(\mathcal{X}) \vert }) \vert  R(t_{N-\vert \mathcal{V}(\mathcal{X}) \vert  +1}) \text{ }...\text{ }   R(t_{N-1})   O \rangle.
\end{equation}
Consequently, 
\begin{equation}
 \vert  p(\mathcal{X})\vert  e^{\vert \mathcal{V}(\mathcal{X})\vert}  \leq \left\{ \begin{array}{ll} e^2 \| O \| (e\varepsilon_{0})^{\vert \mathcal{V}(\mathcal{X}) \vert-2} u_{N-\vert \mathcal{V}(\mathcal{X}) \vert +1} & \quad 2 \leq \vert \mathcal{V}(\mathcal{X}) \vert  \leq N,\\[8pt]
 \| O \| (e\varepsilon_{0})^{N}e  &   \quad \vert \mathcal{V}(\mathcal{X}) \vert =N+1, \end{array} \right.
\end{equation}
and 
  \begin{equation}
  \label{Rset1}
 \sum_{\underset{\vert \mathcal{B}(\mathcal{X}) \vert=0}{\mathcal{X} \in \mathbb{P}_N, \text{ }(N-1,\cdot), \text{ }(N,R) \in \mathcal{V}(\mathcal{X})}}   \vert  p(\mathcal{X})  \vert   e^{\vert \mathcal{V}(\mathcal{X})\vert}  \leq  \| O \| \big( e^2 \sum_{k=1}^{N-1} (e\varepsilon_{0})^{k-1} \text{ } u_{N-k}+ e (e\varepsilon_{0})^{N} \big).
  \end{equation}
  \vspace{2mm}
  
\noindent \textbf{Class 2}: $ \vert \mathcal{B}(\mathcal{X}) \vert \neq 0$.\\
The polymer $\mathcal{X}$ in \eqref{222} must contain the vertex $(N-1,\cdot)$ and an edge. The color of $(N-1,\cdot)$ is either red ($R$) or blue ($B$), and we treat  differently  these two possibilities.  We use the bound \eqref{borne2} on $p(\mathcal{X})$ and continue our argument as in  the proof of Proposition \ref{clustt}, Property (2); see Section \ref{proo}. We use the same notations as in  Section \ref{proo}. We remind the reader that $K:=2e^{ (4 \tau   \vert    \vert  f \vert   \vert_{L^1}  \text{ }  \| G \|^{2} +1)}$ and that we have defined $\eta_{K}(\mathscr{E}):= K^{2} \eta(\mathscr{E})$ for a single edge, and  
 $\eta_{K}(A)=(e\varepsilon_{0})^{\vert A \vert }$ for a union, $A$, of neighboring $R$-vertices.
 \vspace{3mm}

\noindent   \textit{Case 2.a}:  $(N-1,B) \in \mathcal{V}(\mathcal{X})$. We single out one edge $\mathscr{E}$ of $\mathcal{X}$ such that  $(N-1,B)$ belongs to $\mathscr{E}$.  Following the same arguments as in Section \ref{proo}, we then get that 
\begin{align*}
\label{Bset1}
& \sum_{\mathcal{X} \in \mathbb{P}_N, \text{ } (N-1,B), \text{ }(N,R) \in \mathcal{V}(\mathcal{X})}   \vert  p(\mathcal{X}) \vert   e^{ \vert \mathcal{V}(\mathcal{X}) \vert}\\
 & \quad   \leq \|O  \| e    \sum_{\underset{(N-1, B)  \in \mathscr{E}}{\mathscr{E} \in \mathbb{E}_N}}  \eta_{K}(\mathscr{E})  \Big( 1+\sum_{n=1}^{N} \frac{1}{n!} \underset{ \underset{g(\mathscr{E},A_1,...,A_n) \in \textbf{C}(\mathbb{N}_{n+1})}{A_1,...,A_n  \in \mathbb{A}_N}}{\sum}   \prod_{i=1}^{n} \eta_{K}(A_i)  \Big) \\
 &  \quad  \leq C(\tau,\lambda(0))  \|O  \|    \lambda(t_{N-1})  \lambda(0).
\end{align*}
where the constant $C(\tau,\lambda(0))>0$ depends on the parameters $\tau$ and $\lambda(0)$, but \textit{not} on $N$ and $\lambda(t)$, $t>0$. Furthermore, $C(\tau,\lambda(0))$ decreases when $\lambda(0)$ decreases.
\vspace{3mm}

  \noindent \textit{Case 2.b}: $(N-1,R) \in \mathcal{V}(\mathcal{X})$.  We denote by $A$ the maximal block  of neighboring $R$-vertices of  $\mathcal{X}$ that contains $(N-1,R)$. The block  $A$  \textit{must} be  adjacent to an edge on its left side, because $\mathcal{X}$ is a fusion of adjacent connected graphs, and  because $\mathcal{X}$ contains at least one edge. This is also the reason why $(0,R)$ and $(1, R)$ cannot belong to $A$. We use these remarks  to  extract a factor $\lambda(t_{m(A)})^{\mu}$, $\mu \in (0,1)$,  from the edge attached to the left side of $A$;  ($m(A)$ has been defined in  \eqref{mmm}). We proceed as follows. We define $\eta_{\mu,K}(\mathscr{E}):=(\lambda(t_i)\lambda(t_j))^{-\mu} \eta_{K}(\mathcal{E})$ for every edge $\mathcal{E}=((i,B);(j,B))$ and  $\eta_{\mu,K}(A'):=\eta_{K}(A')$ for every block $A'$ of neighboring $R$-vertices. It is easy to check that the estimates carried out in Section \ref{proo} remain almost the same  if we replace the weights $\eta_{K}$ by the new weights $\eta_{\mu,K}$.  Singling out the maximal block $A$ that contains $(N-1,R)$ and extracting a factor $\lambda(t_{m(A)})^{\mu}$ from the edge attached to it, we get that 
 \begin{align*}
 &\sum_{\mathcal{X} \in \mathbb{P}_N, \text{ } (N-1,R), \text{ }(N,R) \in \mathcal{V}(\mathcal{X})}   \vert  p(\mathcal{X}) \vert   e^{ \vert \mathcal{V}(\mathcal{X}) \vert}\\
& \quad \leq    \|O\| e  \sum_{\underset{(N-1,R)  \in A,   \text{ } (0,\cdot),(1,\cdot), (N,R)  \notin A}{A \in \mathbb{R}_N}}  (e\varepsilon_{0})^{\vert A \vert} \lambda^{\mu}(t_{m(A)}) \text{ } \sum_{n=1}^{N} \frac{1}{n!}  \text{ } \underset{ \underset{g(A,A_1,...,A_n) \in \textbf{C}(\mathbb{N}_{n+1})}{A_1,...,A_n  \in \mathbb{A}_N}}{\sum}   \prod_{i=1}^{n} \eta_{\mu,K}(A_i).
\end{align*}
Using similar calculation as in Section \ref{proo}, we can bound the last line of the previous equation by 
\begin{equation}
\label{Bset2}
  C(\tau,\lambda(0))  \|O\|  \sum_{k=1}^{N-2}  (e\varepsilon_{0})^{k} \lambda^{\mu}(t_{N-k-1}), 
\end{equation}
where $C(\tau,\lambda(0))>0$ is independent of $N$ and $\lambda(t)$, $t>0$, and decreases when $\lambda(0)$ decreases. The right side of Inequality \eqref{Rset1} and the bound \eqref{Bset2} are  of the form
\begin{equation}
\label{cccsa}
\Sigma_N:=\sum_{k=1}^{N} \varepsilon^{k} v_{N-k},
\end{equation}
where $$(v_n)_{n=1}^{\infty}$$ is a sequence of positive numbers converging to zero,  and $0<\varepsilon<1$.  All $v_{n}$'s  are bounded by some positive constant $C$ , and  $\Sigma_N$  is bounded  by 
$C \frac{\varepsilon}{1-\varepsilon}$, for all $N$. We therefore conclude that  $\Sigma_N \rightarrow 0$, as $N$ tends to $\infty$. Applying this result to  $ \eqref{Rset1}$ and $\eqref{Bset2}$, we finally find that $Z_N(O) \rightarrow 0$, as $N \rightarrow \infty$.

\vspace{3mm}

\section{ Extensions of Theorem \ref{clu}}
\label{S5}
\subsection{Extension to initial  field states with a finite number of photons}
We generalize Theorem \ref{clu} to initial field states with a finite number of photons.  We assume that the system $S \vee E$ is initially in the state $  \Psi = \varphi   \otimes  \varphi_E  $, where $   \varphi_E     = \Phi(f_1)....\Phi(f_{n_0})  \Omega  $  for a fixed number  $n_0 \in \mathbb{N}$. We assume that the functions $f_i$, $i=1,...,n_0$, satisfy
\begin{equation} \label{5.1}
\langle f_i, \phi_t \rangle_{L^2}  \propto \frac{1}{(1+t)^{\alpha}}
\end{equation}
for all $i=1,...,n_0$, where $\alpha > 2$ is the same number as in Assumption \ref{Co}. We also assume that $ \varphi_E  $ is normalized.
\begin{corollary} 
\label{normals}
We choose $n_0$ functions $f_i \in L^2(\mathbb{R}^3)$, $i=1,...,n_0$. Suppose that  assumptions  \eqref{5.1}, \ref{Fe} and  \ref{Ev} are satisfied.  Then there  is  a constant $\lambda_c>0$ such that, for any $0<\lambda(0)<\lambda_c$, 
\begin{equation}
\label{resultat}
\langle \Psi(t) \vert  (O \otimes 1) \Psi(t) \rangle  \underset{t \rightarrow \infty}{\longrightarrow}  \langle \varphi_1 \vert O  \varphi_1  \rangle,
\end{equation}
for all   $O\in \mathcal{B}(\mathcal{H}_S)$ and  for all initial states $  \Psi =\varphi     \otimes  \varphi_E $. $ \varphi_1 $ is the  ground state of  $H_S$ (unique up to a phase) corresponding to the eigenvalue $E_1$.
\end{corollary}

\subsection{ Thermalization at  positive temperature}
The method we used to prove Theorem \ref{clu} works in a  similar way at  positive temperature. We explain below how to show that  the  system $S$  thermalizes in the limit  $t \rightarrow +\infty$ if the field is initially in  thermal equilibrium at temperature $T>0$.   

We work directly in the thermodynamic limit. We  consider the Hilbert space
\begin{equation}
\mathfrak{h}:= L^{2}(\mathbb{R}^3, d^3k) \cap L^{2}(\mathbb{R}^3, \vert k \vert^{-1} d^3k).
\end{equation}
$\text{Im}  \langle f,g \rangle_{L^2}  $ is a symmetric non-degenerate symplectic bilinear form on $\mathfrak{h}$ and the  $C^*$-algebra $\mathcal{U}(\mathfrak{h})$ generated by the Weyl operators 
\begin{align*}
W(-f)&=W(f)^*, \quad f \in \mathfrak{h},\\
W(f)W(g)&= e^{-i\text{Im}  \langle f,g \rangle_{L^2}/2 } W(f+g), \quad  f,g \in \mathfrak{h},
\end{align*}
 is unique up to a $^*$-isomorphism; see e.g. \cite{Brat}. $\mathcal{U}(\mathfrak{h})$ is the algebra of field observables. Time-evolution on $\mathcal{U}(\mathfrak{h})$ is given by the one-parameter group  of $^*$-automorphism, $\{\alpha_{t}^{E}\}_{t \in \mathbb{R}}$, defined by
\begin{equation}
\alpha^{E}_{t}(W(f)):=W(e^{i \omega t} f)
\end{equation}
for all $f \in \mathfrak{h}$ and for all $t \in \mathbb{R}$, where $\omega(k)=\vert k \vert$. It is well-known that $\alpha^{E}_{t}$ is not norm continuous ($\|W(f)- \mathds{1} \|=2$ if  $f \neq 0$),  and the dynamical properties of the interacting  system  $S \vee E$ can  only be understood in a representation dependent way. We consider the KMS state at temperature $1/\beta>0$ defined on $\mathcal{U}(\mathfrak{h})$ by 
\begin{equation}
\rho_{\beta}(W(f))=\exp \Big(-\frac{1}{4} \int_{\mathbb{R}^3}d^3k \text{ }  \frac{1+e^{-\beta \vert k \vert }}{1-e^{-\beta \vert k \vert }}  \vert f (k) \vert^2 \Big), \quad f \in \mathfrak{h}.
\end{equation}
 The function $t \mapsto \rho_{\beta}(W(tf))$ is real analytic and it is  possible to make sense of the infinitesimal generators $\Phi_{\rho_{\beta}}(f)$ of the one-parameter group of unitary transformations  $t \mapsto \pi_{\rho_{\beta}}(W(tf))$ in the GNS representation $(\mathcal{H}_{\rho_{\beta}},\pi_{\rho_{\beta}}, \Omega_{\rho_{\beta}})$ of $(\mathcal{U}(\mathfrak{h}), \rho_{\beta})$; see \cite{Brat}.  The two-point correlations are  given by
  \begin{equation}
  \rho_{\beta} (\Phi_{ \rho_{\beta}}(f) \Phi_{ \rho_{\beta}}(g))=\langle f , (1-e^{- \beta  \omega} )^{-1} g  \rangle_{L^2} + \langle g , e^{-\beta \omega}(1-e^{- \beta  \omega})^{-1} f \rangle_{L^2}
\end{equation}
for all $f,g \in \mathfrak{h}$, and easy calculations show that the state $\rho_{\beta}$ is quasi-free.  The one parameter group  $\{\alpha_{t}^{E}\}_{t \in \mathbb{R}}$  is represented on $\pi_{\rho_{\beta}}$ by 
\begin{equation}
\pi_{\rho_{\beta}}(\alpha^{E}(t)(O_E))=U^{*}(t) \pi_{\rho_{\beta}}(O_E) U(t),
\end{equation}
where $\{U(t)\}_{t \in \mathbb{R}}$ is the one-parameter group of unitary transformations defined by
$$U(t) \pi_{\rho_{\beta}}(O_E)  \Omega_{\rho_{\beta}}:=\pi_{\rho_{\beta}}(\alpha^{E}(-t)(O_E))  \Omega_{\rho_{\beta}}, \qquad  U(t) \Omega_{\rho_{\beta}}=\Omega_{\rho_{\beta}},$$
for all $O_E \in \mathcal{U}(\mathfrak{h})$.  Time  translation of the  operators  $\Phi_{ \rho_{\beta}}(f)$ is given by  $\Phi_{ \rho_{\beta}}(f)(t)= \Phi_{ \rho_{\beta}}(e^{it\omega}f)$, for all $t \in \mathbb{R}$.

 We compose the field $E$ with the atomic system $S$ and we consider the C$^*$-algebra $\mathcal{A}=\mathcal{B}(\mathcal{H}_S) \otimes \mathcal{U}(\mathfrak{h})$ equipped with the projective C$^*$  cross-norm; see \cite{Tak}. The free dynamics on $\mathcal{A}$ is generated by the one-parameter group of $^*$-automorphisms  $\lbrace \alpha^{0}_{t} \rbrace_{t \in \mathbb{R}}$, where $ \alpha^{0}_{t}$ is determined by
\begin{equation}
 \alpha^{0}_{t}(O \otimes O_E)=e^{itH_S} O e^{-itH_S} \otimes \alpha_{t}^{E}(O_E)
 \end{equation}
 for all $O \in \mathcal{B}(\mathcal{H}_S)$ and all $O_E \in \mathcal{U}(\mathfrak{h})$. 
 We now turn on the interaction between the atom and the field. The dynamics of the interacting system is defined through a  Dyson series. One has  to be careful  here because  $ \alpha_{t}^{E}$ is not norm continuous, and, hence, the Dyson series only makes sense in  a representation dependent way.  A rigorous construction  of the  interacting dynamics on $\mathcal{A}$ as the limit of a regularized and norm-continuous dynamics on a regularized algebra  can be found in \cite{Therm}. We avoid these complications here since we are only interested in the time evolution of observables of the form $O \otimes \mathds{1}$. We work directly  in the  representation $(\mathcal{H}_S \otimes \mathcal{H}_{\rho_{\beta}}, \mathds{1} \otimes \pi_{\rho_{\beta}})$.  We consider the self-adjoint and densely defined operator on $\mathcal{H}_S \otimes \mathcal{H}_{\rho_{\beta}}$,
 $$H_{I} :=G \otimes \Phi_{\rho_{\beta}}(\phi),$$
  where $\phi \in  \mathfrak{h}$ is the form factor of the interaction Hamiltonian of the last sections. The  interaction Hamiltonian translated at time $t$ is given by 
 $$H_{I}(t)= e^{itH_S} G e^{-itH_S} \otimes \Phi_{\rho_{\beta}}(e^{it \omega} \phi).$$ 
 Let $O \in \mathcal{B}(\mathcal{H}_S)$. The quadratic form 
  \begin{equation*}
  \begin{split}
q_{O}^{t,s}(\Psi,\Psi'):=\sum_{n=0}^{\infty} & i^{n} \int_{s}^{t} dt_1 ...   \int_{s}^{t_{n-1}} dt_n  \text{ } \lambda(t_1)...\lambda(t_n) \langle \Psi \vert (e^{-isH_S } \otimes U(s)) \\
&  [ H_I(t_n),...,[ H_I(t_1), e^{itH_S} O e^{-itH_S}]...]  (e^{isH_S} \otimes U^{*}(s)) \Psi' \rangle
\end{split}
\end{equation*}
is well-defined for all $\Psi,\Psi' \in \mathcal{H}_S \otimes F( \mathcal{H}_{\rho_{\beta}})$, where $F( \mathcal{H}_{\rho_{\beta}}):=\{ \Phi(f_1) ... \Phi(f_n) \Omega_{\rho_{\beta}} \mid n \in \mathbb{N}, \text{ }f_i \in \mathfrak{h}\}$.  The quadratic form $q_{O}^{t,s}$  induces a unique operator  $\mathcal{Z}^{t,s}_{\beta}(O) \in  \mathcal{B}(\mathcal{H}_S)$,  defined  by
\begin{equation}\label{tempi}
\langle \varphi \vert \mathcal{Z}^{t,s}_{\beta}(O) \psi \rangle:=q_O^{t,s}(\varphi \otimes \Omega_{\rho_{\beta}},\psi \otimes \Omega_{\rho_{\beta}}), \qquad \forall \varphi,\psi \in \mathcal{H}_S.\end{equation}
The expression of  $\mathcal{Z}^{t,s}_{\beta}(O)$ is similar to  \eqref{DD1'}. The only  change consists in the replacement of the correlation function $f$ in \eqref{DD1'}  by the correlation function
\begin{equation}
 f_{\beta}(t):= \rho_{\beta} (\Phi(\phi_t) \Phi(\phi))=\langle \phi _{t}, (1-e^{- \beta  \omega} )^{-1} \phi  \rangle + \langle \phi , e^{-\beta \omega}(1-e^{- \beta  \omega})^{-1} \phi_{t} \rangle
\end{equation}
at temperature $1/\beta$. 
\vspace{2mm}

\begin{corollary} 
\label{ther}
Let $T>0$. Suppose that Assumptions \ref{Co} ( with $f$  replaced by $ f_{\beta}$), \ref{Fe} and \ref{Ev} are satisfied. Then there exists a   constant $\lambda_c>0$, such that, for any $0<\lambda(0)<\lambda_c$, 
\begin{equation}
 \underset{ t \rightarrow \infty}{\lim} \text{ } \langle \varphi \vert\mathcal{Z}^{t,0}_{\beta}(O) \varphi \rangle= \text{Tr}_{\mathcal{H}_S} (\rho_{S,\beta} O), 
\end{equation}
 for all $\varphi \in \mathcal{H}_S$ with  $\|\varphi \|=1$.  The state $\rho_{S,\beta}:=e^{-\beta H_S}/ \text{Tr}_{\mathcal{H}_S} (e^{-\beta H_S})$ is the Gibbs equilibrium state of S at temperature $T=1/\beta$.
\end{corollary}

\newpage
\begin{appendix}
\section{ Cluster expansions}
\label{clustersec}
 We review some standard features of  cluster expansions.  The reader  is referred to  \cite{Proca}, \cite{Bry}, \cite{Ue}, \cite{FPr} for more details.  We mainly  follow the exposition in \cite{Ue} and \cite{Kno}. A set of  polymers  is  a measurable set $(\mathbb{X},\Sigma, \mu)$ where  $\mu$ is a complex measure with finite total variation $\vert \mu \vert(\mathbb{X})$.  An element  $x \in \mathbb{X}$ is  called  a ``polymer".  Let  $\xi: \mathbb{X} \times \mathbb{X} \rightarrow \mathbb{R}$ be a symmetric function with the property that
\begin{equation}
\vert 1 + \xi( x , y) \vert \leq 1, \qquad \forall x,y  \in \mathbb{X}.
\end{equation}
$\xi$ encodes an adjacency relation $\sim$, i.e. a symmetric and irreflexive binary  relation. For hardcore polymer models, $\xi(x,y)=-1$ if  $x \sim y$, and $0$ otherwise.  For  the  polymer set $\mathbb{P}_N$ introduced in  Section \ref{graphph},  $\mathcal{X} \sim \mathcal{X}'$ if $\text{dist}(\mathcal{X},\mathcal{X}') \leq 1$. We  consider the partition function
 \begin{equation}
 \label{ccll}
 Z:=1+\sum_{n \geq 1} \frac{1}{n!} \int d\mu(x_1)...d\mu(x_n) \prod_{1 \leq i < j \leq n} \left(1+ \xi(x_i,x_j) \right).
 \end{equation}
Formula  \eqref{ccll} is a cluster expansion.  Under certain circumstances, the right side of \eqref{ccll} can be rewritten as the argument of an exponential. To do so, we define
 \begin{equation}
 \mathbb{N}_n:=\lbrace 1,...,n \rbrace.
 \end{equation}
 For every $A \subset \mathbb{N}$, we denote by $\textbf{G}(A)$ the set of graphs with vertex set  $A$ and with edges pairs $(i,j)$ with $i \neq j$ and $i,j \in A$.  Among those graphs, the connected ones are denoted by $\textbf{C}(A)$, and the trees are denoted by $\textbf{T}(A)$. To make the distinction with the set $\mathbb{P}_{N}$ (see Section \ref{graphph}), we denote the graphs in  $\textbf{G}(A)$ with small letters, i.e. $g,f,...$. The set of edges of the graph $g$ is denoted by $\mathcal{E}(g)$.
 One has that
 \begin{equation}
 \prod_{1 \leq i <j \leq n} \left(1+ \xi(x_i,x_j) \right)=\sum_{g \in \textbf{G}( \mathbb{N}_n)} \prod_{(i,j)\in \mathcal{E}(g)} \xi(x_i,x_j).
 \end{equation}
The connected part of  $\underset{g \in \textbf{G}( \mathbb{N}_n)}{\sum}  \underset{(i,j)\in \mathcal{E}(g)}{\prod} \xi(x_i,x_j)$ is given by 
$$\sum_{g \in \textbf{C}( \mathbb{N}_n)} \prod_{(i,j)\in \mathcal{E}(g)} \xi(x_i,x_j).$$
We introduce the ``Ursell functions"
\begin{equation}
\varphi^{T}(x_1,...,x_n):= \left\{  \begin{array}{ll}\qquad  1 & \mbox{ if } n=1,\\[5pt] 
\underset{g \in \textbf{C}( \mathbb{N}_n)}{\sum}  \underset{(i,j)\in \mathcal{E}(g)}{\prod} \xi(x_i,x_j)& \mbox { if } n \neq 1. \end{array} \right.
\end{equation}

\noindent If sums and integrals can be exchanged, we get that 
\begin{eqnarray*}
Z&=&1+\sum_{n \geq 1} \frac{1}{n!} \int d\mu(x_1)...d\mu(x_n) \sum_{g \in \textbf{G}( \mathbb{N}_n)} \prod_{(i,j)\in \mathcal{E}(g)} \xi(x_i,x_j)\\
&=&1+\sum_{n \geq 1} \frac{1}{n!} \int d\mu(x_1)...d\mu(x_n)  \sum_{k=1}^{n}\frac{1}{k!} \sum_{A_1 \cup ... \cup A_k=  \mathbb{N}_n} \prod_{l=1}^{k}  \left(\sum_{g \in \textbf{C}( A_l)} \prod_{(i,j)\in \mathcal{E}(g)} \xi(x_i,x_j) \right)\\
&=& 1+ \sum_{n \geq 1} \frac{1}{n!}  \sum_{k=1}^{n}\frac{1}{k!} \sum_{A_1 \cup ... \cup A_k=  \mathbb{N}_n}   \int d\mu(x_1)...d\mu(x_n) \prod_{l=1}^{k}  \left(\sum_{g \in \textbf{C}( A_l)} \prod_{(i,j)\in \mathcal{E}(g)} \xi(x_i,x_j) \right).
\end{eqnarray*}

\noindent To go from the first to the second line, we have decomposed every graph $g$  into its connected components. Furthermore,  $A_1\cup ... \cup A_k$ is a partition of  $ \mathbb{N}_n$ such that $A_l \neq \emptyset$ for all $l$. We write 
$$d\mu(x_{A_l})=\prod_{x \in A_l} d\mu(x).$$
Then, 
 $$\int d\mu(x_{A_l})  \left(\sum_{g \in \textbf{C}( A_l)} \prod_{(i,j)\in \mathcal{E}(g)} \xi(x_i,x_j) \right)$$
  depends only on the number of elements in $A_l$. There are $\frac{n!}{m_1!...m_k!}$ partitions of $\mathbb{N}_n$ in $k$ subset $A_l$ with $m_l$ elements,  and we deduce that 
\begin{eqnarray*}
Z&=& 1+ \sum_{n \geq 1} \frac{1}{n!}  \sum_{k=1}^{n}\frac{1}{k!} \sum_{A_1 \cup ... \cup A_k=  \mathbb{N}_n}   \prod_{l=1}^{k}   \int d\mu(x_{A_l}) \left(\sum_{g \in \textbf{C}( A_l)} \prod_{(i,j)\in \mathcal{E}(g)} \xi(x_i,x_j) \right)\\
&=& 1+ \sum_{n \geq 1} \frac{1}{n!}  \sum_{k=1}^{n}\frac{1}{k!} \sum_{m_1+...+m_k=n} \frac{n!}{m_1!...m_k!}   \prod_{l=1}^{k}   \int d\mu(x_{\mathbb{N}_{m_l}}) \left(\sum_{g \in \textbf{C}( \mathbb{N}_{m_l})} \prod_{(i,j)\in \mathcal{E}(g)} \xi(x_i,x_j) \right)\\
&=&1+\sum_{k=1}^{\infty}  \frac{1}{k!} \left( \sum_{m_1=1}^{\infty} \frac{1}{m_1!} \int d\mu(x_{\mathbb{N}_{m_1}}) \left(\sum_{g \in \textbf{C}( \mathbb{N}_{m_1})} \prod_{(i,j)\in \mathcal{E}(g)} \xi(x_i,x_j) \right)  \right)^{k}\\
&=&\exp \left( \sum_{n \geq 1}  \frac{1}{n!}  \int d \mu(x_1)...d\mu(x_n) \varphi^{T}(x_1,...,x_n) \right).
\end{eqnarray*} 

Our calculations are formal and  the exchange of sums and integrals  must  be justified. This  exchange can be done if the function $\xi$ and the measure $\mu$ satisfy  specific criteria  such that the series above are absolutely convergent; see e.g. \cite{FPr}. In this paper, we  use the  Kotecky-Preiss criterion stated below.

\begin{proposition} (KP criterion, see e.g.\cite{Kote}, \cite{Ue}) \label{Ue}
Let us assume that there is  a non-negative function $a: \mathbb{X} \rightarrow \mathbb{R}_+$ such that
\begin{equation}
\label{crite}
\int d\vert \mu \vert (x') \vert \xi(x,x') \vert e^{a(x')} \leq a(x) \qquad  \forall x \in \mathbb{X},
\end{equation}
and $\int d\vert \mu \vert(x) e^{a(x)} < \infty$.  Then
\begin{equation}
Z=\exp \left( \sum_{n \geq 1}  \frac{1}{n!}  \int d \mu(x_1)...d\mu(x_n) \varphi^{T}(x_1,...,x_n) \right),
\end{equation}
and combined sums and integrals converge absolutely. Furthermore, for all $x_1 \in \mathbb{X}$,
\begin{equation}
\label{boon}
1+ \sum_{n \geq 2} \frac{1}{(n-1)!}  \int d \vert \mu \vert (x_2)...d\vert \mu \vert (x_n)  \vert \varphi^{T}(x_1,...,x_n)  \vert \leq e^{a(x_1)}.
\end{equation}
\end{proposition}

We work here with a finite polymer set, $\mathbb{P}_N$,  and the integral over $\mathbb{X}$  has to be  replaced by  a finite sum:
$$\int d\mu(x) \leftrightarrow \sum_{\mathcal{X} \in \mathbb{P}_N} p(\mathcal{X}),$$
where $p(\mathcal{X})$ is  the weight of the polymer $\mathcal{X}$. 
\vspace{3mm} 

\section{Proofs of the Lemmas stated in Section \ref{S2}}
\vspace{2mm}
\subsection{Proof of Lemma \ref{Dyson}}
\label{A1}
\noindent We introduce  the operator
\begin{equation}
\label{Dyson1}
\tilde{U}(t,s) :=1 +\sum_{k=1}^{\infty} (-i)^{k} \int_{s}^{t}d u_{k} ...\int_{s}^{u_{2}} du_{1} \lambda(u_k) H_{I}(u_k)... \lambda(u_1) H_{I}(u_1)
\end{equation}
for all $t,s \in \mathbb{R}$.  $H_I(t)=e^{itH_{0}} H_{I} e^{-it H_{0}}$
for all $t \in \mathbb{R}$. We denote by $F(L^{2}(\mathbb{R}^3)) \subset \mathcal{F}_+(L^{2}(\mathbb{R}^3))$ the subspace of finite particle vectors. We show that $F(L^{2}(\mathbb{R}^3)) \subset D(\tilde{U}(t,s))$  and that  \eqref{Dyson1} converges  strongly  on $F(L^{2}(\mathbb{R}^3))$, for all $t,s \in \mathbb{R}$. Let $n \in \mathbb{N}$ and let  $\varphi^{(n)}:=\varphi_S \otimes \psi^{(n)}$, with $\varphi_S \in \mathcal{H}_S$ and  $\psi^{(n)} \in \mathcal{F}^{(\leq n)}_{+}(L^{2}(\mathbb{R}^3))$.   An easy calculation shows that 
\begin{equation}
\label{Hit}
H_{I}(t)=G(t)\otimes \Phi(\phi(t))
\end{equation}
where $\Phi$, $G(t)$ and $\phi(t)$ have been defined in (\ref{PHII}), (\ref{Gt}) and (\ref{phihi}), respectively. Therefore, 
\begin{equation}
\label{ineq1}
\| H_{I}(t) \varphi^{(n)}  \| \leq  2 \| G  \| (n+1)^{1/2} \| \phi \|_{L^{2}} \text{ } \|   \varphi^{(n)} \|. 
\end{equation}
Inserting (\ref{ineq1}) into \eqref{Dyson1}, we get that
\begin{eqnarray*}
\| \tilde{U}(t,s)\varphi^{(n)} \| &\leq&  \|  \varphi^{(n)} \|    +\sum_{k=1}^{\infty} \frac{1}{k!} (2\lambda(s) \| G \| \text{ }  \| \phi \|_{L^2}  \text{ }  \vert t-s \vert)^{k}  (n+1)^{1/2}...(n+k)^{1/2} \|   \varphi^{(n)} \| \\
&\leq&\|  \varphi^{(n)} \|  \left( 1+  \sum_{k=1}^{\infty} \frac{1}{\sqrt{k!}} (4 \lambda(s) \| G \| \text{ }  \| \phi \|_{L^2} \text{ }  \vert t-s \vert)^{k} \right) 2^{-n/2}  \prod_{p=1}^{n} \left(\frac{n}{p}+1\right)^{1/2},
\end{eqnarray*}
\normalsize
which clearly converges for all $t,s \in \mathbb{R}$. To go from the first to the second line, we have  used that
\begin{eqnarray*}
 \frac{(n+1)^{1/2}}{1^{1/2}}... \frac{(n+k)^{1/2}}{k^{1/2}} &=&  (n+1)^{1/2}...  (n/k+1)^{1/2}\\
 &\leq&  \prod_{p=1}^{n} \left(\frac{n}{p}+1\right)^{1/2} 2^{-(n-k)/2}
 \end{eqnarray*}
 for all  $k \leq n$, and that
\begin{eqnarray*}
 \frac{(n+1)^{1/2}}{1^{1/2}}... \frac{(n+k)^{1/2}}{k^{1/2}} &\leq&  \prod_{p=1}^{n} \left(\frac{n}{p}+1\right)^{1/2} 2^{(k-n)/2}
 \end{eqnarray*}
  for all $k>n$. This shows that the series defining $\tilde{U}(t,s)$ converges strongly  on $F(L^2(\mathbb{R}^3))$.
  
 \vspace{2mm}

\subsection{Proof of Lemma \ref{lem3}}
\label{A22}
\noindent We  prove (\ref{DD1'}).   We rewrite  \eqref{D1} with  the notations introduced in \eqref{simpl} and \eqref{ordon}.
\begin{equation}
\label{simply1}
U(t,s) =e^{-i(t-s)H_0} +\sum_{k=1}^{\infty} (-i)^{k} \int_{\Delta^{k}\left[s,t\right]} d \underline{u}  \text{ }\lambda(\underline{u}) \text{ }   e^{-i t H_0}  (H_{I}(\underline{u}))^{*}    e^{i s H_0}
\end{equation}
for all $t,s \in \mathbb{R}_+$. We plug \eqref{simply1}  and its adjoint into  \eqref{def_O}. We get that
\begin{equation*}
\begin{split}
&\mathcal{Z}^{t,s}(O) P_{\Omega} = \\
& \sum_{k_1,k_2=0}^{\infty}  (-1)^{k_2} i^{k_1+k_2}  \int_{\Delta^{k_1}\left[s,t\right] \times \Delta^{k_2}\left[s,t\right] } d \underline{u}  d \underline{u}' \text{ }     \lambda(\underline{u})   \lambda(\underline{u}')  \text{ }P_{\Omega}  e^{-i s H_S}   H_{I}(\underline{u}) O(t) (H_{I}(\underline{u}'))^{*}  e^{i s H_S} P_{\Omega}, 
\end{split}
\end{equation*}
where $O(t)= e^{i t H_S}    O  e^{-i t H_S}$; see \eqref{Gt}. Formula \eqref{Hit}  implies that
\begin{equation*}
H_{I}(\underline{u})  O(t) (H_{I}(\underline{u}'))^{*} =G(\underline{u})  O(t) [G(\underline{u}')]^{*} \otimes \Phi(\phi(u_1))... \Phi(\phi(u_{k_1}))  \Phi(\phi(u'_{k_2}))... \Phi(\phi(u'_{1}) ).
\end{equation*}
We glue the time coordinates $\underline{u}$ and $\underline{u}'$ together and introduce the new coordinate 
$$\underline{x}:=(x_1,...,x_{k_1+k_2}):=(u_1,...,u_{k_1},u'_{k_2},...,u'_{1}).$$
Wick's theorem (see \eqref{qfree}) implies that   
\begin{equation}
P_{\Omega}     H_{I}(\underline{u})  O(t) (H_{I}(\underline{u}'))^{*}   P_{\Omega} = P_{\Omega}    \sum_{\underset{i<j}{\text{pairings } \pi}} G(\underline{u})   O(t)  [G(\underline{u}')]^{*}  \prod_{(i,j)\in \pi}   f(x_i-x_j) 
\end{equation}
 if $k_1+k_2$  is even. We assign a number $r_i \in \{0,1 \}$ to every time $x_i \in \underline{x}$ and set $r_i=0$ if $i\leq k_1$ and $r_i=1$ if $i >k_1$. We write $(x_{i},r_i)$ and we use this new index to take into account the fact that the operator $G(x_i)$ multiplies $O$ from the left if $r_i=0$ and from the right if $r_i=1$. Using \eqref{F3} and \eqref{F2},  we get that  
 \begin{equation}
\label{better}
 (-1)^{k_2} i^{k_1+k_2} P_{\Omega}    H_{I}(\underline{u})  O(t) (H_{I}(\underline{u}'))^{*}  P_{\Omega} = P_{\Omega}     \sum_{\underset{i<j}{\text{pairings } \pi}} \mathcal{T}_S \big( \prod_{(i,j)\in \pi}  \textbf{F}(x_i,r_i;x_j,r_j)  \big)\left[O(t) \right].
\end{equation}

\noindent Let $k_1+k_2=2 k$, and let $((x_{i_1},x_{j_1}),..., (x_{i_{k}},x_{j_{k}}))$  be a tuple of $k$ pairs ($i_l, j_l \in \{1,...,2k\}$).  We classify the pairs in  $((x_{i_1},x_{j_1}),..., (x_{i_{k}},x_{j_{k}}))$  in  increasing order.  There exists a unique permutation   $\sigma$ of $\lbrace 1,...,2k\rbrace$, such that  the $k$-tuple $((x_{i_1},x_{j_1}),..., (x_{i_{k}},x_{j_{k}}))$ can be rewritten as
\begin{equation}
\label{pipi}
\left( (x_{\sigma(1)},x_{\sigma(2)}),...,(x_{\sigma(k_1+k_2-1)},x_{\sigma(k_1+k_2)}) \right), \qquad x_{\sigma(1)}<x_{\sigma(3)}<...<x_{\sigma(k_1+k_2-1)},
\end{equation}
and $x_{\sigma(2i-1)}< x_{\sigma(2i)}$, for all $i=1,...,k$.  Every tuple of $k$ pairs arises $4^k$ times by summing over the indices $r_i \in \{0,1\}$.   Using a change of variables  for each permutation $\sigma$ and summing over all possible permutations, we get that 
 \begin{align*}
&e^{is H_S}\mathcal{Z}^{t,s}(O) e^{-is H_S} \\
&= \sum_{k=0}^{\infty} \text{ } \underset{s<x_1<x_{3}<...<x_{ 2k-1}<t}{\int} d \underline{x} \text{ } \lambda( \underline{x})  \sum_{ \underline{r}  \in \lbrace 0,1 \rbrace^{2k}} \mathcal{T}_S \Big[ \text{ }\prod_{i=1}^{k}  \big( \chi(x_{2i-1}< x_{ 2i}) \text{ } \textbf{F}(x_{2i-1},r_{2i-1} ;x_{2i},r_{2i})  \big) \Big] \left[ O(t) \right].
\end{align*}
 Introducing $(u_i,r_i):=(x_{2i-1},r_{2i-1})$, $(v_i,r'_i):=(x_{2i},r_{2i})$, $w_i=(u_i,r_i;v_i,r'_i)$, and using the measure (\ref{mumu}),  we finally get that
\begin{equation*}
e^{is H_S}\mathcal{Z}^{t,s}(O) e^{-is H_S} = \sum_{k=0}^{\infty} \underset{[s,t]^{2k}}{\int} d\mu_k(\underline{w}) \text{ } \lambda( \underline{w})     \mathcal{T}_S \Big[ \prod_{i=1}^{k}    \textbf{F}(u_{i},r_i ;v_{i},r'_i)  \Big] \left[ O(t) \right].
\end{equation*}
\vspace{2mm}

We now show that the series  converges strongly.  If $u_i,v_i \in [t,s]$, $u_i<v_i$,  we remind the reader that $\lambda(v_i)<\lambda(u_i)\leq\lambda(s)$.  One has that
\begin{equation}
\label{b1}
 \lambda( \underline{u}) \lambda(\underline{v} )  \big \| \mathcal{T}_S \Big[ \prod_{i=1}^{k}   \textbf{F}(u_{i},r_i;v_{i},r'_i)  \Big] \left[ O(t) \right] \big \| \leq  \lambda^{2k} (s) \text{ }  \| O  \| \text{ } \| G  \|^{2k}   \prod_{i=1}^{k} \vert f(v_i-u_i) \vert .
\end{equation} 

\noindent We plug this bound into (\ref{DD1'}), and  we get that
\begin{equation*}
\begin{split}
\underset{[s,t]^{2k}}{\int} d\mu_k(\underline{w}) & \big \|  \mathcal{T}_S \Big[ \prod_{i=1}^{k}   \textbf{F}(u_{i},r_i;v_{i},r'_i)  \Big] \left[ O(t) \right] \big \|  \\
&\leq \lambda^{2k} (s) \text{ }   \underset{[s,t]^{2k}}{\int} d\mu_k(\underline{w})  \text{ }  \|  O  \| \text{ }  \| G  \|^{2k}   \prod_{i=1}^{k} \vert f(v_i-u_i) \vert .
\end{split}
\end{equation*} 
Then we integrate over the $v's$, which leads us to
\begin{equation*}
\begin{split}
\underset{[s,t]^{2k}}{\int} d\mu_k(\underline{w}) &   \Big\|  \mathcal{T}_S \Big[ \prod_{i=1}^{k}   \textbf{F}(u_{i},r_i;v_{i},r'_i)  \Big] \left[ O(t) \right] \Big\|  \\
&\leq \lambda^{2k} (s)\text{ }   4^{k}  \underset{s<u_1<u_{2}<...<u_k<t}{\int} d \underline{u}  \text{ } \| O \| \| G \|^{2k}     \vert  \vert f \vert  \vert_{L^1}^{k} .
\end{split}
\end{equation*} 
Finally, we can integrate over the $k$-dimensional simplex and sum over $k$ to obtain that
\begin{equation}
\begin{split}
\sum_{k=0}^{\infty} \underset{[s,t]^{2k}}{\int} d\mu_k(\underline{w}) & \Big \|  \mathcal{T}_S \Big[ \prod_{i=1}^{k}   \textbf{F}(u_{i},r_i;v_{i},r'_i)  \Big] \left[ O(t) \right]\Big \|  \\
&\leq \sum_{k=0}^{\infty} \frac{1}{k!} \lambda^{2k} (s) \text{ }  4^{k}  \vert t-s \vert^{k}  \| G \|^{2k}     \vert  \vert f \vert  \vert_{L^1}^{k}  \| O \| .
\end{split}
\end{equation} 
\vspace{2mm}

\subsection{Proof of Lemma \ref{spectrum}}
\label{A3}

\noindent We  compute $ \mathcal{K}_s  \vert \Pi_{i j} \rangle$, where $ \Pi_{i j}=\vert \varphi_i \rangle \langle \varphi_j \vert \in \mathcal{B}(\mathcal{H}_S)$, and $\varphi_1,...,\varphi_n$ are the normalized eigenvectors of $H_S$. We get that
\begin{eqnarray*}
\mathcal{K}_s  \vert \Pi_{i j} \rangle&=&f(s) \mathbf{R}(G) e^{is \mathcal{L}_S} \mathbf{L}(G)  \vert \Pi_{i j} \rangle + f(-s) \mathbf{L}(G) e^{is \mathcal{L}_S} \mathbf{R}(G) \vert \Pi_{i j} \rangle\\
&-&f(-s) \mathbf{L}(G) e^{is \mathcal{L}_S} \mathbf{L}(G)  \vert \Pi_{i j} \rangle- f(s) \mathbf{R}(G) e^{is \mathcal{L}_S} \mathbf{R}(G) \vert \Pi_{i j} \rangle.
\end{eqnarray*}

\noindent Using the equality
\begin{eqnarray*}
 \Pi_{i j} G&=&   \sum_{k,  l } G_{k l}    \Pi_{i j} \Pi_{k l}=   \sum_{k, l } G_{k l}   \vert \varphi_{i} \rangle \langle \varphi_{j} \vert  \varphi_{k} \rangle \langle \varphi_{l} \vert = \sum_{l}  G_{j l}      \Pi_{i l},  \\
 G \Pi_{i j} &=&   \sum_{k, l } G_{k l}    \Pi_{k l} \Pi_{i j} =   \sum_{k, l } G_{k l}   \vert  \varphi_{k} \rangle \langle \varphi_{l} \vert  \varphi_{i} \rangle \langle \varphi_{j} \vert= \sum_{k}  G_{k i}      \Pi_{k j} ,
\end{eqnarray*}

\noindent  we  obtain that
\begin{eqnarray*}
\mathcal{K}_s  \vert \Pi_{i j} \rangle&=&f(s) \sum_{k,m}  e^{is \epsilon_{k j}}  G_{k i} G_{j m}    \vert \Pi_{k m} \rangle + f(-s)   \sum_{l,m}    e^{is \epsilon_{i l}}  G_{j l}  G_{m i}     \vert \Pi_{m l} \rangle\\
&-&f(-s)   \sum_{k,m}   e^{is \epsilon_{k j}} G_{m k}  G_{k i}    \vert \Pi_{m j} \rangle- f(s)   \sum_{l,m}   e^{is \epsilon_{i l}}  G_{j l}  G_{ l m}    \vert \Pi_{i m} \rangle.
\end{eqnarray*}

\noindent  If $i \neq  j$, then
\begin{eqnarray*}
P_{\epsilon_{i j}} \mathcal{K}_s  \vert \Pi_{i j} \rangle &=&\left( f(s)  e^{is \epsilon_{i j}}  G_{i i} G_{j j}   + f(-s)     e^{is \epsilon_{i j}}  G_{j j}  G_{i i}  \right) \vert \Pi_{i j} \rangle\\
&-&\Big( f(-s)   \sum_{k}   e^{is \epsilon_{k j}} G_{i k}  G_{k i}  + f(s)   \sum_{l}   e^{is \epsilon_{i l}}  G_{j l}  G_{ l j} \Big)\vert \Pi_{i j} \rangle
\end{eqnarray*}
 and  we deduce that 
\begin{eqnarray*}
\mathcal{M} \vert \Pi_{i j} \rangle &=& 2\int_{0}^{\infty} ds \text{ } \Re(f(s))  \text{ } G_{i i} G_{j j}   \vert \Pi_{i j} \rangle\\
&-&\Big(   \sum_{k}    \int_{0}^{\infty} ds \text{ }  f(-s)   e^{is \epsilon_{ k i}} \vert G_{i k} \vert^{2}   +  \sum_{l}  \int_{0}^{\infty} ds \text{ } f(s) e^{is \epsilon_{j l}}   \vert G_{j l}   \vert^{2}  \Big)\vert \Pi_{i j} \rangle.
\end{eqnarray*}

\noindent If $i=j$, we get that
\begin{eqnarray*}
\mathcal{M} \vert \Pi_{i i} \rangle &=& \int_{0}^{\infty} ds \text{ }f(s) \sum_{k }  e^{is \epsilon_{k i}}  G_{k i} G_{i k}    \vert \Pi_{k k} \rangle +  \int_{0}^{\infty} ds \text{ } f(-s)   \sum_{l }    e^{is \epsilon_{i l}}  G_{i l}  G_{l i}     \vert \Pi_{ll} \rangle\\
&-&  \int_{0}^{\infty} ds \text{ }f(-s)   \sum_{k}   e^{is \epsilon_{k i}} G_{i k}  G_{k i}    \vert \Pi_{i i} \rangle-   \int_{0}^{\infty} ds \text{ }f(s)   \sum_{l }   e^{is \epsilon_{i l}}  G_{i l}  G_{ l i}    \vert \Pi_{i i} \rangle\\
&=& 2 \sum_{k }  \int_{0}^{\infty} ds \text{ }  \Re(f(s) e^{is \epsilon_{k i}})  \vert G_{k i}  \vert^{2}    \vert \Pi_{k k} \rangle -2 \sum_{k}     \int_{0}^{\infty} ds \text{ }  \Re(f(-s) e^{is \epsilon_{k i}}) \vert  G_{k i}   \vert^{2}     \vert \Pi_{i i} \rangle.
\end{eqnarray*}

\noindent Since
\begin{eqnarray*}
f(t)&=& \int_{\mathbb{R}^{3}} d^{3}k \text{ } \vert \phi(k) \vert^{2} e^{-it \omega(k)},
\end{eqnarray*}
we deduce that 
\begin{eqnarray}
\label{Re1}
\Re \left( \int_{0}^{\infty} ds \text{ }   f (s) e^{is \epsilon_{ j i}} \vert G_{i j} \vert^{2} \right)&=& \pi  \int_{\mathbb{R}^{3}} d^3k \text{ } \vert \phi(k) \vert^{2}  \delta(\epsilon_{ j i}-\omega(k)) \vert G_{i j} \vert^{2}\\
\label{Re2}
\Re \left( \int_{0}^{\infty} ds \text{ }   f (-s) e^{is \epsilon_{ j i}} \vert G_{i j} \vert^{2} \right)&=& \pi  \int_{\mathbb{R}^{3}} d^3k \text{ } \vert \phi(k) \vert^{2}  \delta(\epsilon_{ j i}+\omega(k)) \vert G_{i j} \vert^{2}.
\end{eqnarray}

 \noindent The Fermi golden rules  in  \eqref{Fer}  imply that
$$\mathcal{M} \vert \Pi_{i j}  \rangle  =m_{i j}  \vert \Pi_{i j} \rangle$$
 for all  $i \neq j$, with
\begin{equation}
\begin{split}
\label{Re3}
\Re (m_{i j}) &= -   \int_{0}^{\infty} ds \text{ } \Re  (f (s))    (  G_{i i}   -  G_{j j} )^{2}\\
&-\Re \Big(   \sum_{k < i}    \int_{0}^{\infty} ds \text{ }  f (-s)   e^{is \epsilon_{ k i}}  \vert  G_{k i}  \vert^{2} +  \sum_{l < j}  \int_{0}^{\infty} ds \text{ } f(s) e^{is \epsilon_{j l}}  \vert G_{j l}  \vert^{2}  \Big).
\end{split}
\end{equation}
Plugging (\ref{Re1}) and (\ref{Re2}) into (\ref{Re3}), we deduce that 
\begin{equation}
\label{Re4}
\Re (m_{i j}) \leq - \pi \int_{\mathbb{R}^{3}} d^3k  \Big(   \sum_{k < i}  \text{ } \vert \phi(k) \vert^{2}  \delta(\epsilon_{ k  i}+\omega(k)) \vert G_{i k } \vert^{2}  +     \sum_{l < j}   \vert \phi(k) \vert^{2}  \delta(\epsilon_{ j l}-\omega(k)) \vert G_{l  j} \vert^{2} \Big). 
\end{equation}
Equations (\ref{Re4}) and \eqref{Fer} show that $\Re (m_{i j}) <0$  for all $i \neq j$. The eigenvalue $m_{i j}$ of $\mathcal{M}$ associated to the normalized eigenvector $\vert \Pi_{i j} \rangle $ has therefore  a strictly negative real part, for all $i \neq j$. 
\vspace{2mm}

\noindent If $i=j$,
\begin{eqnarray*}
\mathcal{M} \vert \Pi_{i i} \rangle &=& 2 \sum_{i <k }  \int_{0}^{\infty} ds \text{ }  \Re(f (s) e^{is \epsilon_{   k i}}) \vert  G_{k i} \vert^{2}   \vert \Pi_{k k} \rangle -2 \sum_{k < i}     \int_{0}^{\infty} ds \text{ }  \Re(f (-s) e^{is \epsilon_{  k i}}) \vert  G_{k i} \vert^{2}   \vert \Pi_{i i} \rangle,
\end{eqnarray*}
which we can rewrite using \eqref{Re1} and \eqref{Re2} as 
\begin{equation}
\label{Re5}
\mathcal{M} \vert \Pi_{i i} \rangle = 2 \pi \int_{\mathbb{R}^{3}} d^3k  \Big(   \sum_{i <k }    \text{ } \vert \phi(k) \vert^{2}  \delta(\epsilon_{ k  i}-\omega(k)) \vert G_{i k } \vert^{2}  \vert \Pi_{k k} \rangle  -   \sum_{k < i}  \vert \phi(k) \vert^{2}  \delta(\epsilon_{ k i}+\omega(k)) \vert G_{k  i} \vert^{2}   \vert \Pi_{i i} \rangle  \Big). 
\end{equation}

Using \eqref{Re4} and \eqref{Re5}, we represent  $\mathcal{M}$  as a  $n^2 \times n^2$ bloc matrix in the basis $ (\Pi_{i j})$. It takes the form
\begin{equation}
\label{M}
\mathcal{M}=\left( \begin{array}{cc} \mathcal{M}_D&\textbf{0}\\ \textbf{0}  & \mathcal{M}_T \end{array} \right),
\end{equation}
where $\mathcal{M}_D$ is  the  $(n^2-n)\times (n^2-n)$ diagonal matrix
\begin{equation} 
\mathcal{M}_D= \left( \begin{array}{cccc} m_{12}&0&...&0\\0 &m_{13}&0&0\\0&...&^{.}.&0\\0&0&...&m_{n(n-1)} \end{array} \right),
\end{equation}
and $\mathcal{M}_T$ is the lower triangular  $n \times n$ matrix given by 
\begin{equation*}
\label{M'}
\mathcal{M}_T= \left( \begin{array}{ccccc} 0&0&...&...&0\\(\mathcal{M}_T)_{21}&-(\mathcal{M}_T)_{21}&0&...&0\\(\mathcal{M}_T)_{31}&(\mathcal{M}_T)_{32}&-(\mathcal{M}_T)_{31}-(\mathcal{M}_T)_{32}&0&...\\...&...&...&...&0\\(\mathcal{M}_T)_{n1}&(\mathcal{M}_T)_{n2}&...&(\mathcal{M}_T)_{n(n-1)}&-\sum_{i=1}^{n-1} (\mathcal{M}_T)_{ni} \end{array} \right).
\end{equation*}
The coefficients $(\mathcal{M}_T)_{ij}$ are  positive; see \eqref{Re5}. They satisfy $\sum_{j=1}^{i-1}(\mathcal{M}_T)_{ij}>0$  because of the rules \eqref{Fer}.
 \vspace{2mm}
 
\section{Proof of  Proposition \ref{clustt}} \label{AppC}
\noindent We   rewrite  (\ref{DDt2}) as  $$\langle \Psi(t_N) \vert O \Psi(t_N) \rangle=\sum_{\mathcal{G}, \text{ } (N,R) \in \mathcal{G}}   p(\mathcal{G})$$
where the sum is carried out over all  decorated graphs $\mathcal{G}$ on $\{0,...,N\}$  with $(N,R) \in \mathcal{V}(\mathcal{G})$ (see Section \ref{PN}) and, 
\begin{equation}
\label{DDt3}
\begin{split}
 & p(\mathcal{G})= \sum_{k=0}^{\infty}     \underset{[0,t_N]^{2k} }{\int} d\mu_{k} (\underline{w})  \lambda( \underline{w})    \text{ }   \text{ F}(\underline{w})  \chi_{\mathcal{G}}(\underline{w} )   \langle \Pi(t_{-1}) \vert \\
 & \text{ } \mathcal{T}_S \Big( \underset{ (l,B) \in \mathcal{V} (\mathcal{G})}{\prod} \Big[ e^{-it_{l} \mathcal{L}_S} \big( \prod_{(t,r) \in \underline{w} \cap {\bf{I}}_l }  (i G)(t,r) \big) e^{it_{l+1} \mathcal{L}_S} \Big]     \underset{(j,R)  \in  \mathcal{V} (\mathcal{G}) }{\prod} R(t_j)   \underset{(m,\cdot) \notin  \mathcal{V}(\mathcal{G}) }{\prod} P(t_m)  \Big) \vert 1_S \rangle.
 \end{split}
\end{equation} 
The function $ \chi_{\mathcal{G}}$ is used to select pairings $\underline{w}$ whose correlation lines only  join intervals that are linked by an edge in $\mathcal{G}$, and that are such, that, for each edge $((i,B);(j,B)) \in \mathcal{E}(\mathcal{G})$, there is $(u,r;v,r') \in \underline{w}$ with  $u \in I_i$ and $v \in I_j$. We then use that  $P(t_i)$  is a one-dimensional projection and that 
\begin{equation}
\label{RR}
P(t_i) P(t_j)=\vert 1_S \rangle \langle \Pi(t_{j}) \vert.
\end{equation}
Equation \eqref{RR} implies that   $P(t_i) P(t_{i+1})...P(t_j)=\vert 1_S \rangle \langle \Pi(t_{j}) \vert.$ As  mentioned in  Section \ref{secon}, we can  decompose the set $\mathcal{V}(\mathcal{G})$   into maximal blocks of neighboring vertices.   Any   $U \in  \mathcal{U}(\mathcal{G})$ is  surrounded by a projection $P(t_{m(U)})$ on its left,  and  by a  projection $P(t_{\max(U)+1})$ on its right.  Together with (\ref{RR}), this implies that
\begin{equation}
\label{DDt4}
p(\mathcal{G}) =\sum_{k=0}^{\infty}      \underset{[0,t_N]^{2k} }{\int} d\mu_k (\underline{w})  \lambda( \underline{w})    \text{ }\chi_{\mathcal{G}}(\underline{w}) \text{ F}(\underline{w}) \prod_{U \in \mathcal{U}(\mathcal{X})} h_{U}(\underline{w}),
\end{equation}
where $h_U$ has been  defined in \eqref{hU}. By construction of the polymer set $\mathbb{P}_N$,  $\mathcal{G}=\mathcal{X}_1 \vee ...\vee \mathcal{X}_n$  for some $n$, where $\mathcal{X}_i \in \mathbb{P}_N$ and $\text{dist}(\mathcal{X}_i,\mathcal{X}_j) \geq 2$ for all $i \neq j$. Since the polymers $\mathcal{X}_i$ are surrounded by one dimensional projections $P(\cdot)$,  we see using \eqref{RR} that we can factorize
\begin{equation*}
\prod_{U \in \mathcal{U}(\mathcal{X})} h_U(\underline{w})=\prod_{i=1}^{n} \left( \prod_{U \in \mathcal{U}(\mathcal{X}_i)} h_U(\underline{w} \cap \mathcal{V}(\mathcal{X}_i)) \right)
\end{equation*}
for every pairing $\underline{w}$. The integral over pairings can be splitted into the product of $n$ integrals, and we   deduce that 
\begin{equation}
\label{DDt6}
p(\mathcal{G})=  p(\mathcal{X}_1) \text{ ... }p(\mathcal{X}_n).
\end{equation}
\vspace{3mm}

\section{Proofs of Corollaries \ref{normals} and  \ref{ther} (sketch) } \label{sketch}
Some modifications have to be done  to adapt the proof of Theorem \ref{clu} to Corollaries \ref{normals} and \ref{ther}. The construction of connected graphs has to be modified to prove Corollary \ref{normals}. The projection $P=\vert 1_S \rangle \langle \Pi_{11} \vert$ in Lemma \ref{spectrum}  has to be replaced by $P_{\beta}:= \vert 1_S \rangle \langle  \rho_{S,\beta} \vert$ to prove Corollary \ref{ther}, where $\rho_{S,\beta} $ is the Gibbs equilibrium state of  $S$ at temperature $T=1/\beta$.

\subsection{Corollary \ref{ther}}
 The  operator $\mathcal{Z}^{t,s} $ of Section  \ref{S2}  has to be replaced by the new operator $\mathcal{Z}_{\beta}^{t,s} \in \mathcal{B}(\mathcal{B}(\mathcal{H}_S))$ defined in \eqref{tempi}.  The calculations carried out in Sections 1,2,3,4  remain valid with $f(t)$ replaced by   $f_{\beta}(t)$.  The only  difference occurs in Paragraph \ref{2.4.2}, as the Lindbladian $\mathcal{M}$ now depends on the inverse temperature $\beta$. $\mathcal{M}_{\beta}$ is   a block matrix of the form
 \begin{equation}
\label{M2}
\mathcal{M}_{\beta}=\left( \begin{array}{cc} \mathcal{M}_{\beta D}&\textbf{0}\\ \textbf{0}  & \tilde{\mathcal{M}}_{\beta} \end{array} \right),
\end{equation}
where $\mathcal{M}_{\beta D}$ is a $(n^2-n)\times (n^2-n)$ diagonal matrix,
\begin{equation}
\mathcal{M}_{\beta D}= \left( \begin{array}{cccc} m_{\beta 12}&0&...&0\\0 &m_{\beta 13}&0&0\\0&...&^{.}.&0\\0&0&...&m_{\beta n(n-1)} \end{array} \right),
\end{equation}
and $\tilde{\mathcal{M}}_{\beta}$ is  a $n \times n$ matrix,
\begin{equation*}
\tilde{\mathcal{M}}_{\beta}= \left( \begin{array}{ccccc} - \sum_{i \neq 1 } a_{i1}&a_{21}&...&a_{n1}\\e^{\beta \epsilon_{21}} a_{21}&-a_{21} e^{\beta \epsilon_{ 21}} - \sum_{i >2  } a_{i2} &...&a_{n2}\\...&...&...&...\\e^{\beta \epsilon_{n1}} a_{n1}&e^{ \beta \epsilon_{n2}} a_{n2}&...& -\sum_{i=1}^{n-1}e^{\beta \epsilon_{ni}} a_{ni} \end{array} \right).
\end{equation*}
\normalsize
The off-diagonal entries of  $\tilde{\mathcal{M}}_{\beta}$ are positive,  and $\sum_{j \neq i} \tilde{\mathcal{M}}_{\beta;ij}>0$, for all $i=1,...,n$; see \eqref{Fer}.  It is clear that $ \tilde{\mathcal{M}}_{\beta} \vert 1_S \rangle=0$. The reader can check  that  the Gibbs equilibrium state at temperature $T=1/\beta$, $  \rho_{S,\beta}  :=\frac{1}{tr(e^{-\beta H_S})}\sum_{i=1}^{n}   e^{- \beta \epsilon_{i} }   \Pi_{ii} $,  satisfies $$\langle  \rho_{S,\beta} \vert \tilde{\mathcal{M}}_{\beta}=0.$$
 We use a Perron-Frobenius argument to show that  any  $ z \in \sigma(\tilde{\mathcal{M}}_{\beta}) \setminus \lbrace 0 \rbrace$    satisfies  $\Re(z) <0$, and that $0$ is a non-degenerate eigenvalue of $\tilde{\mathcal{M}}_{\beta}$. We introduce the matrix $$\mathcal{M}':=\tilde{\mathcal{M}}_{\beta} + x_M \mathds{1}_{n \times n},$$ where $x_M:= \text{max}_{i \geq 1} (-\tilde{\mathcal{M}}_{\beta;ii})$.  $\mathcal{M}'$ is irreducible non-negative. This  follows from  \eqref{Fer} and from the characterization of irreducible matrices with strongly connected directed graphs; see \cite{Meyer}.  The  theorem  of Perron-Frobenius for non-negative irreducible matrices implies that the maximal eigenvalue of $\mathcal{M}'$ is unique and  that it  is equal to $x_M$ (because $\sum_{j=1}^{n}\tilde{\mathcal{M}}_{\beta;ij}=0$, for all $i=1,...,n$.). Furthermore,   the left- and right eigenspaces of $\mathcal{M}'$ associated to $x_M$ are one-dimensional.  We deduce that $0$ is a non-degenerate eigenvalue of $\tilde{\mathcal{M}}_{\beta} $, and that $   \rho_{S,\beta}   $ is the only   left-eigenvector of $\tilde{\mathcal{M}}_{\beta}$ with associated eigenvalue $0$  and trace one. The rest of the spectrum of $\tilde{\mathcal{M}}_{\beta}$ lies on the left  side of the imaginary axis in the complex plane. The projection $P$ in  Lemma   \ref{spec2}  must be  replaced by $P_{\beta}:= \vert 1_S \rangle \langle  \rho_{S,\beta} \vert$, and the analysis is then completely similar to what has been done in Sections 2,3, and 4.
\vspace{2mm}

\subsection{Corollary \ref{normals}}
We only sketch the modifications that need to be done to adapt the proof presented in Sections 2-4.
\vspace{2mm}

\noindent \textit{Modifications in Section \ref{graphph}}.  Let  $   \varphi_f   = \Phi(f_1)....\Phi(f_{n_0}) \Omega  $ be the initial state of the field. We add a discrete set of points $I_{-1}:=\lbrace \tilde{t}_1, ...,\tilde{t}_{n_0}\rbrace$ to the time axis to represent the contribution of $  \varphi_f  $ to  the Dyson expansion.
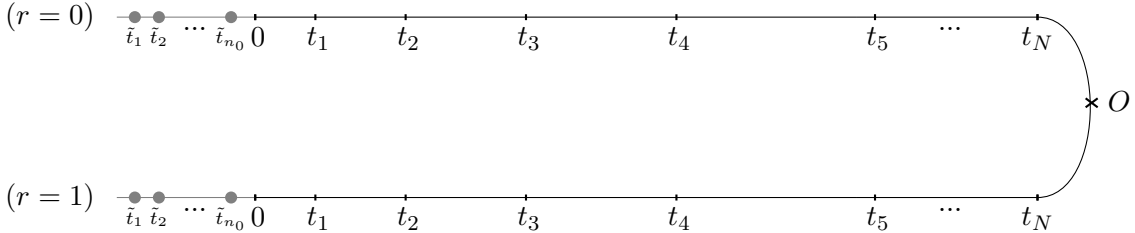
\begin{center}
\begin{tikzpicture}[scale=0.8]
\draw[-] (0,0) -- (7.0,0);
\draw[-] (10,0) -- (13.0,0);
\draw[-,] (0,-3) -- (1,-3);
\draw[-] (1,-3) -- (4.5,-3);
\draw[-] (10,-3) -- (13.,-3.0);
\draw[-] (13,0.0) to[bend left=90] (13.0,-3);

\draw[-](7,0) -- (10.0,0);
\draw[-](4.5,-3) -- (10.0,-3);

\draw[-,thick] (0.0,-0.05) -- (0.0,0.05);
\draw[-,thick]  (13,-0.05) -- (13,0.05);

\draw[-,thick] (0.0,-3.05) -- (0.0,-2.95);
\draw[-,thick]  (13,-3.05) -- (13,-2.95);

\draw[-,thick]  (13.8,-1.5) -- (14,-1.35);
\draw[-,thick]  (13.8,-1.35) -- (14,-1.5);
\draw(14,-1.4) node[right] {$O $};

\draw[-,thin,color=gray] (-2.3,0) -- (0.0,0);
\draw[-,thin,color=gray] (-2.3,-3) -- (0.0,-3);

 \draw (-2.5,0) node[left] {$  (r=0)$};
 \draw (-2.5,-3) node[left] {$(r=1)$};

\draw[-,thick] (0.0,-0.05) -- (0.0,0.05);
\draw[-,thick]  (1,-0.05) -- (1,0.05);
\draw[-,thick]  (2.5,-0.05) -- (2.5,0.05);
\draw[-,thick]  (4.5,-0.05) -- (4.5,0.05);
\draw[-,thick]  (7,-0.05) -- (7.0,0.05);
\draw[-,thick]  (10.3,-0.05) -- (10.3,0.05);
\draw[-,thick]  (13,-0.05) -- (13,0.05);

\draw[-,thick] (0.0,-3.05) -- (0.0,-2.95);
\draw[-,thick]  (1,-3.05) -- (1,-2.95);
\draw[-,thick]  (2.5,-3.05) -- (2.5,-2.95);
\draw[-,thick]  (4.5,-3.05) -- (4.5,-2.95);
\draw[-,thick]  (7,-3.05) -- (7.0,-2.95);
\draw[-,thick]  (10.3,-3.05) -- (10.3,-2.95);
\draw[-,thick]  (13,-3.05) -- (13,-2.95);

 \draw (0.05,0) node[below] {$0$};
 \draw (1.05,0) node[below] {$t_1$};
 \draw (2.55,0) node[below] {$t_2$};
 \draw (4.55,0) node[below] {$t_3$};
 \draw (7.05,0) node[below] {$t_4$};
 \draw (10.35,0) node[below] {$t_5$};
 \draw (11.55,0) node[below] {$...$};
 \draw (13.0,0) node[below] {$t_N$};

 \draw (0.05,-3) node[below] {$0$};
 \draw (1.05,-3) node[below] {$t_1$};
 \draw (2.55,-3) node[below] {$t_2$};
 \draw (4.55,-3) node[below] {$t_3$};
 \draw (7.05,-3) node[below] {$t_4$};
 \draw (10.35,-3) node[below] {$t_5$};
 \draw (11.55,-3) node[below] {$...$};
 \draw (13.0,-3) node[below] {$t_N$};

 \draw (-2,0) node[below] {\tiny $\tilde{t}_1$};
 \draw (-1.6,0) node[below] {\tiny $\tilde{t}_2$};
 \draw (-1,0) node[below] {$ \tiny  ...$};
 \draw (-0.4,0) node[below] { \tiny $\tilde{t}_{n_0}$};
  \draw (-2,-3) node[below] {\tiny $\tilde{t}_1$};
 \draw (-1.6,-3) node[below] {\tiny $\tilde{t}_2$};
 \draw (-1,-3) node[below] {$ \tiny  ...$};
 \draw (-0.4,-3) node[below] { \tiny $\tilde{t}_{n_0}$};

\fill[gray] (-0.4,0) circle (0.1cm);
\fill[gray] (-1.6,0) circle (0.1cm);
\fill[gray] (-2.0,0) circle (0.1cm);
\fill[gray] (-0.4,-3) circle (0.1cm);
\fill[gray] (-1.6,-3) circle (0.1cm);
\fill[gray] (-2.0,-3) circle (0.1cm);

 \end{tikzpicture}
\end{center}

We introduce  7   Feynman rules corresponding to  contractions involving  the field operators $\Phi(f_i)$.
\begin{center}
\begin{tikzpicture}[scale=0.6]

\draw[-,color=gray] (0,0) -- (2.0,0);
\draw[-,color=gray] (4.0,0) -- (6.0,0);
\draw[-,color=gray] (4,-3) -- (6,-3);
\draw[-,color=gray] (9,-3) -- (11,-3);
\draw[-,color=gray] (9,0) -- (11,0);

 \draw (-0.5,0) node[left] {$  (r=0)$};
 \draw (-0.5,-3) node[left] {$(r=1)$};
 
 \draw[decorate, decoration={snake, segment length=1mm, amplitude=0.4mm}] (0.2,0) to[bend left=90](1.5,0);
 \draw[decorate, decoration={snake, segment length=1mm, amplitude=0.4mm}] (4.8,0) to[bend left=10](4.9,-3);
   \draw[decorate, decoration={snake, segment length=1mm, amplitude=0.4mm}] (9.2,-3) to[bend right=90](10.5,-3);

\fill[gray] (0.2,0) circle (0.1cm);
\fill[gray] (1.5,0) circle (0.1cm);
\fill[gray] (4.8,0) circle (0.1cm);
\fill[gray] (4.9,-3) circle (0.1cm);
\fill[gray] (9.2,-3) circle (0.1cm);
\fill[gray] (10.5,-3) circle (0.1cm);

 \draw (0.2,0) node[below] {$\tilde{t}_i$};
 \draw (1.5,0) node[below] {$\tilde{t}_j$};

  \draw (4.8,0) node[above] {$\tilde{t}_i$};
 \draw (4.9,-3) node[below] {$\tilde{t}_j$};

   \draw (9.2,-3) node[above] {$\tilde{t}_i$};
 \draw (10.5,-3) node[above]{$\tilde{t}_j$};
 
  \draw (1,-4) node[below] {\textbf{(a')}};
    \draw (5,-4) node[below] {\textbf{(b')}};
      \draw (9.5,-4) node[below] {\textbf{(c')}};

\end{tikzpicture}
\end{center}
\begin{center}
\begin{tikzpicture}[scale=0.7]

\draw[-] (0,0) -- (2.0,0);
\draw[-] (4.0,0) -- (6.0,0);
\draw[-,] (4,-3) -- (6,-3);
\draw[-] (8,-3) -- (10,-3);
\draw[-] (8,0) -- (10,0);
\draw[-] (12,-3) -- (14,-3.0);

 \draw (-0.5,0) node[left] {$  (r=0)$};
 \draw (-0.5,-3) node[left] {$(r=1)$};
 
 \draw[decorate, decoration={snake, segment length=1mm, amplitude=0.4mm}] (0.2,0) to[bend left=90](1.5,0);
 \draw[decorate, decoration={snake, segment length=1mm, amplitude=0.4mm}] (4.2,0) to[bend left=10](5.5,-3);
  \draw[decorate, decoration={snake, segment length=1mm, amplitude=0.4mm}] (8.2,-3) to[bend left=10](9.5,0);
   \draw[decorate, decoration={snake, segment length=1mm, amplitude=0.4mm}] (12.2,-3) to[bend right=90](13.5,-3);

 \draw (0.2,0) node[below] {$\tilde{t}_i$};
 \draw (1.5,0) node[below] {$v_i$};

  \draw (4.2,0) node[above] {$\tilde{t}_i$};
 \draw (5.5,-3) node[below] {$v_i$};
 
  \draw (8.2,-3) node[below] {$\tilde{t}_i$};
 \draw (9.5,0) node[above] {$v_i$};
 
   \draw (12.2,-3) node[above] {$\tilde{t}_i$};
 \draw (13.5,-3) node[above] {$v_i$};
 
  \draw (1,-4) node[below] {\textbf{(e')}};
    \draw (5,-4) node[below] {\textbf{(f')}};
      \draw (9,-4) node[below] {\textbf{(g')}};
        \draw (13,-4) node[below] {\textbf{(h')}};

\fill[gray] (0.2,0) circle (0.1cm);
\fill[gray] (4.2,0) circle (0.1cm);
\fill[gray] (8.2,-3) circle (0.1cm);
\fill[gray] (12.2,-3) circle (0.1cm);

\end{tikzpicture}
\end{center}
\begin{align}
\textbf{(a')}&:=\langle \Phi(f_i)  \vert  \Phi(f_j)   \rangle, \label{D1} \\
\textbf{(b')}&:=\langle \Phi(f_i)  \vert  \Phi(f_j)   \rangle, \\
\textbf{(c')}&:=\langle \Phi(f_j)  \vert  \Phi(f_i)   \rangle, \\
\textbf{(e')}&:=\lambda(v_i) \langle   \Phi(f_i) \vert    \Phi(\phi(v_i))  \rangle  \textbf{L}(iG(v_i)),   \\
\textbf{(f')}&:=\lambda(v_i) \langle    \Phi(f_i)  \vert   \Phi(\phi(v_i))   \rangle  \textbf{R}(iG(v_i)),\\
\textbf{(g')}&:=\lambda(v_i) \langle    \Phi(\phi(v_i)) \vert  \Phi(f_i)   \rangle  \textbf{L}(iG(v_i)) \\
\textbf{(h')}&:=\lambda(v_i) \langle    \Phi(\phi(v_i)) \vert  \Phi(f_i)   \rangle  \textbf{R}(iG(v_i)) \label{D2}. 
\end{align}
The function $\textbf{F}$ in \eqref{F2}  has to be  modified to take the rules \eqref{D1}-\eqref{D2} into account. The set of  polymers $\mathbb{P}_N$ is constructed from the pairings as  in Paragraph \ref{presum}.  The set of vertices of a polymer can  now  contains the vertex $(-1,B)$, corresponding to contractions with the initial field operators $\Phi(f_{i})$.  When $s=0$, Formula \eqref{DD1'} is replaced by
\begin{equation}
\label{DD13'}
\mathcal{Z}^{t,0}(O)  = \sum_{k=0}^{\infty}  \underset{ (I_{-1} \cup [0,t])^{2k} }{\int} d\mu_k(\underline{w}) \lambda( \underline{w}) \mathcal{T}_S \left[ \prod_{i=1}^{k}   {\bf{F}}(u_{i}, r_{i}; v_{i}, r'_{i})  \right] \left[ O(t) \right],
\end{equation}

The  integral over  $\textbf{I}_{-1}$ is an abuse of notations. It is actually a discrete sum and  the measure $\mu_k(\cdot)$ is modified such that  pairs $(\tilde{t}_{i}, \tilde{r}_i;\tilde{t}_{j},\tilde{r}_j)$   in \eqref{DD13'} are classified  in lexicographic order:   $(\tilde{t}_{i},\tilde{r}_i) < (\tilde{t}_{j},\tilde{r}_j)$  iff $(i,\tilde{r}_i)<(j, \tilde{r}_j)$ in the lexicographic sense.  The sum over isolated intervals sketched  in Section \ref{suba}  remains the same, up to some change of notations similar to \eqref{DD13} $\rightarrow$ \eqref{DD13'}.  The contribution   of the isolated vertex $\textbf{I}_{-1}$ to the Dyson series corresponds to the left multiplication  by the  identity operator,  because $\varphi_f $ is normalized.  The vertex $(-1, \cdot)$ can carry two decorations: it is decorated with a $B$ if  a correlation line starts in $\textbf{I}_{-1}$ and ends in another time interval;  or it is decorated with  a $P$ if $\textbf{I}_{-1}$ is isolated. In the latter case, we set $P(t_{-1}):=\mathds{1}_{\mathcal{B}(\mathcal{B}(\mathcal{H}_S))}$. The integral over  pairings  in \eqref{p(G)}  must be modified as \eqref{DD13} $\rightarrow$ \eqref{DD13'} if $(-1,B) \in \mathcal{V}(\mathcal{X})$. 
\vspace{2mm}

\noindent \textit{Modifications in Section \ref{laco}}
The function $\eta(\mathscr{E})$ has to be modified to  take  edges  that start from $(-1,B)$ into account. We set
\begin{equation}
\label{gLLL}
\eta(\mathscr{E}):= \left\{ \begin{array}{cc} 4 \| G \|^{2} \int_{t_{i}}^{t_{i+1}} du \int_{t_{j}}^{t_{j+1}} dv \vert f(v-u) \vert \lambda(u) \lambda(v) & \text { if  }\mathscr{E}=((i,B); (j,B)), \text{ } i,j \neq -1, \\[7pt] 4 \| G \|  \int_{t_{j}}^{t_{j+1}} dv \vert \langle f_j \vert \phi({v}) \rangle \vert  \lambda(v) & \text{ if } \mathscr{E}=((-1,B); (j,B)).  \end{array} \right.
\end{equation}
The bound  \eqref{borne2} remains true with $e^{4 \tau \| f \|_{L^1} \|G\|^2 \vert \mathcal{B}(\mathcal{X}) \vert}$ replaced by $M_{\tau,n_0}^{ \vert \mathcal{B}(\mathcal{X}) \vert}$,
where
$$M_{\tau,n_0}:= \text{max}(e^{4 \tau \| f \|_{L^1} \|G\|^2} ,C (n_0)) $$ and $C(n_0)>0 $ is a constant  that depends on the absolute values of the scalar products $( f_i , f_j )_{L^2}$ and  $n_0$. The rest of Section \ref{laco} is mainly unchanged (even if we loose a factor $\lambda(\cdot)$ for correlations involving the vertex $(-1,B)$) and the convergence of the cluster expansion as $N \rightarrow \infty$ can be carried out by inspection, following the proofs given in Sections \ref{proo} and \ref{conv}.
\vspace{4mm}

\end{appendix}

\nocite{*}
\bibliographystyle{plain}
\bibliography{main}

\end{document}